\documentclass{ws-ijmpa-arXiv}

\usepackage[super]{cite}
\usepackage{graphicx}
\usepackage{textcomp}
\usepackage{url,units}
\usepackage{slashed}
\usepackage{paralist,xspace,hyperref}
\usepackage[caption=false]{subfig}

\DeclareGraphicsExtensions{.pdf,.png,.jpg}
\graphicspath{{figures/}}

\newcommand{\be}{\begin{equation}}
\newcommand{\ee}{\end{equation}}
\newcommand{\bea}{\begin{eqnarray}}
\newcommand{\eea}{\end{eqnarray}}
\newcommand{\ifb}{\mbox{fb$^{-1}$}\xspace}
\newcommand{\ipb}{\mbox{pb$^{-1}$}\xspace}
\newcommand{\tev}{\mbox{TeV}\xspace}
\newcommand{\gev}{\mbox{GeV}\xspace}
\newcommand{\gd}{\ensuremath{g_{\textrm D}}\xspace}
\newcommand{\dedx}{\mbox{\ensuremath{dE/dx}}\xspace}
\newcommand{\half}{\nicefrac{1}{2}\xspace}

\begin{document}

\markboth{N.~E.\ Mavromatos and V.~A.\ Mitsou}
{Magnetic Monopoles Revisited}

\catchline{KCL-PH-TH/2019-{\bf 94}}{IFIC/20-12}{}{}{}

\title{MAGNETIC MONOPOLES REVISITED: MODELS AND SEARCHES AT COLLIDERS AND IN THE COSMOS}

\author{\footnotesize NICK E.\ MAVROMATOS}

\address{King's College London, Department of Physics, \\Theoretical Particle Physics and Cosmology Group, Strand, London, WC2R 2LS, UK\\
nikolaos.mavromatos@kcl.ac.uk}

\author{\footnotesize VASILIKI A.\ MITSOU}

\address{Instituto de F\'isica Corpuscular (IFIC), CSIC -- Universitat de Val\`encia, \\ 
C/ Catedr\'atico Jos\'e Beltr\'an 2, E-46980 Paterna (Valencia), Spain\\
vasiliki.mitsou@ific.uv.es
}

\maketitle

\pub{Received (Day Month Year)}{Revised (Day Month Year)}

\begin{abstract}
In this review, we discuss recent developments in both the theory and the experimental searches of magnetic monopoles in past, current and future colliders and in the Cosmos. The theoretical models include, apart from the standard Grand Unified Theories, extensions of the Standard Model that admit magnetic monopole solutions with finite energy and masses that can be as light as a few TeV. Specifically, we discuss, among other scenarios,  modified Cho-Maison monopoles and magnetic monopoles in (string-inspired, higher derivative) Born-Infeld extensions of the hypercharge sector of the Standard Model. We also outline the conditions for which effective field theories describing the interaction of monopoles with photons are valid and can be used for result interpretation in monopole production at colliders. The experimental part of the review focuses on, past and present, cosmic and collider searches, including the latest bounds on monopole masses and magnetic charges by the ATLAS and MoEDAL experiments at the LHC, as well as prospects for future searches.
 
\keywords{Magnetic monopoles; electromagnetism; theory; experimental techniques; searches; LHC; ATLAS; MoEDAL; IceCube; ANTARES.}
\end{abstract}

\ccode{PACS Nos.: 14.80.Hv, 12.10.Dm, 11.25.-w, 34.80.Dp, 85.25.Dq, 29.40.Ka, 29.40.Mc}

\tableofcontents

\section{Introduction}

The existence of magnetic poles, that would ``symmetrize'' Maxwell's equations of electromagnetism, has always been an interesting subject that fascinated physicists since the end of the $19^{\rm th}$ Century, although it was Dirac~\cite{Dirac:1931kp,Dirac:1948um} who put it in a modern quantum field theory concept and demonstrated that the existence of magnetic monopoles are consistent with quantum theory, provided a quantization rule is valid that connects the magnetic with the electric charges, in a sort of weak/strong coupling duality \emph{(electromagnetic duality).} The existence of monopoles in Grand Unified gauge Theories (GUTs), as the pioneering work of 't Hooft and Polyakov~\cite{tHooft:1974kcl,Polyakov:1974ek} has demonstrated, prompted Guth~\cite{Guth:1980zm} to discuss their dilution during the inflationary epoch of the Universe, as one of the fundamental problems that the inflationary model would solve.

Despite extensive experimental searches, both of cosmic origin and at colliders, the monopoles (and extensions thereof, such as dyons, carrying both electric and magnetic charges, as suggested by Schwinger~\cite{Schwinger:1969ib}) remain elusive. Unfortunately, in Dirac's model the monopole mass \emph{cannot} be determined. This is because they are merely treated as background sources of magnetic charge without further structure, thus viewed as a new kind of elementary particles~\cite{Vento:2013jua}. On the other hand, in the context of 't Hooft-Polyakov model~\cite{tHooft:1974kcl,Polyakov:1974ek} and its extensions to encompass modern non-Abelian supersymmetric GUT gauge theories~\cite{Rossi:1982fq,Weinberg:2006rq}$^{\text{ and references therein}}$ as well as strings~\cite{Wen:1985qj}, the monopole mass was initially expected to be near the GUT scale with masses in the ballpark of $10^{14}$--$10^{16}$~\gev. This makes their detection in colliders practically impossible~\cite{Fairbairn:2006gg}, leaving as the only possibility the cosmic searches~\cite{Giacomelli:2003yu,Burdin:2014xma,Patrizii:2015uea},  even though difficult, given their dilution during inflation, whose typical Hubble scale lies near the GUT scale. 

However, in the recent literature, there has been systematic attempts to reduce such masses. Detailed models achieving this purpose do exist in several instances, such as topological extensions of the Higgs sector of the Standard Model (SM) of Particle Physics~\cite{Cho:1996qd,Cho:2013vba,Ellis:2016glu}, or extensions of the SM by larger (grand unifying) groups with specific breaking patterns to the SM group~\cite{Kephart:2017esj}, by higher derivatives, say in the hypercharge sector~\cite{Arunasalam:2017eyu,Ellis:2017edi}, or through embeddings in brane/string models~\cite{Arai:2018uoy,Mavromatos:2016mnj,Mavromatos:2018drr}. All such models, as we shall discuss in this work, predict monopoles with masses as light as a few \tev, thus susceptible to the production at current or future colliders. Such models revived experimental collider searches for magnetic monopoles in recent years. The latter, being highly ionizing particles, are searched for alongside other such avatars of new physics~\cite{Acharya:2014nyr}. 

It is the purpose of this review, first to outline the latest developments in the field, by presenting recent theoretical models predicting such light monopoles, and then to discuss current searches, both cosmic and at colliders, by describing the state of the art in search strategies and giving the most recent bounds on the monopole mass, velocity  and magnetic charge, consistent with the current data pointing towards non-observation of magnetic monopoles so far. Detection prospects for future experiments, both cosmic and terrestrial are also given, in view of novel techniques and ideas that have been presented recently, such as thermal (Schwinger-like) production of magnetic monopole--antimonopole pairs from the vacuum at finite temperatures and/or strong magnetic fields (conditions characterizing neutron stars or heavy-ion collisions~\cite{Gould:2017fve,Gould:2017zwi}), as well as studies of experimental data on light-by-light scattering at colliders~\cite{Ellis:2017edi}. From such points of view, this review might then be considered as somewhat complementary to other excellent, and more detailed, reviews existing in the literature on the theory and searches of magnetic charges~\cite{Shnir:2005xx,Rossi:1982fq,Weinberg:2006rq,Milton:2006cp,Rajantie:2012xh,Patrizii:2015uea}, where we refer the interested reader for further study. 

The structure of the article is as follows. In Sec.~\ref{sec:history} we give a brief historical perspective of magnetic sources, which goes as far back as the end of the $19^{\rm th}$ Century. In Sec.~\ref{sec:theory} we discuss detailed theoretical models of magnetic monopoles which imply topologically nontrivial solutions with structure, while Sec.~\ref{sec:pheno} is dedicated to the phenomenological implications of such models. Section~\ref{sc:techniques} provides a brief review of the monopole detection techniques. Sections~\ref{sc:cosmics} and~\ref{sc:colliders} present a historical account of experimental searches of monopoles, of cosmic origin and in colliders, respectively. In the latter, emphasis is given to latest results from the Large Hadron Collider (LHC)~\cite{Evans:2008zzb} experiments ATLAS~\cite{Aad:2008zzm} and MoEDAL~\cite{Pinfold:2009oia}. We describe search strategies and experimental prospects for detection, distinguishing model-independent searches of pointlike (Dirac) monopoles, without structure, but also without underlying concrete theory, from structured solutions characterizing concrete models, whose production at colliders is expected to be strongly suppressed. A summary and an outlook with future prospects is given in Sec.~\ref{sc:outlook}.

\section{Historical Overview \label{sec:history}}	

In the 1873 seminal work of Maxwell on the basic formalism (dynamical equations) unifying Electricity and Magnetism, thus leading to the first (relativistic actually!) Gauge Unified theory, that of Electromagnetism, magnetic monopoles did not appear in the magnetic Gauss' law since Nature appeared to allow isolated electric yet no magnetic charges. Nonetheless, the concept of a magnetic charge, that would ``symmetrize'' Maxwell equations under an ``electric--magnetic duality'', has puzzled and simultaneously fascinated physicists as early as the end of the $19^{\rm th}$ Century. 

\begin{figure}[ht]
\centering
     \subfloat[Birkenland's arrangement\label{fig:birkeland}]{%
       \includegraphics[width=0.8\textwidth]{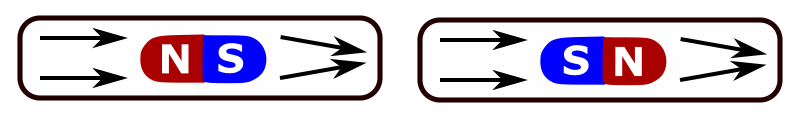}}
       \hfill
     \subfloat[Poincar\'e's case $\Rightarrow$ isolated magnetic poles \label{fig:poincare}]{%
       \includegraphics[width=0.8\textwidth]{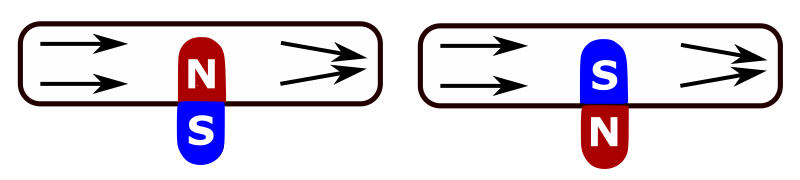}}
\caption{Schematic illustration of the Birkeland experiment~\protect\subref{fig:birkeland} and Poincar\'e's arrangement~\protect\subref{fig:poincare}. Poincar\'e explained the focusing of the cathodic beam in a Crook's tube in the presence of a magnet making use of the geodesics of charged particles under the influence of a magnetic field due to an isolated magnetic pole, which was formally similar to the one produced by a magnetic monopole. \label{fig:birkeland-exp}}
\end{figure}

Pierre Curie was the first to suggest (1894) the existence of a magnetic charge and thus a magnetic current on grounds of symmetry of Maxwell's equations~\cite{Curie:1894}. Soon after, Henri Poincar\'e~\cite{hp1896crh}, in an attempt to explain the result of the Birkeland experiment~\cite{birkeland1,birkeland2} (1896), namely the focusing of the cathodic beams\footnote{This is how physicists called the electrons at the time, since Thomson's demonstration that cathodic beams were electrons came a year later (1897).} in a Crook's tube in the presence of a magnet (see Fig.~\ref{fig:birkeland}), ascribed this effect to the force of a magnetic pole at rest on a moving electric charge, thus postulating in a sense the existence of magnetic monopoles (see Fig.~\ref{fig:poincare}). 

\begin{figure}[htb]
\centerline{\includegraphics[width=4.0in]{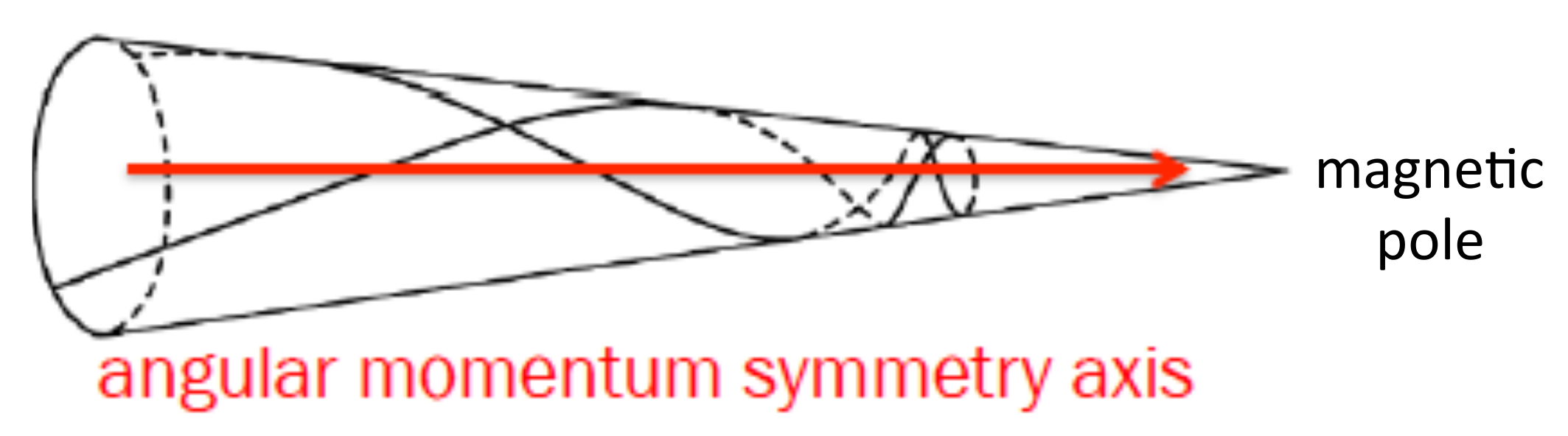}}
\vspace*{8pt}
\caption{The geodesics of charged  particles (continuous and partially dashed curved lines) moving under the influence of the magnetic field due to a magnetic pole lie on a cone, whose symmetry axis coincides with the direction of the total angular momentum of the particles. Poincar\'e used this to explain the focusing of the beam in the arrangement of Fig.~\protect\ref{fig:birkeland}. 
\protect\label{fig:cone}}
\end{figure}

Indeed, the path (geodesic) of the electrons under the influence of such a force lies on an \emph{axially symmetric Poincar\'e cone}, the symmetry axis of the cone being the direction of the angular momentum of the electron (see Fig.~\ref{fig:cone}). For an electron of mass $m$, electric charge $e$ and velocity $\vec v = \frac{d \vec r }{d t} $ (with $\vec r$ the position vector), in the presence of a magnetic field due to an isolated magnetic pole (at rest) with ``magnetic charge'' $g$, the Lorentz force law 
\be\label{magfield}
\vec B_{\rm mono} = g\, \frac{\vec r}{r^3}~,
\ee 
implies an acceleration (in the non-relativistic case, as the analysis was carried out before the discovery of special relativity by Einstein (1905), who, by the way, was motivated by the behavior of Maxwell's equations in different frames):
\be\label{accel}
\frac{d^2\vec r }{d t^2}= \frac{e g}{mc} \frac{1}{r^3} \frac{d\vec r}{d t}\times \vec r ~,
\ee
where $c$ is the speed of light \emph{in vacuo}. The solution for the electron path $\vec{r}(t)$ as a solution of \eqref{accel} then leads to the conical geodesic demonstrated in Fig.~\ref{fig:cone}, and thus to an explanation of the focusing effect in the Birkeland experiment~\cite{birkeland1,birkeland2} (see Fig.~\ref{fig:birkeland-exp}). 

As demonstrated by Thomson a few years later (1904)~\cite{thomson1904}, the total classical angular momentum of the electron in this system is 
\bea
\vec L =  m  \vec r  \times  \frac{d \vec r}{d t} - \frac{e g}{c}  \frac{\vec r}{r}~,
\eea
where the last part on the right-hand side is the contribution due to the interaction of the electron with the magnetic field pole, which is initially expressed as a volume integral involving the Poynting vector~\cite{Shnir:2005xx}
\be\label{poynt}
 \vec L_{eg} = \frac{1}{4\pi c} \int d^3 r^\prime \Big[ \vec r^{\,\prime}  \times \big(\vec E \times \vec B_{\rm mono} \big) \Big] = - \frac{g}{4 \pi  c} \int  d^3 r^\prime \frac{\vec r^{\,\prime}}{r^\prime}  \Big(\vec \nabla^\prime \cdot \vec E \Big) = -\frac{ e g}{c} \frac{\,\vec r}{r}~,
 \ee
where $\vec E$ denotes the electric field of the classical pointlike electric charge of the electron, and in \eqref{poynt} we used integration by parts, assuming vanishing fields at infinity, as well as Maxwell's equation to write $\vec \nabla^\prime \cdot \vec E = 4\pi e  \delta^{(3)}(\vec r - \vec r^{\, \prime})$.

It is worth noting at this stage that by invoking the usual quantization rule of the angular momentum for its electromagnetic field part~\eqref{poynt},  i.e.\ considering the case where the electron and the magnetic monopole are both at rest, $|\vec L_{eg}|$ is required to take integer or half-integer values: 
\be\label{dirac}
e g = \frac{n}{2} \hbar  c ~, \quad n \in {\mathbb Z}~.
\ee
This is nothing other than the Dirac quantization condition (DQC) of the monopole charge (!), derived by Dirac properly more than two decades later (1931), when he presented his quantum (field) theory of relativistic electrons interacting with a magnetic monopole source~\cite{Dirac:1931kp,Dirac:1948um}.

Indeed, in Dirac's theory~\cite{Dirac:1931kp,Dirac:1948um}, the magnetic monopole (corresponding, say, to the ``north magnetic pole'')  finds itself at one end of an infinite solenoid which carries away the magnetic flux (\emph{Dirac string}).\footnote{At this stage we remark that, to make contact with Maxwell's theory, in which an isolated magnetic pole cannot exist, one might think of the monopole solution as an extreme case of a magnet, in which formally an antimonopole (``south pole''), with opposite-sign magnetic charge, lies on the other end of the solenoid at spatial infinity, and thus out of experimental reach, so the magnetic lines emanating from the monopole end on the antimonopole at infinity. This is a set up which allowed condensed matter physicists to talk about construction of ``Dirac monopoles in spin-ice materials''~\cite{Castelnovo:2007qi} by properly creating (through magnetic frustration) ``magnetic-pole-like'' quasiparticle excitations  at the ends of magnetic dipoles (defect--antidefect pairs), separated by relatively large Dirac-like magnetic flux tubes. However, this is \emph{not} the picture envisaged by Dirac. Such spin-ice monopoles are \emph{not}  fundamental magnetic monopoles, given that the Dirac string in the latter is not physical, which leads to the DQC \eqref{dirac}. For a recent article making clear this distinction, see Ref.~\refcite{Bender:2014xva}.} In Dirac's theory, the magnetic monopole is viewed as a \emph{new elementary particle} and the associated Dirac string is not physical, which means that as one goes around it, the (static part of the) quantum wave function $\psi (\vec r)$ of the electron in the static magnetic field of a monopole should be single valued. Dirac associated the magnetic field pole with a \emph{singular} electromagnetic potential $\vec A$, using the standard definition:\footnote{From now on, unless otherwise stated, we work in units where $\hbar = c = 1$, and the electric constant (permittivity) of the vacuum is $4\pi \epsilon_0 =1$.}
\be\label{pot}
\vec B_{\rm mono} = g\, \frac{\vec r}{r^2} = \vec \nabla  \times \vec A (\vec r)~,
\ee
which in general would lead to multivaluedness of the electron wave function, due to the singular nature of $\vec A$ at $r=0$ (monopole center). In fact, it can be easily seen~\cite{Shnir:2005xx} that the solution for $\vec A(\vec r)$ in \eqref{pot} is proportional to a \emph{singular gauge} transformation $U$
\be\label{gauge}
\vec A(\vec r) = (1 + \cos\theta ) \frac{i}{e}  U^{-1} \vec \nabla  U, \quad U=e^{-ie g \phi}~,
\ee
where $\theta$, $\varphi$ are appropriate spherical polar angles. This potential has a singularity along the Dirac string (in the positive $z$ semi-axis, say), which needs to be properly \emph{regularized}. In doing so~\cite{Shnir:2005xx}, one obtains, instead of  \eqref{magfield}, a regularized magnetic field for the monopole
\be\label{regul}
\vec B_\text{mono-reg} = B_{\rm mono} + B_{\rm sing} \equiv g\, \frac{\vec r}{r^3} - 4 \pi g \, \widehat \eta \, \theta (z)  \delta (x)  \delta(y) ~,
\ee
where the unit vector $\widehat \eta = (0,0,1)$ is directed along the $z$ axis (parallel to the Dirac string) and $\theta (z)$ denote the Heaviside function. The importance of using the regularized $\vec B_\text{mono-reg}$ field is that the resulting flux of this field over a closed surface (with the monopole at the center $\vec r=0$) vanishes, as a consequence of a cancellation of the respective contributions from the two terms in the right-hand side of \eqref{regul}, which is a self-consistent result.

We next observe that, as one goes around the Dirac string in a close loop $\ell$, at radial distances $r$ far away from the monopole center, the wave function of the electron changes by a phase as 
\be\label{changes}
\psi(\vec r) \to \psi(\vec r) \, \exp{\left(i e \oint _{\ell} \vec A \cdot d\vec x\right)} = \psi(\vec r) \, \exp{\left(i e \int_{\sigma (\ell)}  d {\mathbf \sigma} \cdot B_{\rm sing} \right) },
\ee
where $B_{\rm sing}$ is the singular part of the magnetic field of the monopole defined in \eqref{regul}, and we used Stoke's theorem, with $\sigma (\ell)$ a surface that has the loop $\ell$ as a boundary. From~\eqref{regul} we obtain $\frac{e}{\hbar  c} \int_{\sigma (\ell)}  d {\mathbf \sigma} \cdot B_{\rm sing} = 4\pi e g$, and thus, singlevaluedness of the wave function in~\eqref{changes}, 
that is the requirement that $\frac{e}{\hbar c} \oint _{\ell} \vec A \cdot d\vec x = 2\pi  n, \: n \in {\mathbb Z}$, implies the DQC \eqref{dirac}. 

For completeness and future use, we now mention that one may write~\eqref{dirac} as:
\be\label{dirac2}
\frac{e^2 g}{\hbar c } = \frac{n}{2} e, \quad  n \in {\mathbb Z} \qquad \Longrightarrow  \qquad g = {68.5}  e  n \equiv \gd  n~,  \quad n \in {\mathbb Z},
\ee 
where 
\be\label{fdc}
\gd \equiv \frac{e}{2 \alpha} = 68.5 e,
\ee
is the so-called \emph{fundamental Dirac magnetic charge unit}, with the quantity $\alpha \equiv \frac{e^2}{\hbar  c} = \frac{1}{137}$ denoting the fine structure constant of electromagnetism, at zero energy scale (in units in which the permittivity of the vacuum is $\epsilon_0 =\frac{1}{4\pi}$). As a consequence of the rule \eqref{dirac}, we observe that the existence even of a single magnetic monopole in the whole Universe would explain the observed \emph{electric charge quantization} in a fundamental way.\footnote{Of course, the latter property might also be a consequence of some group structure of a GUT without monopoles, which the Standard Model of Particle Physics is embedded to, although standard GUTs are known to contain monopoles, as we shall review below~\cite{tHooft:1974kcl,Polyakov:1974ek}.}

The presence of the (non-local) Dirac string singularity, even if invisible after the DQC \eqref{dirac2}, has important consequences when one attempts to formulate effective field theories for magnetic monopoles. As we shall discuss later on in the review, its presence leads to apparent Lorentz-invariance violations, when one is calculating, for instance,  the scattering of an electric charge off a magnetic monopole via a one-photon-exchange graph~\cite{Weinberg:1965rz}. To avoid using Dirac strings, Cabibbo and Ferrari~\cite{Cabibbo:1962td}, Salam~\cite{Salam:1966bd} and Zwanziger~\cite{Zwanziger:1970hk} developed a two-potential formulation of a classical field theory for magnetic monopoles. In particular, Salam postulated that the second potential corresponds to a ``photon'' with different $C$ (charge conjugation) and $P$ (parity) characteristics as compared to the normal photon. In the approach of Zwanziger~\cite{Zwanziger:1970hk}, extra conditions have been imposed on among these two potentials, so that the physical degrees of freedom correspond to only those of the real electromagnetic photon, as observed in Nature. Subsequently, Singleton~\cite{Singleton:1995cc,Singleton:2011ru} discussed spontaneous symmetry breaking (SSB) \`a la Higgs in  the context of the two-potential formulation of Dirac monopoles (which can also be applied to the Abelian Wu-Yang monopoles, discussed below), with the phenomenologically interesting conclusion that a massive dual (``dark'') photon emerges from the formalism, under some proper phase transitions. This could be of relevance to dark matter searches, in which dark photons are represented by such SSB massive  ``magnetic'' photons, as we shall discuss in Sec.~\ref{sc:dark}.

\begin{figure}[htb]
\centerline{\includegraphics[width=0.58\textwidth]{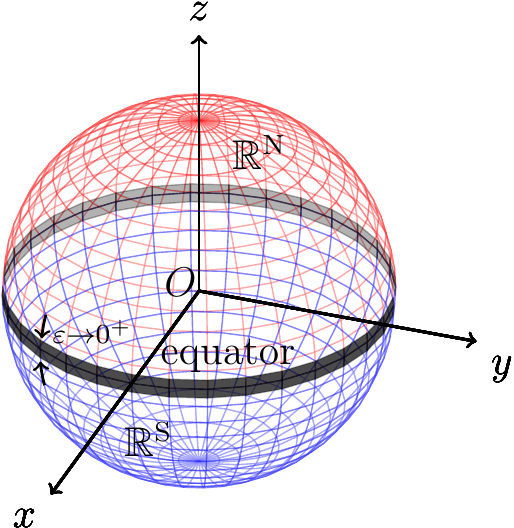}}
\caption{Patchwise coverage of the three-space ${\mathbb R}^3/\{0\}$ surrounding a Wu-Yang magnetic monopole, located at the origin $O$. The gauge potentials $\mathbf A$ are well defined on each patch ($\mathbf A^{\rm N}$ in red and $\mathbf A^{\rm S}$ in blue) and on the equator (black band) ${\mathbb R}^{\rm N}  \cap {\mathbb R}^{\rm S}$ (of width $\varepsilon \to 0^+$), where they are connected to each other by a gauge transformation.}
\label{fig:wy}
\end{figure}

Wu and Yang~\cite{Wu:1975es,Wu:1976ge}, developed an elegant pointlike magnetic monopole solution of the field equations for the $SU(2)$ Yang-Mills gauge theory, {\it without} Dirac strings. The solution is characterized by a spherically symmetric magnetic field, which for all distances $r$ measured from the origin, where the monopole is located, is given by the form \eqref{magfield}. This solution is based on differential geometry concepts that characterize gauge theories. To arrive at their string-less construction, Wu and Yang exploited the fact that the direction of the Dirac string is ambiguous, in the sense that it is defined only up to a gauge transformation. In this sense, one can avoid the Dirac-string singularity by parametrizing the three-space surrounding the monopole, excluding the origin, ${\mathbb R}^3/\{0\}$, by two overlapping hemispheres, a northern ${\mathbb R}^{\rm N}$ and a southern ${\mathbb R}^{\rm S}$ (see Fig.~\ref{fig:wy}). The ``equator'' is defined as the overlap region ${\mathbb  R}^{\rm N} \cap {\,\mathbb R}^{\rm S}$. Gauge potentials are well defined on the hemispheres, and are regular on the equator, where they are connected through non-Abelian gauge transformations. Wu and Yang also proposed a {\it classical} Lagrangian for describing the non-quantum-mechanical interaction of their monopole with a pointlike charge via the electromagnetic field, where any strings are absent~\cite{Wu:1976qk}.

There is a subtlety, however, in the non-Abelian Wu-Yang monopole, in that its magnetic charge cannot be easily identified. Being a solution of a pure Yang-Mills gauge theory, without a Higgs field, it does not possess a topological charge that could be interpreted as a magnetic charge, in contrast to the 't Hooft-Polyakov monopole~\cite{tHooft:1974kcl,Polyakov:1974ek}, as we shall discuss later. Another subtlety, which was recognized by Wu and Yang themselves~\cite{Wu:1975es}, concerns the non-satisfaction of the Bianchi identity for the electromagnetic field strength at the origin, which would imply that the corresponding gauge field does not behave as a proper gauge field there. 

Such subtleties have only been partially addressed in the past~\cite{Lanyi:1977kp} and it is only recently that the authors of Ref.~\refcite{Constantinidis:2016wjo} provided a resolution of the Bianchi identity issue via the postulation of a magnetic point source at the origin. The latter is required for the existence of well-defined solutions of the pertinent Yang-Mills differential field equations everywhere in space. By ``regularizing the singularity'' of the Wu-Yang monopole at the origin, through appropriate application of distribution theory to the Yang-Mills $SU(2)$ system, these authors also managed to identify a dynamically conserved non-vanishing magnetic charge. Indeed, they demonstrated that the magnetic field $\mathbf B$ of the Wu-Yang monopole solution of the Yang-Mills theory should satisfy 
\be\label{WYreg}
\vec{\mathcal D} \cdot \mathbf B  = - \frac{1}{e} \frac{\vec r \cdot {\vec T}} {r^3} \, \delta (r) .
\ee
In the above formula, $\delta{(r)}/r^2 $ denotes the radial part of the three-dimensional $\delta$-function distribution $\delta^{(3)}(\vec r)$; 
 $T^i, \, i=1,2,3$, are the $SU(2)$ generators; $\mathcal D_\mu (\dots) = \partial_\mu (\dots) + i e [ A_\mu\, ,  \,  \dots ]$ is the $SU(2)$ gauge covariant derivative; and $e$ is the $SU(2)$ coupling, which, as evidenced from \eqref{WYreg}, plays the role of the ``electric'' coupling, and its inverse is proportional to the magnetic charge, satisfying the DQC \eqref{dirac2}. The magnetic field $\mathbf B$ is defined through $B_i  = -\frac{1}{2} \epsilon_{ijk}\, F^{jk}$, $i,j,k=1,2,3$, where $\epsilon_{ijk}$ is the totally antisymmetric symbol in three dimensions, $\epsilon_{123}=+1$, etc., and $F^{\mu\nu} = \partial_\mu A_\nu - \partial_\nu A_\mu + i\, e [A_\mu, \, A_\nu]$ is the $SU(2)$ Yang-Mills strength. 

It is important to note that, in order to arrive at \eqref{WYreg},  the authors of Ref.~\refcite{Constantinidis:2016wjo} made use of the so-called integral form of the Yang-Mills equations~\cite{Ferreira:2012aj,Ferreira:2011ed}, which involve a second type of a ``global magnetic field'', $B_i^W$, in addition to the local one $B_i$ mentioned above. The ``global'' field $B_i^W$ is defined through Wilson loop operators $W$, $B_i^W = W\, B_i \, W^{-1}$, $i=1,2,3$, where $W$ is defined on a path originating from a reference point in space and ending at the point where $B_i$ is evaluated. It is the use of this global field that helps identifying a magnetic charge for the non-Abelian Wu-Yang monopole. It is {\it not} possible to identify the magnetic charge of the non-Abelian Wu-Yang monopole using only the local form of the magnetic field $B_i$. One has to use the definition of dynamically conserved charges that appear in the integral form of the Yang-Mills equations, which then reduce to the form \eqref{WYreg}, after appropriate regularization, using distribution function theory.

To demonstrate the basic features of the Wu-Yang solution, and the way the DQC \eqref{dirac2} arises, once a magnetic charge is defined, it suffices to consider explicitly the Abelian version of the Wu-Yang monopole~\cite{Wu:1976ge}, where the magnetic charge definition is straightforward~\cite{Shnir:2005xx}. In that case, the magnetic field assumes the radial form \eqref{magfield} at any point in space, at distance $r$ from the origin, and thus it is singular at the origin. As already mentioned, the electromagnetic potentials are well defined separately on each of the two hemispheres of Fig.~\ref{fig:wy} ($\mathbf A^{\rm N}$, $\mathbf A^{\rm S}$), and they are also regular in the overlapping region, ${\mathbb  R}^{\rm N} \cap {\mathbb R}^{\rm S}$, where they are connected to each other by a gauge transformation.

The DQC for the Abelian Wu-Yang monopole is obtained~\cite{Shnir:2005xx,Wu:1976ge} by noting that the wave function of an electron with charge $e$, in the presence of a point source for the singular magnetic field \eqref{magfield}, is defined patchwise. Considering a closed loop $\ell$ that lies entirely in the ``equator region'' of Fig.~\ref{fig:wy}, the phase of the electron wave function is $e \oint_\ell d\ell \cdot \mathbf A^{\rm S,N} $.  By applying Stokes's theorem in each patch, we may write (taking into account the 
opposite orientation of the normal vectors of the surface elements $d\mathbf S$ of the two hemispheres) :
\bea
e \oint_\ell d\ell \cdot \mathbf A^{\rm N} &=& e \int_{{\mathbb R}^{\rm N}}\, d{\mathbf S} \cdot (\nabla \times \mathbf A^{\rm N}) = 
e \int_{{\mathbb R}^{\rm N}} \, d{\mathbf S} \cdot \mathbf B, \nonumber \\
e \oint_\ell d\ell \cdot \mathbf A^{\rm S} &=&-e \int_{{\mathbb R}^{\rm S}} \, d{\mathbf S} \cdot (\nabla \times \mathbf A^{\rm S}) = 
-e \int_{{\mathbb R}^{\rm S}} \, d{\mathbf S} \cdot \mathbf B, 
\eea
where $\mathbf B$ is the magnetic field of the monopole. The action $S$ is therefore defined up to a term
\be\label{actionWY}
\Delta S = e \int_{\mathbb R^{\rm N} \cup \, \mathbb R^{\rm S}} d{\mathbf S} \cdot \mathbf B = \int_V\, d^3 r \, \nabla \cdot \mathbf B = 4\pi e  g,
\ee
where in the last two equalities we used Gauss's law over the entire space volume $V$, and Maxwell's equations for the magnetic field \eqref{magfield} generated by the monopole. Dirac's charge quantization rule \eqref{dirac2} is then obtained by requiring that the action change \eqref{actionWY} should not affect any physical observables, which means that one must have (in units $\hbar =1$):  $\Delta S = 2\pi  n, \; n \in \mathbb Z$.
Alternatively, one may arrive at the same result by considering the gauge dependence of the electron wave function in a monopole field~\cite{Shnir:2005xx,Wu:1976ge}.
 
Schwinger postulated in 1969 the existence of {\it dyons}, which are hypothetical elementary particles that carry {\it both} electric and magnetic charge~\cite{Schwinger:1969ib}. As Schwinger demonstrated, a system of two dyons, with electric and magnetic charges $q_{e\,i}$, and $g_{m\, i}$, $i=1,2$, respectively, is consistent with quantum mechanics if the following {\it (Schwinger) quantization rule for the charges} is obeyed:
\be\label{dyon}
q_{e 1} \, g_{m 2} - q_{e 2} \, g_{m 1}  = - m^\prime  ~, \quad m^\prime \in {\mathbb Z}.
\ee
Notice that the dyon has a fundamental magnetic charge \emph{twice} the Dirac charge $2\gd$. A dyon with a zero electric charge yields the magnetic monopole, while a dyon with zero magnetic charge corresponds to an ordinary electrically charged particle. However, in such a case, to obtain the DQC \eqref{dirac} from \eqref{dyon} one has to take $m^\prime$ half integer. In this review, we shall restrict ourselves to magnetic monopoles and shall not discuss dyons in detail.

Unfortunately, in Dirac's or Schwinger's models, the mass and spin of the magnetic monopole or dyon cannot be estimated, given that the respective configuration is viewed only as a source for the singular magnetic field, with no attempt to discuss detailed dynamics. Knowledge of the monopole or dyon mass is an essential missing ingredient in order to assess the producibility of such objects in principle at colliders~\cite{Milton:2006cp}. 

't Hooft and Polyakov (in 1974)~\cite{tHooft:1974kcl,Polyakov:1974ek}, independently, were the first to give such a microscopic derivation of a monopole solution within the context of concrete gauge theories with simply connected gauge groups, in the presence of Higgs fields, and explain the topologically nontrivial nature of their solutions. Such monopoles have structure and exhibit quite interesting properties,  especially if embedded in supersymmetric theories, where some exact results can be derived~\cite{Rossi:1982fq,Weinberg:2006rq}. It is the point of this work to review briefly  such microscopic constructions, giving emphasis to those aspects relevant for experimental searches. 't Hooft-Polyakov monopoles characterize GUT models, and as such have masses near the GUT scale, outside the reach of colliders, thus susceptible only to cosmic searches. However, recently, attempts to lower significantly their scale within appropriate GUT models have been considered~\cite{Kephart:2017esj}.  Then we shall move on to discuss other detailed solutions predicting light monopoles with masses that can be sufficiently light (e.g.\ at \tev scales) so as to be of relevance to colliders, such as the modified Cho-Maison monopole~\cite{Cho:1996qd}, that can exist in SM extensions~\cite{Cho:2013vba,Ellis:2016glu,Arunasalam:2017eyu,Ellis:2017edi,Arai:2018uoy}, or some string-inspired monopoles in models involving pseudoscalar fields (axions) carrying an axion charge proportional to the magnetic charge. The latter are self-gravitating and have some interesting properties in that the asymptotic spacetime is almost Minkowski but with a conical defect~\cite{Mavromatos:2016mnj,Mavromatos:2018drr}. As we shall discuss, this may lead to interesting effects in the scattering of light off such structures~\cite{Mazur:1990ak,Ren:1993hr,BezerradeMello:1996si,RoderiguesSobreira:1998tb,Lousto:1991rh,Mavromatos:2017qeb}, which imply a special role of such monopoles as ``lenses''~\cite{Mavromatos:2017qeb}, of potential relevance to cosmic searches for such defects. 

\section{Theoretical Models of Magnetic Monopoles \label{sec:theory}}

We next proceed to review concrete models, embedded in specific theories beyond the Standard Model, including strings, which admit magnetic monopole solutions. As already mentioned, in this review we shall restrict ourselves to solutions with magnetic charges only, leaving out dyons. Our discussion will be brief, focusing only on those aspects of the solution relevant for the main theme of this review, which is the relevance of these configurations for collider physics and cosmic searches. More detailed reviews on the theory of such objects can be found in the current literature~\cite{Rossi:1982fq,Weinberg:2006rq}, where we refer the interested reader.

\subsection{Magnetic monopoles in non-Abelian gauge theories \label{sec:hp}}

The first example of a detailed monopole solution, which is a classical solution to the equations of motion of the pertinent Lagrangian, has been presented by 't Hooft  and Polyakov independently~\cite{tHooft:1974kcl,Polyakov:1974ek}, for gauge theories on \emph{compact} groups exhibiting spontaneous symmetry breaking via a Higgs field.  
Such solutions are known as the \emph{'t Hooft-Polyakov (HP) monopole}.
't Hooft used as an explicit example the so-called Georgi-Glashow $SU(2)$ gauge theory model~\cite{Georgi:1974sy}, where the symmetry breaking is implemented by means of an adjoint Higgs triplet field. The case of such $SU(2)$ gauge monopoles can be  generalized to  phenomenologically realistic GUTs, such as $SU(5)$ gauge theory models~\cite{Rossi:1982fq,Weinberg:2006rq}. The GUT monopoles have masses near the GUT scale ($\sim 10^{14}$--$10^{15}$~\gev), relevant for cosmic searches, which though are expected to have been diluted by inflation~\cite{Guth:1980zm}. 

We commence our discussion on such monopole solutions with the Georgi-Glashow $SU(2)$ monopole of 't Hooft, which encompasses the basic topologically nontrivial features we want to emphasize here, that can be generalised to the GUT monopole cases, to be discussed later. The fields in the HP $SU(2)$ gauge model~\cite{tHooft:1974kcl,Polyakov:1974ek}  are a scalar field $\phi^{a}(t,\vec{x})$ and a gauge field $A_{\mu}^{a}(t,\vec{x})$, where $a =1,2,3$ is an $SU(2)$ index. The Lagrangian density $\mathcal{L} (t,\vec{x})$ is 
\be\label{eq:1}
\mathcal{L}(t,\vec{x}) = -\frac{1}{4}F_{\mu\nu}^{a}F^{a\mu\nu} + \frac{1}{2}\left(D_{\mu}\phi^{a}\right)\left(D^{\mu}\phi^{a}\right) -\frac{1}{4}\lambda\left(\phi^{a}\phi^{a}-\eta^{2}\right)^2.
\ee
The field tensor $F_{\mu\nu}^{a}$ is given by
\be\label{eq:2}
F_{\mu\nu}^{a} = \partial_{\mu}A_{\nu}^{a}-\partial_{\nu}A_{\mu}^{a}+\tilde g\epsilon^{abc}A_{\mu}^{b}A_{\nu}^{c},
\ee
where $\epsilon^{abc}$ is the antisymmetric Levi-Civita symbol. The covariant derivative $D_{\mu}\phi^{a}$ is defined as 
\be\label{eq:3}
D_{\mu}\phi^{a} = \partial_{\mu}\phi^{a}+ \tilde g\epsilon^{abc}A_{\mu}^{b}\phi^{c}.
\ee
The parameters of the model are $\tilde g,\lambda(>0)$ and $\eta$.  The equations of motion derived from $\mathcal{L}$ read
\be\label{eq:4}
D_{\mu}F^{a\mu\nu} = \tilde g\epsilon^{abc}\left(D^{\nu}\phi^{b}\right)\phi^{c}
\ee
and 
\be\label{eq:5}
D_{\mu}D^{\mu}\phi^{a}=-\lambda\left(\phi^{b}\phi^{b}\right)\phi^{a}+\lambda\eta^{2}\phi^{a}.
\ee
For the static solutions of \eqref{eq:4} and \eqref{eq:5}, pertinent to the monopole configuration, in the gauge $A_{0}^{a}(\vec{x})=0$, the following ans\"atze are assumed
\be\label{eq:6}
\phi^{a}(\vec{x}) = \delta_{ia}\left(\frac{x^{i}}{r}\right)\, F(r), 
\quad
A_{i}^{a}(\vec{x}) = \epsilon_{aij}\left(\frac{x^{j}}{r}\right)W(r),
\ee
where $a,i,j=1,2,3$ and $r=\left|\vec{x}\right|$. The boundary conditions adopted read
\be\label{eq:8}
F(r) \rightarrow \eta \quad \text{and} \quad W(r) \rightarrow 1/\tilde gr
\ee
as $r\rightarrow\infty.$ In 't Hooft's and Polyakov's solution  one has
\be
\tilde grW(r) = 1-\frac{r \tilde g\eta}{\sinh\left(\tilde g\eta r\right)} \quad \text{and} \quad \tilde grF(r) \sim \frac{r \tilde g\eta}{\tanh\left(\tilde g\eta r\right)}-1.\label{eq:9}
\ee
 The electromagnetic field tensor $f_{\mu\nu}$ is defined as~\cite{tHooft:1974kcl,Polyakov:1974ek} 
\be
f_{\mu\nu}=\hat{\phi}^{a}F_{\mu\nu}^{a}-\frac{1}{g}\epsilon^{abc}\hat{\phi}^{a}D_{\mu}\hat{\phi}^{b}D_{\nu}\hat{\phi}^{c},
\label{eq:10}
\ee
where $\hat{\phi}^{a}=\phi^{a}/|\vec{\phi}|$ and $|\vec{\phi}|=\left(\sum_{a=1}^{3}\phi^{a}\phi^{a}\right)^{1/2}$, which defines a ``sphere'' in the internal $SU(2)$ group space.
The magnetic field strength, $B_{k}=\frac{1}{2}\epsilon_{kij}f_{ij}$, which has the asymptotic behavior 
\be
\vec{B}(\vec{x}) \rightarrow \frac{\vec{x}}{\tilde g\, r^{3}} \quad \text{as} \quad r\rightarrow\infty ,
\label{eq:11}
\ee
corresponds to a magnetic monopole of magnetic charge  $1/\tilde g$. Moreover, as $r \to \infty$, where $\phi^a \to \eta \frac{x^a}{r}$ \eqref{eq:8}, one can show that~\cite{Shnir:2005xx}
\be
\frac{1}{2}\epsilon_{\mu\nu\rho\sigma}\partial^{\nu}f^{\rho\sigma}=\frac{1}{2g}\epsilon_{\mu\nu\rho\sigma}\epsilon_{abc}\partial^{\nu}\hat{\phi}^{a}\partial^{\rho} \, \hat{\phi}^{b} \partial^{\sigma} \, \hat{\phi}^{c}\equiv \frac{k_{\mu}}{g},\label{eq:12}
\ee
where $k_{\mu}$ is a topological current. As a consequence of the \emph{simple-connectedness} of the non-Abelian gauge group $SU(2)$, the topological charge $Q=\int d^{3}x\, k_{0}$
is quantized to be an integer $n^\prime$ and the monopole magnetic charge is
\be\label{hpcharge}
g_{\rm HP}= \frac{n^\prime}{\tilde g}\, , \quad  n^\prime \in {\mathbb Z}~.
\ee
The HP monopole has $n^\prime=1$.  Upon the identification of the coupling $\tilde g$ with the electron charge $e$, \eqref{hpcharge} becomes Schwinger's quantization rule \eqref{dyon} with $n^\prime=m^\prime = 1$, corresponding to a fundamental magnetic charge which is twice the Dirac charge $\gd$ \eqref{dirac2}. However, 
if one considers spin-\half representations of the $SU(2)$ group of the Georgi-Glashow model~\cite{tHooft:1974kcl}, which 
describe particles with charges $q =\pm\frac{1}{2}\, e$, the condition \eqref{hpcharge} becomes the DQC \eqref{dirac}, $q g = \frac{1}{2}$, with integer $n=1$.

It should be noted that $f_{_{\mu\nu}}$ as defined in \eqref{eq:10} does \emph{not} satisfy the Bianchi identity.  For completion we mention at this point that, one can construct a  version $f_{\mu\nu}^{\rm reg}$ of the 't Hooft electromagnetic tensor which, unlike \eqref{eq:10}, is not singular at the zeros of the Higgs triplet, and is finite everywhere~\cite{Shnir:2005xx}, 
\begin{align}\label{newdef}
F_{\mu\nu}^a = f_{\mu\nu}^{\rm reg}  \frac{\phi^a}{\eta} \, \Rightarrow \,  f_{\mu\nu}^{\rm reg} = \frac{\phi^a}{\eta} F_{\mu\nu}^a & = \partial_\mu {\mathcal A}_\nu - \partial_\nu {\mathcal A}_\mu + \frac{1}{\eta^3\, g} \epsilon_{abc}\, \phi^a \partial_\mu \phi^b \partial_\nu \phi^c ~, \nonumber \\ \phi^a \phi^a & = \eta^2.
\end{align}
The presence of the vector potential ${\mathcal A}_\mu$ owes its existence to the fact~\cite{Shnir:2005xx} that a general solution of the equation $D_\mu \phi^a =0$ for $\phi^a \phi^a = \eta^2$ implies: $A_\mu^a = \frac{1}{\eta^2 g} \epsilon_{abc}\, \phi^b\, \partial_\mu \phi^c + \frac{1}{\eta} \phi^a \, {\mathcal A}_\mu$, with ${\mathcal A}_\mu$ an arbitrary four-vector, which can be identified with the electromagnetic potential $A_\mu^{\rm em}$, since for $\phi^a \phi^a = \eta^2$, one has ${\mathcal A}_\mu = A_\mu^a \frac{\phi^a}{\eta}$. Upon substituting in \eqref{eq:2}, then, 
one obtains the expression \eqref{newdef} for $f_{\mu\nu}^{\rm reg}$. Both definitions, \eqref{eq:10} and \eqref{newdef}, coincide at the spatial boundary $r \to \infty$. In the topologically trivial sector, the $\phi$-dependent terms on the right-hand side of the definition of $f_{\mu\nu}^{\rm reg}$ in \eqref{newdef} vanish, and thus one obtains the standard expression for the ``electromagnetic'' field strength $f_{\mu\nu}^{\rm reg} $ in terms of regular gauge potentials. The Bianchi identity is satisfied in that case. However, in the presence of monopoles, one obtains a violation of the Bianchi identity for the dual of $f_{\mu\nu}^{\rm reg}$, \eqref{eq:12}.  This is a consequence of the fact that the electromagnetic tensor can be formally expressed in terms of singular potentials at the monopole center, $r \to 0$, using  the construction of Ref.~\refcite{Halpern:1978ik}.

One of the most important, and relevant for our purposes in this review, properties of the HP monopoles is that their total energy ((rest) mass  ${\mathcal M}$, in flat spacetime)  is \emph{finite}. Indeed, in flat space times, where the HP monopole was considered initially~\cite{tHooft:1974kcl,Polyakov:1974ek}, the total energy $E$ (Hamiltonian) functional of this \emph{static} configuration is minus the integral of the Lagrangian density $\mathcal L$ over space: 
\be
E = -\int d^3 x \, {\mathcal L}~.
\ee
For the HP solution~\cite{tHooft:1974kcl,Polyakov:1974ek} this leads to a \emph{finite} expression 
\be\label{hpmass}
E =  {\mathcal M} = \frac{1}{\tilde g^2} \, m_V~, \quad m_V^2 = \tilde g^2 \, v^2 ~,
\ee
where $m_V$ is the mass of the (massive) gauge bosons of the Georgi-Glashow (or, in general, the massive gauge bosons of the generalized gauge theory over a compact gauge group), and $v^2$ is the Higgs field vacuum expectation value.

\begin{figure}[htb]
\centerline{\includegraphics[width=0.58\textwidth]{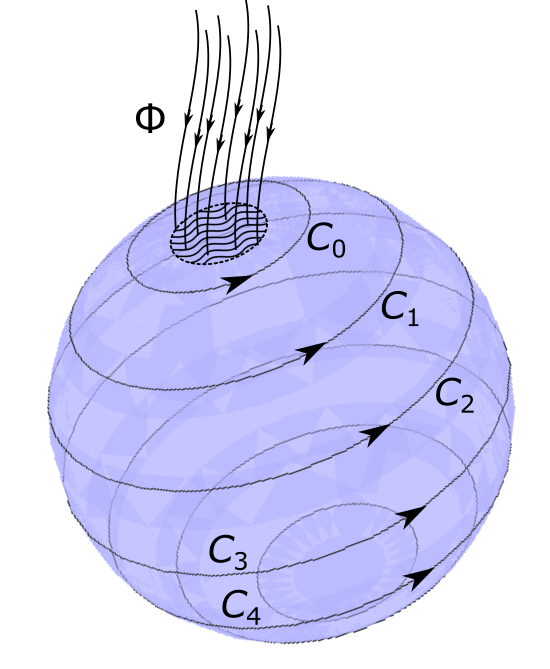}}
\caption{A magnetic flux ending at a monopole located on the north pole of a sphere in the case of a non-Abelian gauge group with compact cover. The contour $C_0$ that surrounds the monopole can be displaced continuously to $C_1$ and then the $C_1$ is moved smoothly until it shrinks at the south pole, assuming there is no singularity there. There is no other point of the sphere where the flux $\Phi$ exists if the gauge group is a compact non-Abelian group. \protect\label{ndstr}}
\end{figure}

We remark at this stage that in the case of monopoles arising in compact gauge groups there is \emph{no} Dirac string. This can be understood easily, as demonstrated by 't Hooft~\cite{tHooft:1974kcl,Polyakov:1974ek}, by considering a magnetic flux $\Phi$ ending on a monopole which is located on the north pole of a sphere (see Fig.~\ref{ndstr}). Consider a contour $C_0$  surrounding the monopole. Then, along the contour, by using Stokes theorem, one would generate a gauge vector potential $\mathbf{A}$ such that $\oint d\vec \ell \cdot \mathbf{A} = \Phi $. This potential is obtained as a pure gauge from a \emph{multivalued} ``large'' gauge transformation $\Lambda$,  $\mathbf{A} = \mathbf{\nabla } \, \Lambda$. All fields, $\varphi $ transform under this transformations as $\varphi \rightarrow \varphi \, e^{i n  \Lambda}$, and hence the requirement of their single-valuedness results in the flux $\Phi $ being an integer multiple of $2\pi $, which implies a full gauge rotation along the contour $C_0$ in Fig.~\ref{ndstr}.  

If the gauge group were Abelian, then the flux $\Phi$ should come out from another spot on the sphere, because in that case the rotation $2\pi n$, with $n \in {\mathbb Z}$, cannot change into a constant as we displace the contour towards the south pole of the sphere. However in a non-Abelian theory, with a gauge group that has a compact cover, it is possible to change continuously the $2\pi n$ gauge rotation so as it becomes a constant, and in this case one can displace the contour $C_0$ in Fig.~\ref{ndstr} to $C_1$ and then continue to move it until it shrinks at the south pole, where there is no singularity. Hence, in such non-Abelian theories, as is the $O(3)$ case of the Georgi-Glashow model, where the initial HP monopole was considered~\cite{tHooft:1974kcl,Polyakov:1974ek}, the associated monopole solutions are regular solutions of the field equation and no Dirac strings emerge. The quantization \eqref{hpcharge} then is a general feature of compact groups of $SU(N)$, which follows from the homotopy property $\Pi_N (SU(N)) = {\mathbb Z}$. In the HP case, the homotopy property is $\Pi_2(SU(2))$ and the integer $n$ defines topological sectors, corresponding to how many times a sphere $S_2$ wraps around the internal sphere defined by the gauge group $SU(2)$. 

\subsection{Grand Unified Theory monopoles}\label{sc:gut}

Unfortunately, the Georgi-Glashow model discussed above does not correspond to the phenomenologically correct model describing Nature. That role is played by the Glashow-Weinberg-Salam SM formulated on the \emph{non-compact} gauge group $SU(3)_c \otimes SU(2)_L \otimes U_Y(1)$, and as a result, according to our discussion in the previous session, finite-energy monopoles are not expected (see, however, discussion below for some SM extensions maintaining the group structure, where such solutions exist). 

Embedding the SM group to larger Grand Unified Theory groups, e.g.\ $SU(5)$, implies the existence of GUT monopoles. The estimate of their mass parallels that of the Georgi-Glashow model, discussed in the previous section, and hence,  the finite energy of the GUT monopole is given by \eqref{hpmass}, but now $\tilde g$ is the GUT gauge group coupling, and $m_V$ is a vector boson mass of a spontaneously broken GUT theory, much heavier than the electroweak gauge bosons. Typical such scales of GUT theories lie in the range of $10^{14}$--$10^{16}~\gev$. Hence such monopoles are out of the production reach of current colliders, and one can only search cosmically for them~\cite{Patrizii:2015uea}, but the reader should bear in mind that their density will be extremely dilute due to inflation~\cite{Guth:1980zm}, whose Hubble scale lies in the above range. Otherwise, without inflation, a GUT universe would be overclosed, by a huge monopole energy density, typically $10^{11}$ times larger than the critical density of the Universe (!),  $\rho_{\rm mon} \simeq 10^{11} \rho_c$. 

We now remark that the existence of monopoles in such GUT requires only that the $U(1)$ of electromagnetism is embedded in a compact larger group, so the pertinent gauge subgroup, resulting from a breaking of a GUT group,  could even have a cross-product structure. This may lead to significantly lighter monopoles than those in a typical GUT $SU(5)$ theory. An example is provided by the following symmetry-breaking pattern of an initial $SO(10)$ GUT group:
\bea
SO(10) \; \xrightarrow{10^{15}~{\rm GeV}}  \; SU(4) \otimes SU(2) \otimes SU(2) \; \xrightarrow{10^{9}~{\rm GeV}} \; SU(3)_c \otimes SU(2) \otimes U_Y(1)~, \nonumber \\
\eea
where the first breaking of the GUT symmetry group $SO(10)$  into the cross product group $SU(4) \otimes SU(2) \otimes SU(2)$, which contains magnetic monopoles, occurs at GUT scales $10^{15}~\gev$, close to the GUT and inflationary scales, but the second one, where the  $SU(4) \otimes SU(2) \otimes SU(2)$ group breaks into the SM group, occurs at much lower scales  $10^{9}~\gev$. In view of \eqref{hpmass}, when properly applied to this case, one expects the resulting monopole mass to be sufficiently low, of order $10^{9}{\tilde g}$, where $\tilde g$ is the coupling of the  $SU(4) \otimes SU(2) \otimes SU(2)$ group. In this way, one may obtain much lighter monopoles than the original GUT models predicted. There are several examples of such cross-product gauge groups with intermediate-scale monopoles, some of them carrying  multiple quanta of the Dirac magnetic charge, as the ones~\cite{Kephart:2017esj} obtained from intersecting-D-brane-inspired trinification models with gauge groups $SU(3)_c \otimes SU(3)_L \otimes SU(3)_R$, which in addition to monopoles, also predict color-singlet states, possibly accessible at LHC. Whether one can lower these monopoles mass down to a few \tev is not entirely clear, nonetheless such low mass monopoles may not be diluted significantly during inflation, and hence one may have interesting cosmological constraints, arising from requiring absence of their potential contributions to nucleosynthesis, etc. Cosmic searches for such relatively low-mass monopoles  do not differ from the corresponding searches of GUT-mass-scale relativistic ones~\cite{Patrizii:2015uea}, discussed in Sec.~\ref{sc:cosmics}. 

\begin{figure}[htb]
\centerline{\includegraphics[width=0.5\linewidth]{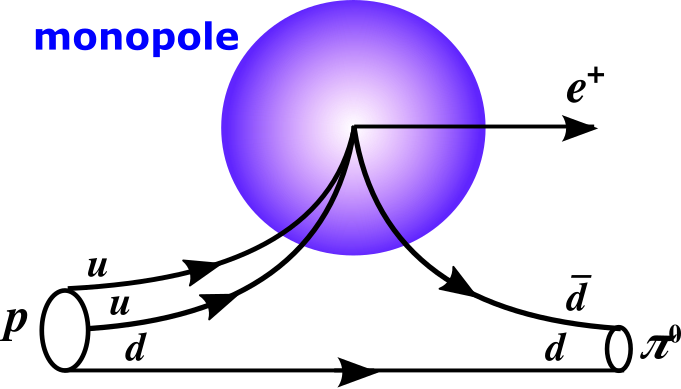}}
\caption{Schematic illustration of monopole catalysis of proton decay $p \rightarrow e^+ + \pi^0$, via the Callan-Rubakov mechanism, whereby baryon number violation is mediated by super heavy gauge bosons of the pertinent GUT theory in the presence of a monopole. \label{cr}}
\end{figure}

An important physical effect of a GUT monopole, which defines some of the corresponding search strategies, is the fact that in its presence proton decay is \emph{catalyzed} (see Fig.~\ref{cr}), due to the induced baryon number non-conservation, the so-called \emph{Callan-Rubakov effect}~\cite{Rubakov:1981rg,Rubakov:1983sy,Callan:1982ac,Preskill:1984re}$^{\text{ and references therein}}$.  A few years before the work of Refs.~\refcite{Rubakov:1981rg,Rubakov:1983sy,Callan:1982ac}, Dokos and Tomaras~\cite{Dokos:1979vu} have already noticed that scattering of fermions off GUT $SU(5)$ monopoles or dyons can catalyze processes involving baryon-number non-conservation as a result of the property of $SU(5)$-monopole dyonic excitations to possess baryon-number-violating couplings. However, the authors of Ref.~\refcite{Dokos:1979vu} thought that the dyonic excitation would split from the monopole ground state by an amount equal to the $\alpha M$, where $\alpha$ is the fine structure constant of electromagnetism, and $M$ the mass of the monopole. For monopoles of GUT scale mass $M \sim 10^{15}$--$10^{16}~\gev$, this would imply $\alpha M \gtrsim 10^{13}~\gev$, and hence such baryon-number-violating processes would be extremely suppressed.  This issue was rectified in the work of Callan and Rubakov~\cite{Rubakov:1981rg,Rubakov:1983sy,Callan:1982ac}, and also of Blaer, Christ and Tang~\cite{Blaer:1981ps} and of Wilczek~\cite{Wilczek:1981du}, who have demonstrated that such dyonic excitations are in fact split from the vacuum by an amount of order $\alpha m_f$, where $m_f$ is the mass of the charged scattered fermion off the monopole, e.g.\ electrons. This is a tiny amount of energy, and hence these dyons are easily excited. This was the consequence of the fact that the \emph{anomalous} electric charge $Q=-\frac{e\vartheta}{2\pi}$, that was known~\cite{Witten:1979ey} to characterize monopoles in theories with $CP$ violation due to a vacuum angle $\vartheta $, such as $SU(5)$ GUT theories, vanishes when the fermion mass vanishes, since such an electric charge is smeared over a region of radius of order $m_e^{-1}$. 

Rubakov~\cite{Rubakov:1981rg,Rubakov:1983sy,Callan:1982ac} also stressed the fact that because of the \emph{axial anomaly} the monopole was not an eigenstate of chirality or baryon number, which opens the possibility for large baryon-number- and chirality-violating cross sections in fermion-monopole scattering. Callan on the other hand~\cite{Rubakov:1981rg,Rubakov:1983sy,Callan:1982ac} emphasized that there are \emph{baryon-number-violating boundary conditions}, when one regularizes the monopole solution by a finite monopole-core region, and then let the size of the core going to zero. Such boundary conditions, he argued, reflect the nontrivial physics inside the monopole core and lead to unsuppressed baryon-number-violating effects (cross sections) for the scattering of fermions off the monopole. As emphasized by Preskill~\cite{Rubakov:1981rg,Rubakov:1983sy,Callan:1982ac}, these are two different explanations of the Callan-Rubakov effect, rather than complementary explanations of the same phenomenon. In the literature, most references to the ``Callan-Rubakov effect''  associate it with the boundary conditions.

As a result of the monopole-induced baryon-number-violating processes, proton decay can be catalyzed in the presence of GUT monopole,  in the sense that protons, with a relative velocity $\beta=v/c$ with respect to the monopole, can de induced to decay with a cross section of order $\sigma_B \beta \sim 10^{-27}\, {\rm cm}^{-2}$, giving a line of catalyzed proton decays on the trail of the monopole.  One can thus search for non-relativistic monopoles at water/ice detectors using catalysis, as we shall discuss later in Sections~\ref{sc:callan} and~\ref{sc:nucl-decay}. 

\subsection{The Cho-Maison (electroweak) monopole in the Standard Model\label{cm}}

\begin{figure}[htb]
\centerline{\includegraphics[width=0.65\linewidth]{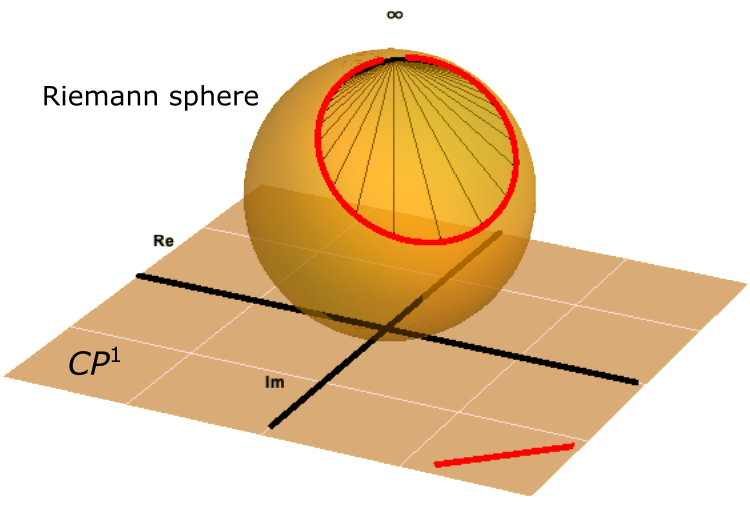}}
\caption{Schematic visualization of how the complex projective line ($CP^1$) is obtained from the Riemann sphere (that is, the complex plane with an added point at infinity $\infty$) through a ``stereographic projection''. The normalized Higgs field in the Cho-Maison (CM) parametrization of the SM plays the role of a complex coordinate with values lying on the $CP^1$ space. Image generated and adapted from Ref.~\protect\refcite{grattoni}.\protect\label{fig:cp1}}
\end{figure}

As mentioned previously, magnetic monopoles were in general not expected in the SM, based on the fact that the associated quotient gauge group $SU_c(2) \otimes SU_L(2) \otimes U_Y(1))/U_{\rm em}(1)$  in the symmetry-broken phase is non-compact and has a trivial second homotopy. However, Cho and Maison argued~\cite{Cho:1996qd} that there is an alternative topological scenario, which has been overlooked, and which justifies the existence of monopoles in the SM. This is associated with the fact that the normalized Higgs doublet field can be viewed as a $CP^1$ field, which is known to be characterized by a nontrivial second homotopy $\Pi_2 (CP^1) = \mathbb Z$. We remind the reader that $CP^1$ is the complex projective line, obtained by a ``stereographic projection'' of the Riemann sphere --- that is the complex plane plus a point at infinity (see Fig.~\ref{fig:cp1}). 

However, the initial CM solution was characterized by infinite energy, due to singular terms at the monopole center  $r \to 0$. Below we shall review the CM monopole and its properties, and see how its initially singular-energy form can be regularized in appropriate extensions of the SM to yield finite-energy solutions. In this sense, although the aforementioned argument of CM in identifying topological reasons for having monopole solutions in scenarios utilizing the standard Higgs doublet of the SM is technically correct, nonetheless, finiteness of the monopole energy requires extensions of the SM Lagrangian. Such extensions, however, keep the SM gauge group structure intact, not requiring embedding in larger compact groups. This, as we shall see, has dramatic consequences in yielding light monopoles, with masses in a range accessible to current or future colliders. 

The CM electroweak monopole~\cite{Cho:1996qd} is a numerical solution of the Weinberg-Salam theory.\footnote{An analytical existence theorem for  such monopole solutions can be established by appropriately adopting arguments by Yang~\cite{Yang:2001bb}.} However, it suffers from a divergence in the energy due to a singularity at the center of the configuration, $r \to 0$, where $r$ is the radial coordinate. Therefore it cannot be considered as physical in the absence of a suitable ultraviolet completion. Also this singularity makes any estimate of the monopole mass impossible, and hence a regularization of the solution is needed before its phenomenological relevance is considered. Cho, Kim and Yoon (CKY) proposed~\cite{Cho:2013vba}  a modification of the electroweak theory that could render the divergence of $F_{\mu\nu}^a$ integrable at the monopole core. This would yield a finite-energy solution that would be physical, but as we shall discuss later on, their modification is not compatible with LHC data phenomenology. There are other modifications, though, which are compatible~\cite{Ellis:2017edi,Arunasalam:2017eyu,Arai:2018uoy}, some of them inspired from string/brane theory, as we shall review them, together with that of Ref.~\refcite{Cho:2013vba}, in the next section. 
 
To be instructive, below we review first the CM monopole solution~\cite{Cho:1996qd}. The pertinent Lagrangian is the one describing the bosonic sector of the Weinberg-Salam electroweak theory, 
\begin{align}\label{LWS}
\mathcal{L} &= -|D_\mu H|^2 - \frac{\lambda}{2}\left(H^\dagger H - \frac{\mu^2}{\lambda}\right)^2 - \frac{1}{4}F^a_{\mu\nu}F^{a\, \mu\nu} - \frac{1}{4}B_{\mu\nu}B^{\mu\nu}  \nonumber \\
&= -\frac{1}{2}(\partial_\mu \rho)^2 - \frac{\rho^2}{2}|D_\mu \xi|^2 - \frac{\lambda}{8}\left(\rho^2 - \rho_0^2\right)^2  
- \frac{1}{4}F^a_{\mu\nu}F^{a\, \mu\nu} - \frac{1}{4}B_{\mu\nu}B^{\mu\nu} \, ,
\end{align}
where the $SU(2)_{L} \times U(1)_{Y}$ gauge-covariant derivative is defined as 
\begin{equation*}
D_\mu \equiv \partial_\mu - i\frac{g}{2}\tau^a A^a_\mu - i \frac{g^\prime}{2}B_\mu \, ,
\end{equation*}
with $A^a_\mu$, $a=1, 2, 3$, the $SU(2)$ non-Abelian gauge field; $F_{\mu\nu}^a$ the corresponding field strength; $B_\mu$ ($B_{\mu\nu}$) the Abelian hypercharge $U_Y(1)$ gauge field (field strength); and $H$ the Englert-Brout-Higgs doublet, responsible for the standard spontaneous symmetry breaking $SU(2) \times U_Y(1) \, \rightarrow \, U_{\rm em}(1)$, with $U_{\rm em}(1)$ the Abelian gauge group of electromagnetism. In the second line of \eqref{LWS} this is written as $H = \frac{1}{\sqrt{2}}\rho\xi$, where $\xi^\dagger \xi = 1$, and we define $\rho_0 = \sqrt{2}\mu^2/\lambda = \sqrt{2}v$. Cho and Maison noticed that the $U(1)_Y$ coupling of $\xi$ is essential for its interpretation as a $CP^1$ field with nontrivial second homotopy, making possible a topologically stable monopole solution of the equations of motion. 

Choosing the following ansatz for the fields in spherical coordinates $(t, r, \theta, \varphi)$~\cite{Cho:1996qd},
\bea\label{ans1}
\rho & = & \rho(r),  \quad  \xi=i\left(\begin{array}{cc} \sin (\theta/2)~e^{-i\varphi}\\
- \cos(\theta/2) \end{array} \right),   \nonumber \\
\vec A_{\mu} & = & \frac{1}{g} A(r)\partial_{\mu}t~\hat r + \frac{1}{g}(f(r)-1)~\hat r \times \partial_{\mu} \hat r, \nonumber\\
B_{\mu} & = & \frac{1}{g'} B(r) \partial_{\mu}t  -\frac{1}{g'}(1-\cos\theta) \partial_{\mu} \varphi,
\eea
one can find spherically symmetric field configurations corresponding to electroweak monopoles and dyons. Notice that \emph{there is} a Dirac string for the CM monopole, which manifests itself in an apparent singularity along the $z$-axis in $\xi$ and $B_\mu$ fields, which however is a gauge artefact and can be removed by an appropriate gauge transformation~\cite{Cho:1996qd}. 

After an appropriate unitary gauge transformation $U$ such that $\xi \to U\xi = \begin{pmatrix}0 \\ 1\end{pmatrix}$, one may obtain the physical gauge fields by rotating through the weak-mixing angle $\theta_W$, 
\begin{align}\label{ans2}
W_{\mu} &=\dfrac{i}{g}\frac{f(r)}{\sqrt2}e^{i\varphi} (\partial_\mu \theta +i \sin\theta_W \partial_\mu \varphi), \nonumber\\
A_{\mu}^{\rm EM} &= e\left( \frac{1}{g^2}A(r) + \frac{1}{g'^2} B(r) \right) \partial_{\mu}t  
- \frac{1}{e}(1-\cos\theta) \partial_{\mu} \varphi,  \nonumber \\
Z_{\mu} &= \frac{e}{gg'} \big(A(r)-B(r)\big) \partial_{\mu}t,
\end{align}
where the electric charge $e = g \sin \theta_W = g^\prime \cos \theta_W$. The simplest nontrivial solution to the equations of motion with $A(r) = B(r) = f(r) = 0$ and $\rho=\rho_0 \equiv \sqrt{2}\mu/\sqrt{\lambda}$ describes a charge $4\pi/e$ point monopole with
\begin{equation*}
A_\mu^{\rm EM} = - \frac{1}{e}(1 - \cos\theta)\partial_\mu\varphi \, .
\end{equation*}

More general electroweak dyon solutions may be obtained by considering nonzero $A, B$ and $f$, e.g.\ with the boundary conditions
\bea\label{bc0}
&\rho(0)=0, \quad f(0)=1, \quad A(0)=0, \quad B(0)=b_0, \nonumber\\
&\rho(\infty)=\rho_0, \quad f(\infty)=0, \quad A(\infty)=B(\infty)=A_0,
\eea
where $0 \leq A_0 \leq e\rho_0$ and $0 \leq b_0 \leq A_0$, we may integrate numerically the equations to obtain solutions representing the CM dyon with electric $q_e$ and magnetic $q_m$ charges
\bea\label{eq:Charge}
&q_e=-\dfrac{8\pi}{e}\sin^2\theta_W \int_0^\infty f^2 A\, dr = \dfrac{4\pi}{e} A_1, \nonumber\\
&q_m = \dfrac{4\pi}{e},
\eea
where $A_1$ is a constant coefficient parametrizing the $1/r$ asymptotic behavior of $A$. Notice that, compared with the Dirac quantization rule \eqref{dirac2}, the CM solution corresponds to a magnetic charge equal to \emph{twice} the fundamental  charge, $2\gd$, as appropriate for the fact that the solution can also incorporate dyons. Since there is, as already mentioned, a Dirac string in the CM monopole, in contrast to the HP monopole, the former is a kind of \emph{hybrid} between the Dirac and HP monopoles. 

However, as already mentioned, the CM electroweak monopole~\cite{Cho:1996qd} suffers from a \emph{non-integrable singularity} in the energy density at the center of the configuration ($r \to 0$). This can be seen by calculating the total energy $E$ of the dyon  configuration, which has the form~\cite{Cho:2013vba}:
\begin{gather}\label{eq:energyint}
E = E_0 +E_1,  \nonumber \\
E_0 = 4\pi\int_0^\infty \frac{dr}{2 r^2} \left\{\frac{1}{g'^2}+ \frac1{g^2}(f^2-1)^2\right\}, \nonumber\\
E_1 = 4\pi \int_0^\infty dr \, \Bigg\{ \frac12 (r\dot\rho)^2 +\frac1{g^2} \left(\dot f^2 +\frac{1}{2}(r\dot A)^2 + f^2 A^2 \right) \nonumber \\
+\frac{1}{2g'^2}(r\dot B)^2 +\frac{\lambda r^2}{8}\left(\rho^2-\rho_0^2 \right)^2 
+\frac14 f^2\rho^2 +\frac{r^2}{8} (B-A)^2 \rho^2 \Bigg\} .
\end{gather}

We see that, with the boundary conditions given by \eqref{bc0}, $E_1$ is finite, but the first term of $E_0$ is divergent at the origin. For future use, we remark that for the realistic parameters of the electroweak theory, including the observed Higgs mass, one has 
\be\label{e1}
E_1 \simeq 4.1~\tev~.
\ee
We next proceed to discuss potential regularizations of $E_0$, that would allow for an estimate of the total monopole energy ((rest) mass), in order to examine the relevance of the solution to collider physics. This requires extending the SM Lagrangian \eqref{LWS} appropriately.

\subsection{Finite-energy Cho-Maison-like monopoles and dyons \label{finite}}

There are two known ways to achieve this, which both involve extensions of the SM Lagrangian, keeping however the gauge group structure $SU_c(3) \otimes SU_L(2) \otimes U_Y(1)$ intact. The first, proposed by CKY~\cite{Cho:2013vba,Cho:2019vzo}, concerns a modification of the Weinberg-Salam theory in such a way that the form of the dielectric ``constant'' in front of the $U(1)_Y$ hypercharge gauge kinetic term (``vacuum permittivity'') changes, by some unspecified dynamics, to become a nontrivial functional of the Englert-Brout-Higgs (EBH) doublet field, $\epsilon(H^\dagger H)$. Such a construction preserves gauge invariance. The second, proposed by Arunasalam and Kobakhidze~\cite{Arunasalam:2017eyu}, involves an extension of the hypercharge sector of the SM Lagrangian to a Born-Infeld-type term, which is inspired by string theory considerations. In both cases, the gauge group of the SM is kept intact. Below we proceed to discuss these approaches briefly and derive the order of magnitude of the monopole mass, in order to assess the feasibility of its production at colliders. 

\subsubsection{Hypercharge vacuum permittivity as a Higgs-dependent function}

Specifically, CKY considered the following form of effective Lagrangian that has a non-canonical kinetic term for the $U(1)_Y$ gauge field 
\be\label{eq:Leff}
{\cal L}_\text{eff} = -|D _\mu H|^2 - \frac{\lambda}{2} \left(H^\dagger H -\frac{\mu^2}{\lambda}\right)^2 
-\frac{1}{4} \vec F_{\mu \nu}^2  - \frac{1}{4} \epsilon \left(\frac{|H|^2}{v^2} \right) B_{\mu \nu}^2,
\ee
where $\epsilon(|H|^2/v^2)$ is a positive dimensionless function of the EBH doublet that approaches one asymptotically as $|H| \to v$. Clearly $\epsilon$ modifies the permittivity of the $U(1)_Y$ gauge field, but the effective action still retains the $SU(2)\times U(1)_Y$ gauge symmetry. Moreover, since $\epsilon \rightarrow 1$ asymptotically, the effective action reproduces the SM when the EBH field adopts its canonical vacuum expectation value: $|H| = v$. However, the factor $\epsilon (|H|^2/v^2)$ effectively changes the $U(1)_Y$ gauge coupling $g'$ to a ``running" coupling $\bar g'=g' /\sqrt{\epsilon}$ that depends on $|H|$. This is because, with the rescaling of $B_\mu \to B_\mu/g'$, $g'$ changes to $g' /\sqrt{\epsilon}$. By choosing $\epsilon$ so that $\bar g' \to \infty$ as $|H| \to 0$, i.e.\ requiring $\epsilon$ to vanish at the  origin, one can regularize the CM monopole. 

Such an {\it ad~hoc} modification of the Standard Model is phenomenologically motivated as a way to render finite the energy integral, leading to a finite mass for the electroweak monopole. We leave open the question of how such a modification may occur in a ``top-down'' approach, and pursue the question how light such a CKY monopole might be. To this end, we first notice that the original proposal~\cite{Cho:2013vba} for regulating the infinite-energy divergence was to consider a functional form 
\be\label{eq:originalepsilon}
\epsilon \simeq \left(\frac{\rho}{\rho_0}\right)^n , \quad n > 4 + 2\sqrt{3}~,
\ee
where the restrictions on $n$ are imposed in order for certain terms in the equations of motion to vanish fast enough as $r \to 0$, so that the energy remains finite. With the boundary conditions \eqref{bc0} the solution at the origin behaves as~\cite{Cho:2013vba}
\begin{align*}
\rho \simeq c_\rho r^{\delta_{-}} , \qquad f \simeq 1 + c_f r^2 , \\
A \simeq c_A r , \qquad B \simeq b_0 + c_B r^{2\delta_{+}} ,
\end{align*} 
where $\delta_{\pm} = \frac{1}{2}(\sqrt{3} \pm 1)$, and behaves asymptotically towards spatial infinity ($r \to \infty$) as 
\begin{align*}
\rho \simeq \rho_0 + \rho_1 \frac{\exp\left(-\sqrt{2}\mu r\right)}{r} , \quad f \simeq f_1 \exp(-\omega r) , \\
A \simeq A_0 + \frac{A_1}{r} , \quad B \simeq A + B_1 \frac{\exp(-\nu r)}{r} ,
\end{align*}
where $\omega = \sqrt{(g\rho_0)^2/4 - A_0^2}$ and $\nu = \frac{1}{2}\sqrt{g^2 + {g^\prime}^2}\rho_0$. 

These behaviors of the fields in the limits can be used together with the pertinent equations of motion to obtain numerical solutions~\cite{Cho:2013vba}. Plugging the simplest $A=B=0$ solution into the energy integral \eqref{eq:energyint}, with the appropriate $\epsilon$ form factor regularization, leads~\cite{Ellis:2016glu} to a monopole mass of $\sim 5.7~\tev$. The nonzero $A, B$ solution yields a larger mass of $\sim 10.8~\tev$ for the dyon. An increase was to be expected, since non-vanishing forms of $A$ and $B$ will always contribute positively to the $E_1$ integral \eqref{eq:energyint}. The topological stability of the lowest-lying monopole is guaranteed by the conservation of magnetic charge~\cite{Cho:1996qd}. However, dyon solutions may be unstable if suitable decays into charged particles and a monopole are kinematically accessible, as is the case in this example.

Unfortunately, however, as demonstrated in Ref.~\refcite{Ellis:2016glu}, the simple power-law  functional form for the $\epsilon$ regulator \eqref{eq:originalepsilon}, chosen in Ref.~\refcite{Cho:2013vba},  is phenomenologically \emph{excluded by data} on Higgs decays to two photons ($H \, \rightarrow \, \gamma \gamma$)~\cite{TheATLASandCMSCollaborations:2015bln}. Indeed, dimension-six operators involving couplings of the Higgs field with the SM gauge sector have been studied~\cite{Ellis:2014jta,Ellis:2014dva},  in an analysis of the data now available from the LHC. Among them, of interest to us here is the operator
\bea\label{eq:cgamma}
\frac{c_\gamma}{\Lambda^2}\mathcal{O}_\gamma \equiv \frac{\bar{c}_\gamma}{M_W^2}{g^\prime}^2|H|^2B_{\mu\nu}B^{\mu\nu}  ,
\eea
where we use the notation of Ref.~\refcite{Ellis:2014jta,Ellis:2014dva} in which constraints are placed on $\bar{c}_\gamma \equiv c_\gamma M_W^2/\Lambda^2$. Based on a global fit to LHC data, mainly from the decay of the Higgs field $H \to \gamma \, \gamma$, the best fit values of $\bar{c}_\gamma $ are in the range of $10^{-3}$ and negative~\cite{Ellis:2014jta,Ellis:2014dva}.

By expanding $\rho$ near its vacuum expectation value $\rho_0 \equiv \sqrt{2}\mu/\sqrt{\lambda}$:
\be\label{rhoexp}
\rho = \rho_0 + \tilde \rho, \qquad \tilde \rho/\rho_0 \ll  1 ,
\ee
we can write the term \eqref{eq:cgamma} as an effective Lagrangian contribution of the form
\begin{align} \label{eq:effcgamma}
 \frac{\bar{c}_\gamma}{M_W^2}{g^\prime}^2|H|^2B_{\mu\nu}B^{\mu\nu} \supset 8\left(\frac{g^\prime}{g}\right)^2\bar{c}_\gamma\frac{\tilde\rho}{\rho_0}B_{\mu\nu}B^{\mu\nu} .
\end{align}

On the other hand, the $\epsilon$-dependent modification \eqref{eq:originalepsilon} of the Lagrangian \eqref{eq:Leff}, when expanded around the vacuum expectation value, yields a term
\begin{align} \label{eq:effepsilon}
-\frac{1}{4}\left(\frac{\rho}{\rho_0}\right)^2B_{\mu\nu}B^{\mu\nu} \supset -\frac{n}{4}\frac{\tilde\rho}{\rho_0}B_{\mu\nu}B^{\mu\nu} ,
\end{align}
where we recall  \eqref{eq:originalepsilon} that finiteness of the monopole total energy/mass then requires for a simple power law that ${\mathbb Z^+} \ni n \ge 8$. 

Comparing~\eqref{eq:effcgamma} with~\eqref{eq:effepsilon}, we see that to linear order in $\tilde{\rho}/\rho_0$, 
\begin{align*}
\bar{c}_\gamma = -\frac{1}{32}\left(\frac{g}{g^\prime}\right)^2 n \simeq -0.1 n  .
\end{align*}

Since $n \geq 8 \Rightarrow \bar{c}_\gamma \lesssim -0.8$ is strongly excluded by the $95\%$-confidence-level observed value $\bar{c}_\gamma \gtrsim 10^{-3}$~\cite{Ellis:2014jta,Ellis:2014dva}, it follows~\cite{Ellis:2016glu} that the simple power-law modification of the $U(1)_Y$ permeability proposed in Ref.~\refcite{Cho:2013vba} cannot be valid all the way from the origin of the EBH field  $\rho \to 0$  up to the region near the expectation value, $\rho \simeq \rho_0$. 

A modification of the SM Lagrangian is, therefore, required of the general form in \eqref{eq:Leff}, but with the $U(1)_Y$ permeability $\epsilon$ an interpolating functional having the following properties~\cite{Ellis:2016glu}:
\bea \label{eq:constraints}
&& \epsilon(\rho) > 0 ~, \nonumber \\
&& \epsilon(\rho)|_{\rho =0} = \epsilon^{(1)} (\rho)|_{\rho =0}= \dots = \epsilon^{(n-1)}(\rho)|_{\rho =0} = 0,   \nonumber \\
&& \epsilon^{(n)}(\rho)|_{\rho = 0} = \frac{n!}{\rho_0^n}  \ne 0, \quad  {\mathbb Z^+}  \ni  n \ge 8~,\nonumber \\
&& \epsilon(\rho)|_{\rho \to  \rho_0} \simeq 1 - 16 \, \overline c_\gamma \left(\frac{g^\prime}{g}\right)^2 \frac{\rho^2}{\rho_0^2}\, , \\ && 
 |\bar{c}_\gamma| \lesssim {\mathcal O}(10^{-3})  ,
\eea 
where the superscript index $(n)$ indicates the $n$-th derivative with respect to $\rho$. In the following we impose the stronger condition $\bar{c}_\gamma = 0$; relaxing this to $|\bar{c}_\gamma| = {\cal O}(10^{-3})$ would not change the results significantly. 

\begin{figure}[htb]
\centering
\includegraphics[width=0.8\textwidth]{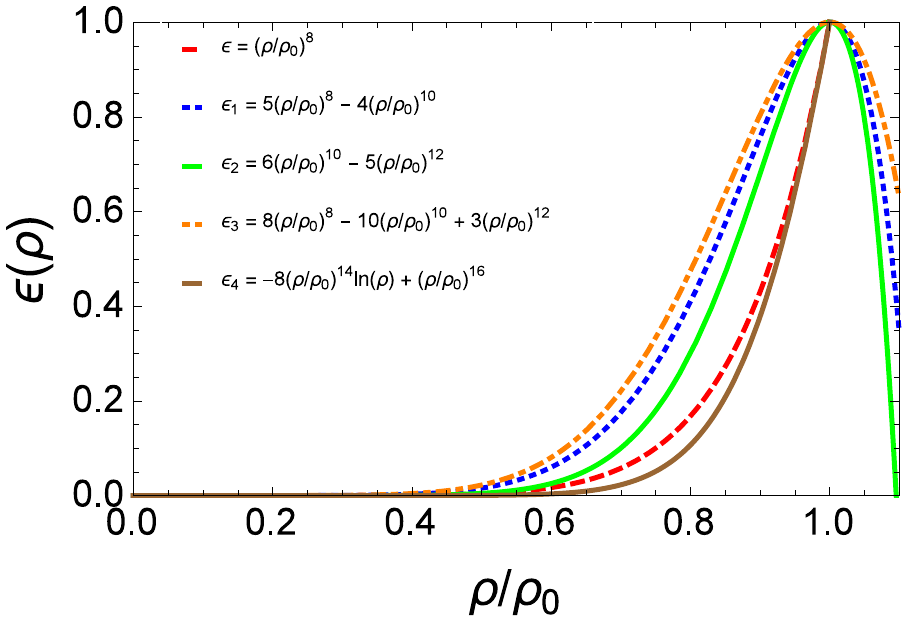}
\caption{Interpolating functions $\epsilon_i(\rho)$, $i=1,2,3,4$, that satisfy the required theoretical and phenomenological properties in solid brown ($\epsilon_4$), solid green ($\epsilon_2$), dotted blue ($\epsilon_1$), and dashed-dotted orange ($\epsilon_3$) lines. The CKY regularization~\protect\cite{Cho:2013vba}, $\epsilon$ that is incompatible with LHC data~\protect\cite{Ellis:2014jta,Ellis:2014dva} is shown in dashed red. Graph taken from Ref.~\protect\refcite{Ellis:2016glu}.}
\label{you2a}
\end{figure}

\begin{figure}[htb]
\centering
\includegraphics[scale=0.31]{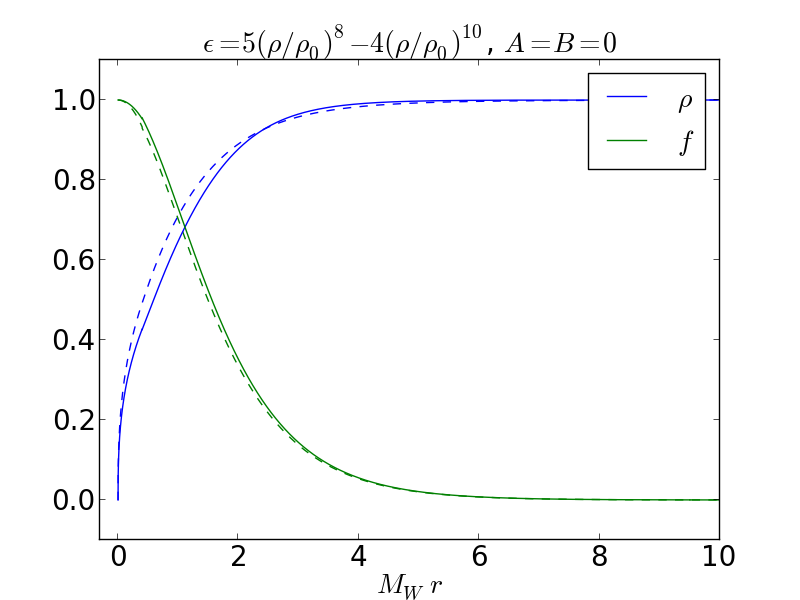} \hfill
\includegraphics[scale=0.31]{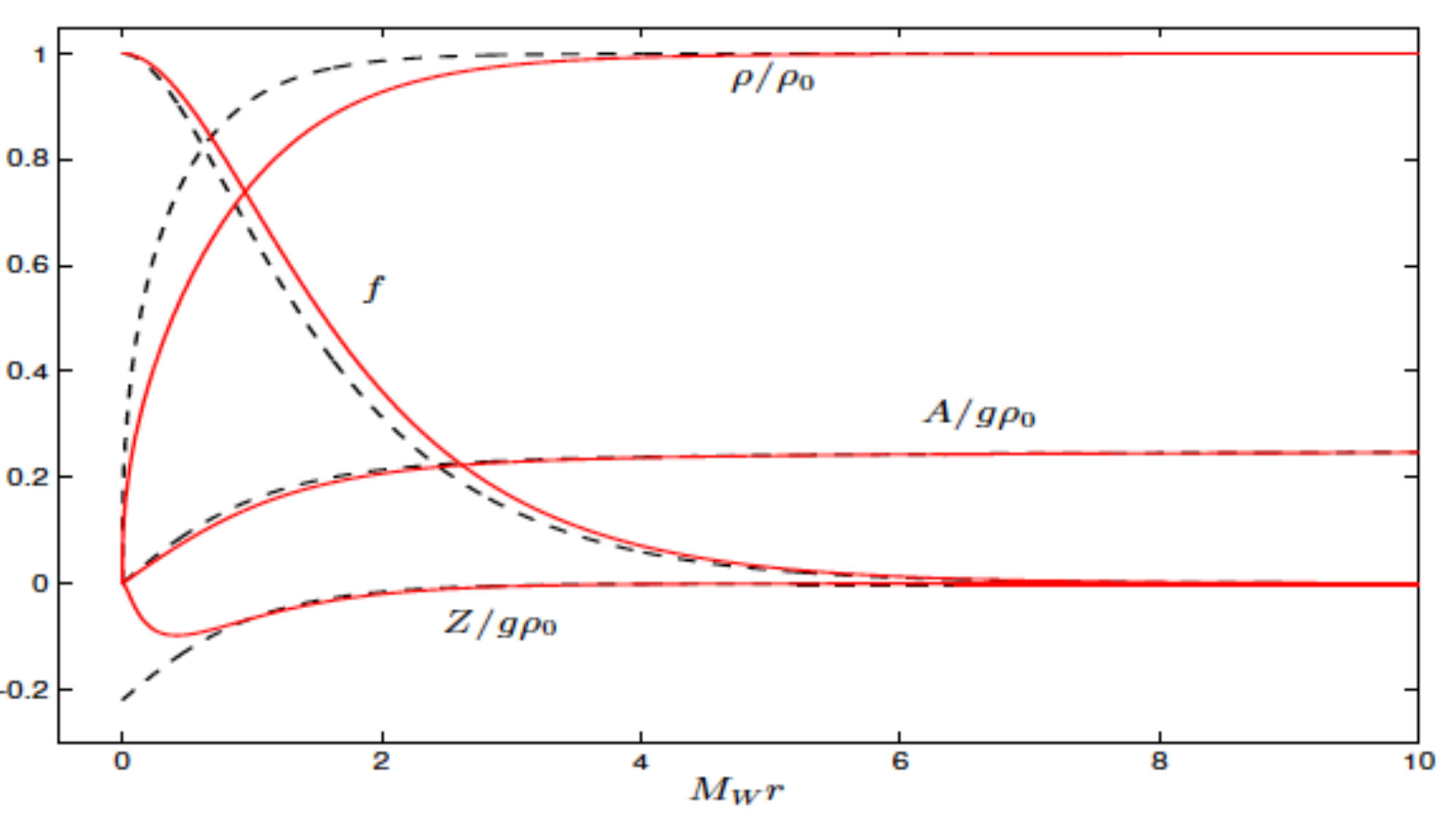}
\caption{{\bf Left:} Finite-energy electroweak monopole solution for the $A=B=0$ case with $\epsilon = 5 (\rho/\rho_0)^8 - 4 (\rho/\rho_0)^8$ that satisfies all the theoretical and phenomenological constraints.  The $\rho$ (normalized by $\rho_0$) and $f$ solutions are represented by solid blue and green lines, respectively. Dashed lines represent the original CKY regularization~\protect\cite{Cho:2013vba}, which is excluded by data. {\bf Right:} The CKY regularization for the electroweak dyon (continuous lines) versus  the original (singular) CM dyon solution (dashed lines), for general $A, B$, with $Z \equiv A-B$,  and the boundary conditions $f(0)= 1$ and $A(\infty) = M_W/2$. We see that there is good agreement between the various regularizations and the original CM theory. Graphs taken from Ref.~\protect\refcite{Ellis:2016glu} (left panel) and Ref.~\protect\refcite{Cho:2013vba} (right panel).}
\label{you2b}
\end{figure}

In Ref.~\refcite{Ellis:2016glu}, several acceptable forms of the interpolating functional $\epsilon$ have been exhibited, which contain algebraic sums of terms corresponding to different powers of $(\rho/\rho_0)$ (see Fig.~\ref{you2a}). They lead to monopole masses ranging from $M \sim 6.8~\tev$ down to $\sim 5.4~\tev$, with a larger mass $\sim 10.8~\tev$ for the dyon case. As remarked in Ref.~\refcite{Ellis:2016glu}, one might expect that a lower-mass monopole could be found in a more exhaustive survey of parameter space, particularly if attention was restricted to powers $n \ge 10$. One may also consider non-polynomial (e.g.\ logarithmic functional forms for $\epsilon$, which converge faster. We note that the associated numerical solutions for the modified finite-energy CM monopoles are pretty close to the original solution of Ref.~\refcite{Cho:1996qd} (see Fig.~\ref{you2b}). Gravitational effects~\cite{Cho:2016npz} in such monopole solutions may lead to further reduction of the mass.

In Ref.~\refcite{Ellis:2016glu} the discussion of the modifications of the SM hypercharge vacuum permittivity was purely phenomenological without any attempt to construct microscopic models. This has been attempted in Ref.~\refcite{Arai:2018uoy}, where by considering  brane world models in higher non-compact dimensions, one may be able to localize the SM on a domain wall, and obtain, upon reduction on four-dimensional spacetime, some of the models discussed in Ref.~\refcite{Ellis:2016glu} with finite-energy monopole solutions. Specifically, the authors of Ref.~\refcite{Arai:2018uoy} proposed a class of five-(non-compact)-dimension theories, where localization of gauge fields is achieved by the  condensation of Higgs field via a non-minimal Higgs-dependent gauge kinetic term in the five-dimensional Lagrangian. The model contains a domain  wall, connecting vacua with unbroken gauge symmetry, which drives the Higgs condensation that provides electroweak symmetry breaking and at the same time gauge field localization. The five-dimensional gauge theory contains a non-minimal coupling of the (five-dimensional) Higgs field with (five-dimensional) gauge field kinetic terms. Such couplings are crucial for the localization mechanism. The effective low-energy four-dimensional theory, then, contains higher-dimensional interaction terms of the form of algebraic sums  ${\mathcal L}_{\rm int} = \sum_{n} c_n \, |H|^{2n}\, (F_{\mu\nu}^a F_{\mu\nu}^a + B_{\mu\nu}\, B^{\mu\nu})$ in the respective Lagrangian, where $c_n$ are appropriate constants, $H$ denotes the four-dimensional Higgs field, $F_{\mu\nu}^a$ the non-Abelian field strength of the electroweak theory, and $B_{\mu\nu}$ the Abelian hypercharge field strength. Depending on the values of the coefficients $c_n$, such terms can lead, as discussed above, to new tree-level contributions to the $H \to \gamma \gamma$ decays and in some cases to finite-energy monopole solutions.  In Ref.~\refcite{Arai:2018uoy}, a model with ${\mathcal L}_{\rm int}$ containing only two terms in the sum, corresponding to the powers $n=8,10$, has been explicitly considered, which is one of the cases of Ref.~\refcite{Ellis:2016glu} leading to finite electroweak monopoles (see Fig.~\ref{you2b}). However, there is no way to provide a more microscopic justification for selecting such cases, other than noticing that  they are consistent with the symmetry-breaking patterns studied in Ref.~\refcite{Arai:2018uoy}. 

\subsubsection{Born-Infeld extension of the Standard-Model hypercharge sector}\label{sc:born-infeld}

The second, more microscopically justifiable, way of obtaining Cho-Maison-like monopole solutions with finite energy is to modify the $U_Y(1)$-hypercharge sector by including higher derivative corrections of the pertinent field strength $B_{\mu\nu}$, as in the case, for instance, of string-inspired Born-Infeld (BI) Lagrangians~\cite{Arunasalam:2017eyu}. In these scenarios, the SM is extended by a non-linear BI-type hypercharge gauge field with a Lagrangian of the form
\bea\label{BI}
\mathcal{L}_{\rm EW} &=& -(D_{\mu}H^{\dagger})(D^{\mu}H)-\frac{\lambda}{2}\Big(H^{\dagger}H-\frac{\mu^2}{\lambda}\Big)^2-\frac{1}{4}F^{a}_{\mu\nu}F^{\mu\nu,a}\nonumber \\ &+& \tilde \beta^2\Big(1-\sqrt{1+\frac{1}{2\, \tilde \beta^2}B_{\mu\nu}B^{\mu\nu}-\frac{1}{16\, \tilde \beta^4}(B_{\mu\nu}\tilde{B}^{\mu\nu})^2}\Big)
\eea
where $D_{\mu}=\partial_{\mu}-i\frac{g}{2}\tau^aA^a_{\mu}-i\frac{g'}{2}B_{\mu}$ is the covariant derivative and $A^a_{\mu}$ and $B_{\mu}$ are the $SU_L(2)$ and $U_Y(1)$ gauge fields, respectively, $H$ is the electroweak Higgs doublet, $B^{\mu\nu}$ is the $U(1)$ field strength tensor with $\tilde{B}^{\mu\nu}$ as its Hodge dual, $F^{\mu\nu,a}$ is the $SU(2)$ field strength tensor, where $a=1,2,3$. The $SU_L(2)$ and $U_Y(1)$ couplings are given by $g$ and $g'$, respectively. The BI parameter $\tilde \beta$ has dimensions of mass squared. It reduces this Lagrangian to the SM Lagrangian for $\tilde \beta\rightarrow\infty$. In string-inspired models, $\tilde \beta $ is proportional to the string mass scale $M_s$. In general, one may consider more complicated modifications, where there are higher derivative terms also for the non-Abelian field strengths, but for the purposes of demonstrating the finiteness of the associated monopole solutions, restricting our attention to the Lagrangian \eqref{BI} suffices. 

Indeed, it can be shown~\cite{Arunasalam:2017eyu}, following the same steps as in the standard CM case discussed in Sec.~\ref{cm}, that with such modifications, a CM-like monopole solution exists in the model \eqref{BI}, which is characterized by a total energy $E$ of the form: 
\bea
E &=& E_0+E_1, \nonumber \\
E_0 & =& \int_0^{\infty}dr\, \tilde \beta^2 \left[\sqrt{ \left(4\pi r^2\right)^2+\frac{h_Y^2}{\tilde \beta^2}}-4\pi r^2 \right] = \frac{h_Y^2}{3\pi}\Gamma\left( \frac{1}{4} \right) \Gamma \left(\frac{5}{4} \right) \sqrt{\frac{\tilde \beta}{h_Y}} = 72.82 \sqrt{\tilde \beta},
 \nonumber \\
E_1 &=&  4\pi\int_0^{\infty}dr \left[ \left( \frac{\dot{f}^2}{g^2}+\frac{ \left( f^4-1 \right)^2}{2g^2r^2} \right) + \left( \frac{f^2\rho^2}{4}+\frac{r^2\dot{\rho}^2}{2}+\notag	\frac{\lambda r^2}{8} \left( \rho^2-\rho_0^2 \right)^2 \right) \right],
\eea
where the notation is the same as in Sec.~\ref{cm}, $h_Y=\frac{4\pi}{g'}$ and $\Gamma(z)=\int_0^{\infty}x^{z-1}e^{-x}dx$.

The integral $E_1$ is similar to the one in the case of the CM monopole, assuming the value \eqref{e1}, for the set of the observed SM parameters. Par contrast to the CM case, however, we now observe that the $E_0$ part of the energy, which diverges in the CM case, is now finite, proportional to the BI parameter $\tilde \beta$, which thus acts as an appropriate cut-off. Thus the total energy (rest mass) of the monopole is \emph{finite}, and for the observed values of the SM parameters becomes 
\be\label{BImass}
E = 4.1 + 72.8 \sqrt{\tilde \beta} \; \tev.
\ee
At this point we remark that a new class of finite-energy magnetic-monopole solutions within this BI framework has been found in Ref.~\refcite{Mavromatos:2018kcd}. This new class was discovered by performing analytic asymptotic analyses of the nonlinear Lagrange differential equations describing the model using Pad\'e approximants, which replaces the shooting method used in numerical solutions to boundary-value problems for ordinary differential equations used in previous analyses~\cite{Cho:1996qd,Arunasalam:2017eyu}. In Ref.~\refcite{Mavromatos:2018kcd}, although the ansatz used to generate static and spherically symmetric  monopole solutions, with finite energy, is the one employed by Cho and Maison~\cite{Cho:1996qd}, nonetheless the so obtained solutions have a different behavior near the monopole core than the standard ones~\cite{Cho:1996qd}, also used in Ref.~\refcite{Arunasalam:2017eyu}. Estimates of the total energy of this new class of BI monopole solutions predict a higher total monopole energy (mass) than \eqref{BImass}: 
\be\label{BImass2}
E= 7.6  + 72.8 \sqrt{\tilde \beta} \; \tev. 
\ee

As discussed in Ref.~\refcite{Ellis:2017edi}, from the recent \emph{measurement} by ATLAS~\cite{Aaboud:2017bwk} and later by CMS~\cite{Sirunyan:2018fhl} Collaborations of light-by-light scattering in LHC Pb--Pb collisions~\cite{dEnterria:2013zqi},  one obtains a lower bound for $\sqrt{\beta} \gtrsim 100~\gev$, which, on account of \eqref{BImass},  implies monopole masses  $E \gtrsim 11~\tev$; for the monopole solution of Ref.~\refcite{Mavromatos:2018kcd}, with mass \eqref{BImass2}, the collider lower limit is $ E \gtrsim 15 \; \tev$. Taking into account that monopoles are produced in pairs with their antiparticles in collisions, they are beyond reach of any LHC experiment, but may be accessible at future colliders~\cite{dEnterria:2016sca,Mangano:2017tke,Zarnecki:2020ics} or cosmic searches~\cite{Patrizii:2015uea}. 

Cosmological consequences of such monopoles, in particular their role in the electroweak phase transition and nucleosynthesis, have been discussed in Ref.~\refcite{Arunasalam:2017eyu}, where we refer the interested reader for details. Here we only mention that nucleosynthesis constraints on the abundance of such monopoles imply $M \lesssim 2.3 \times 10^4~\tev$, while in order for the monopoles to delay the electroweak phase transition and make it more first order, one should have $M \gtrsim 9.3 \times 10^3~\tev$, although this latter property is not necessary. Thus only the upper bound in the monopole mass appears to be rigid. In view of \eqref{BImass}, then, one observes that, in the context of microscopic string theory models, where one expects $\sqrt{\tilde \beta} \sim M_s$, with $M_s$ the string mass scale, we  obtain an upper bound on $M_s \lesssim 2.3 \times 10^4~\tev$, in order to have cosmologically allowed BI-hypercharge monopoles. However, this expectation may be naive, in the sense that the effective low-energy Lagrangians of phenomenologically realistic string theories are actually much more complicated than \eqref{BI}, containing higher-derivative terms also for the non-Abelian gauge fields, which may modify the monopole solution discussed above. Moreover, there might be string-theory brane-Universe models where monopoles are allowed to live in the bulk, and thus their abundance on the brane Universe describing our world may be naturally diminished, thus avoiding the nucleosynthesis constraints. These issues fall beyond the scope of this review. 

\subsection{Kalb-Ramond (self-gravitating) monopoles with finite energy \label{ref:KRmon}} 

Before closing our discussion on theoretical models predicting magnetic monopole solutions with relatively small masses as compared to GUT scale, we would like for completion to briefly describe another type of magnetic monopole inspired by strings. This monopole is associated with a particular type of \emph{axion} fields, termed Kalb-Ramond (KR) axions, which are different from the QCD axions, and have their origin in the massless gravitational multiplet of a string~\cite{Polchinski:1998rq,Polchinski:1998rr}. In fact, the magnetic charge of such configurations is proportional to the axion charge. This model has been proposed in Refs.~\refcite{Mavromatos:2016mnj,Mavromatos:2018drr}. It contains an extra triplet of real scalar fields, $\chi$, beyond the SM, with a Higgs-like potential, associated with a spontaneous breaking of an internal $SO(3)$ symmetry.

The model contains dilatons $\Phi$ and antisymmetric tensor KR fields, $B_{\mu\nu}= - B_{\nu\mu}$, appearing in the Lagrangian only through their field strength, as a result of an underlying $U(1)$ gauge invariance~\cite{Polchinski:1998rq,Polchinski:1998rr}. In four space-time dimensions the KR field strength $H_{\mu\nu\rho}$ is dual to a pseudoscalar (axion) field $b(x)$:\footnote{In string theories~\cite{Polchinski:1998rq,Polchinski:1998rr}, the field strength $H_{\mu\nu\rho}$, in the presence of gauge fields $A_{\mu}$, is no longer given only by the curl of $B_{\mu\nu}$ but contains additional parts proportional to the Chern-Simons three-form $\mathbf{A} \wedge \mathbf{F} $, where $\mathbf{F} = \mathbf{d} \wedge \mathbf{A} + \tilde g^\prime  \mathbf{A} \wedge \mathbf{A}$ is the non-Abelian field strength of the unifying string group with coupling $\tilde g^\prime $. Such terms lead to higher derivative terms in the string effective action \eqref{eq:13}, and have been ignored in Refs.~\refcite{Mavromatos:2016mnj,Mavromatos:2018drr}. Their inclusion for Abelian gauge fields could lead to additional interesting electromagnetic effects~\cite{Das:2000ph,Kar:2000ct,Maity:2005ah,Alexandre:2008ew,Alexandre:2007hs}, which, however, are not of direct interest to this review. }
\be
H_{\mu\nu\lambda}= \partial_{[\mu} \, B_{\nu\rho]} = e^{2\Phi}\, \epsilon_{\mu\nu\lambda\sigma}\partial^{\,\sigma}b ~, \label{eq:21}
\ee
where $[\dots ]$ denotes complete antisymmetrization of the respective indices, $\epsilon_{\mu\nu\rho\sigma}$
is the curved-spacetime Levi-Civita tensor density, 
$\epsilon_{\mu\nu\rho\sigma}=\sqrt{-g}\,\tilde\epsilon_{\mu\nu\rho\sigma}$, with $\tilde\epsilon_{\mu\nu\rho\sigma}$ the flat spacetime antisymmetric symbol ($\tilde \epsilon_{0123} = +1$, etc.)

The four-spacetime-dimensional Lagrangian is given by:
\bea\label{eq:13}
L&=&\left(-g\right)^{1/2}\Big\{ \frac{1}{2}\partial_{\mu}\chi^{a}\partial^{\mu}\chi^{a}-\frac{\lambda}{4}\left(\chi^{a}\chi^{a}-\eta^{2}\right)^{2} - \frac{1}{16\pi \, {\rm G}}\, R \nonumber \\
 &+& \frac{1}{2}\partial_{\mu}\Phi\partial^{\mu}\Phi-V\left(\Phi\right) -\frac{1}{12}\, e^{-2 \Phi}\, H_{\rho\mu\nu}H^{\varrho\mu\nu}-\frac{1}{4}\, e^{-\Phi}\, f_{\mu\nu}f^{\mu\nu}\Big\},  
 \eea
where $\lambda > 0$, as appropriate for the $O(3)$ spontaneous symmetry breaking in the model, ${\rm G}= 1/M^2_{\rm P}$ (in units $\hbar=c=1$) is the Newton (gravitational) constant of the four-dimensional spacetime, with $M_{\rm P} = 2.4 \times 10^{18}~\gev$ the reduced Planck mass, $g=\det\left(g_{\mu\nu}\right)$ is the determinant of the metric tensor, $R$ is the Ricci curvature scalar,  and $f_{\mu\nu}$ is the electromagnetic Maxwell tensor. The SM interactions (in the broken phase of the SM theory, where the photon remains massless) can be added separately to \eqref{eq:13} and, except for the photon, they play no role in the monopole solution. One assumes~\cite{Mavromatos:2016mnj,Mavromatos:2018drr} that a singular gauge field $A_{\mu}$ (up to a gauge transformation) may be associated with $f_{\mu\nu}$, on using  a construction outlined by Halpern~\cite{Halpern:1978ik}. 

In the absence of the electromagnetic, $H$ and $\Phi$ fields, the Lagrangian \eqref{eq:13} reduces to that employed in the self-gravitating \emph{global monopole} case of Ref.~\refcite{Barriola:1989hx}, whose stability is still under debate (see discussion in Refs.~\refcite{Mavromatos:2016mnj,Mavromatos:2018drr} and references therein). The existence of the KR axion $b$ and dilaton fields, which promotes the global monopole to a local one in the presence of an electromagnetic $A_\mu$ potential, is argued in Refs.~\refcite{Mavromatos:2016mnj,Mavromatos:2018drr} to be crucial for stability of the KR magnetic monopole, as a consequence of the fact that its magnetic charge is proportional to the KR axion charge. 

For the scalar triplet field, associated with the spontaneous symmetry breaking of the global $SO(3)$ symmetry,  the following ansatz was made~\cite{Mavromatos:2016mnj,Mavromatos:2018drr}
\be\label{scalar}
{\chi ^A}(r) = \eta f(r)\frac{{{x^A}}}{r}\,, \quad A=1,2,3~,
\ee
where $x^A$, $A=1,2,3$, are Cartesian spatial coordinates, with the asymptotic behavior $f(r \to 0) \simeq   f_0 \, r \to 0$, $f_0 ={\rm constant} \in {\mathbb R}$, $f(r \to \infty )  \to  1$, consistent with the equations of motion, and similar to the scalar field of the HP monopoles, discussed in Sec.~\ref{sec:hp}.

Assuming a static spherically symmetric metric corresponding to an invariance element of the form $ds^2 = B(r) dt^2 - A(r) dr^2 - r^2 \big(d\theta^2 + \sin^2\theta \, d\phi^2 \big)$, with $A  B \simeq 1 $ for both $r \to 0$ and $r \to \infty$, one finds a KR-axion field $b$ of the form~\cite{Mavromatos:2016mnj,Mavromatos:2018drr}:
\be\label{bfield}
b'\left(r\right)=\frac{\zeta}{r^{2}}\sqrt{\frac{A(r)}{B(r)}},
\ee
for the entire range of the radial distance from the center $r$. The quantity $\zeta \in {\mathbb R}$ is the KR-axion charge. The following approximate relations for the entire range of $r$ are adopted for the metric function
\bea\label{ab1}
A(r) B(r) \approx 1~, \quad B(r) = 1 - 8\pi {\rm G}\, \eta^2 - \frac{{2m {\rm G}}}{r} + \frac{16\pi {\rm G}\, \zeta^2}{{{r^2}}} ~, 
\eea
where $m$ is the Schwarzschild mass of the monopole. 

This is an (approximate) \emph{Reissner-Nordstrom (RN) solution} for a \emph{black hole} with ``charge squared'' $Q^2 = 2\, \zeta^2$. In fact, as discussed in Refs.~\refcite{Mavromatos:2016mnj,Mavromatos:2018drr},  this charge is a magnetic one, in the sense that it corresponds to the magnetic charge of the monopole. If one assumes that $\eta \ll M_{\rm P}$, as required in order for such configurations to potentially play a role at collider physics scales, then there are \emph{no horizons}, however there are no naked singularities at $r \to 0$, since the latter are shielded by the monopole core.  Due to the terms proportional to $\eta^2$ in $B(r)$, the spacetime for $r \to \infty$ is \emph{not} asymptotically Minkowski, but is characterized by a \emph{conical deficit angle}, proportional to $8\pi {\rm G}\, \eta^2$. This is a consequence of the fact that the total energy of this type of monopole diverges linearly with distance (in this respect resembling a cosmic string), as is also the case of the global monopole~\cite{Barriola:1989hx}. 
 
To understand the emergence of monopole solutions in the model \eqref{eq:13}, one uses the dilaton equations of motion, assuming that the dilaton is stabilized to a constant value $\Phi=\Phi_0$, which may be guaranteed by an appropriate (string-loop generated) dilaton potential $V(\Phi)$.~\cite{Mavromatos:2016mnj,Mavromatos:2018drr} This implies a relation between the kinetic terms of the $b$ and electromagnetic $A_\mu$ fields, which, on using \eqref{bfield}, leads to a radial magnetic field strength~\cite{Mavromatos:2016mnj,Mavromatos:2018drr} 
\be\label{magnetic}
\mathcal{B}^r  \simeq \frac{\sqrt{2}\, \zeta}{r^2},
\ee
which is consistent with the equations of motion for $A_\mu$. From \eqref{magnetic} it follows that the magnetic charge of the configuration is 
\be\label{magch}
g = \sqrt{2} \, \zeta.
\ee
Hence the KR-axion charge is proportional to the magnetic charge, and thus the charge of the corresponding RN solution black hole \eqref{ab1} is a magnetic charge. As the gauge group considered in this model is the standard Abelian electromagnetic $U_{\rm em}(1)$, there is a Dirac string, according to the arguments discussed in Sec.~\ref{sec:hp}. Hence the DQC \eqref{dirac} would imply a quantization of the KR-axion charge $\zeta$, for the self-gravitating KR monopole to be consistent with quantum theory. 

\begin{figure}[htb]
\centering
\includegraphics[width=0.55\linewidth]{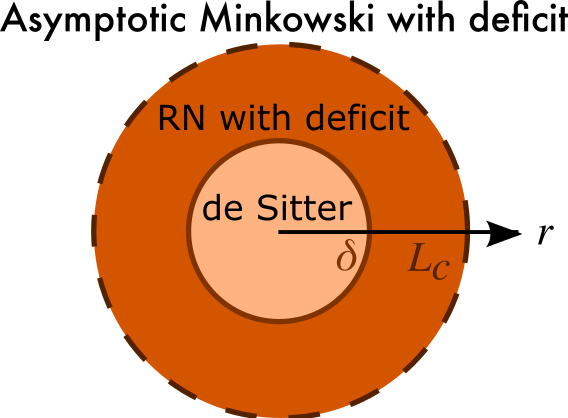} 
\caption{The regularized self-gravitating Kalb-Ramond electromagnetic monopole of Refs.~\cite{Mavromatos:2016mnj,Mavromatos:2018drr}. The dark orange shaded area depicts the shell area where most of the mass of the monopole lies. Only its inner boundary (at $r=\delta$)  is a hard one, requiring matching of the respective space times via Israel conditions. The outer region of the shell (dashed lines) does not require matching and corresponds to an asymptotic Minkowski spacetime with an angular deficit. Its radius $L_c$ is determined by minimization of the total energy functional of the monopole configuration. \label{mass-shell}}
\end{figure}

The presence of an RN-black-hole-curvature singularity at $r \to 0$ in the metric \eqref{ab1} requires \emph{regularization}, which in Refs.~\refcite{Harari:1990cz,Mavromatos:2016mnj,Mavromatos:2018drr} has been performed by cutting off the spacetime around the monopole center at $ r =0$ by a region of radius $\delta$ (see Fig.~\ref{mass-shell}) and replacing it by a de Sitter space time, with positive cosmological constant $\Lambda = \frac{1}{4} \lambda \, \eta^4 > 0$.

We now remark that, if $\zeta = 0$, which is the case of the global monopole~\cite{Barriola:1989hx}, then $m < 0$ and this was interpreted in Ref.~\refcite{Harari:1990cz} as implying repulsive gravitational effects, which might be associated with the instabilities of the global monopole. Par contrast, in the presence of a nontrivial KR-axion charge $\zeta$, the Schwarzschild mass $m$ of the monopole comes out positive~\cite{Mavromatos:2016mnj,Mavromatos:2018drr}, 
\be\label{massm}
m \sim  6.32 \, \pi\, |\zeta| \, \eta  \left(\lambda \, \zeta^2 \right)^{1/4} > 0~,
\ee
and thus normal attractive gravitational effects are expected. This may indicate stability of the KR monopole, in the sense that the repulsive forces induced by the de Sitter region, balance the attractive gravitational forces due to the mass in the outer RN region. However, this still remains an open issue. 

The monopole core of radius $L_c \gg M_{\rm P}^{-1}$ does not constitute a hard boundary requiring metric matching conditions (see Fig.~\ref{mass-shell}), but can be estimated by assuming a ``bag-like'' structure~\cite{Mavromatos:2016mnj,Mavromatos:2018drr}. Most of the total energy of the monopole is assumed concentrated on a shell $\delta \equiv \alpha L_c \le r \lesssim L_c$, with $L_c $ being determined by minimization of the total energy functional with respect to $L_c$ 
\begin{multline}\label{energy} 
{\mathcal E} \simeq  4\pi \int_{\alpha L_c }^{L_c}  dr\,{r^2}   \\   \left[ 
\frac{2W^2(r)}{B(r) r^2} + \frac{b'^2(r)}{4} + \eta^2 \left( \frac{f^2(r)}{B(r) r^2} + \frac{f'^2(r)}{2} \right) + 
\frac{\lambda \, \eta^4}{4 B(r)} \left( f^2(r) - 1 \right)^2  \right]  . 
\end{multline}
The reader should notice that the first two terms on the right-hand side of \eqref{energy}, which are the contributions of the KR axion and electromagnetic fields, yield finite contributions, while the third term, pertaining to the scalars, is responsible for the linear divergence of the energy function  with $L_c$. This feature is shared by the global monopole~\cite{Barriola:1989hx}. The minimization yields for the KR monopole~\cite{Mavromatos:2016mnj,Mavromatos:2018drr}: $0 < \alpha = 1.26 \, \big(\lambda \, \zeta^2 \big)^{-1/2}  \ll 1$, and 
 \be\label{totalM} \mathcal E/ m \sim 1.7~, 
 \ee
where $m$ is given in \eqref{massm}.

The fact that the total energy $\mathcal E$ (total monopole (rest) mass) and the Schwarzschild mass are of the same order is to be expected since both $\delta$ and $L_c$ are much larger than the Planck length, where gravitational effects are expected to be important. In view of $\eta \ll M_{\rm P}$, the monopole mass is much smaller than the Planck scale. However, to be able to make predictions of potential relevance to collider physics, one needs to have a microscopic understanding of this model, in particular how it can extend the SM. This is not discussed in the literature~\cite{Mavromatos:2016mnj,Mavromatos:2018drr}. Nonetheless, having novel finite-energy monopole configurations which can have a potentially small mass  is of phenomenological relevance for collider searches.

Moreover, the presence of a angular deficit $\propto \eta^2$ in the asymptotic spacetime, which characterizes \emph{both} global~\cite{Barriola:1989hx} and KR magnetic~\cite{Mavromatos:2016mnj,Mavromatos:2018drr} monopoles, is known to imply a role for such spacetime defects as particle ``lenses''~\cite{Mazur:1990ak,Ren:1993hr,BezerradeMello:1996si,RoderiguesSobreira:1998tb,Lousto:1991rh,Mavromatos:2017qeb}. The quantum scattering amplitudes of particles of various spins, including photons, off such defects, exhibit  formal infinities at scattering angles equal to the deficit angle. Although the Coulomb repulsion of charged particles of relevance to collider experiments might be too strong, as compared with the gravitational attraction that creates the lensing, to have observable lensing effects at colliders, nevertheless the above lensing phenomenon may be used as an interesting test of the presence of such defects in case they are of cosmic origin, through lensing of cosmic light~\cite{Mavromatos:2017qeb}. Such tests could complement astrophysical and cosmological searches of global monopoles, e.g.\ those using cosmic microwave background radiation~\cite{Lopez-Eiguren:2017dmc}. 

\subsection{D-particle defects and dyonic D-branes}

Following up from the above considerations, we would like to mention that another category of spacetime defects that are characterized by the aforementioned  behavior of spacetime lenses are the so-called \emph{D-particles}. These are stringy defects which appear in the framework of brane theories~\cite{Polchinski:1996na,Polchinski:1998rq,Polchinski:1998rr}. Depending on the type of string theory considered, D-particles are either pointlike D$0$-branes (type~IIA strings), or ``effectively  pointlike'' consisting of compactified branes, wrapped around topologically nontrivial cycles in the compact extra dimensional manifold, which implies their stability (type~IIB strings)~\cite{Li:2009tt}. Embedded in brane Universes, with four uncompactified spacetime dimensions, such defects may play the role of exotic dark matter candidates~\cite{Shiu:2003ta,Ellis:2009vq,Li:2009tt,Mavromatos:2010nk,Mavromatos:2012ha,Elghozi:2015jka}. Their mass is of order $M_s/g$, with $M_s$ the string scale, which is in general different from the four-dimensional Planck mass. Depending on the D-particle mass, they could also be produced at colliders~\cite{Shiu:2003ta,Mavromatos:2010nk}. 

The scattering of stringy matter off such defects~\cite{Kogan:1996zv} results in local distortions of spacetime~\cite{Ellis:1999jf}. If one considers string matter propagating through an ensemble of such defects embedded in a brane Universe then, asymptotically far away from the (recoiling) defects, the distortions resemble the deficit spacetime of a global monopole~\cite{Mavromatos:2009pp}. The deficit parameter in this case is provided by the stochastic variance of the recoil velocity fluctuations in the ensemble of D-particles.

Although the above considerations referred to ``neutral D-brane matter'', nonetheless D-branes~\cite{Polchinski:1996na,Polchinski:1998rq,Polchinski:1998rr,Bachas:1998rg} may be electrically and/or magnetically charged~\cite{Deser:1997se,Deser:1998vc,Sen:1999mg,Bertolini:1998mg}, hence their connection with the main topic of this review. It is a basic property of brane theory that only {\it extended}  non-perturbartive  $(p - 1)$-branes, with $p > 1$,  can be {\it magnetically} charged, as required by extending the electric--magnetic duality appropriately to string/brane theory. This lead to the construction of $(p - 1)$-brane dyons coupled to appropriate Abelian $p$-form potentials in $D=2(p+1)$ spacetime dimensions. We remind the reader that, in a first-quantization framework~\cite{Bachas:1998rg}, a $(p - 1)$-brane, which  sweeps a $(p)$-dimensional world volume ${\mathcal W}_p$, as it propagates in time, couples {\it electrically} to a $(p)$-form Abelian potential ${\mathbf C}^{(p)}$ (Ramond-Ramond (RR)  gauge field~\cite{Polchinski:1996na,Polchinski:1998rq,Polchinski:1998rr}) 
\be\label{elec}
\mu_{p-1} \int_{\mathcal W_p} \mathbf C^{(p)}\, ,
\ee
where $\mu_{p-1}$ plays the role of the brane ``electric charge $e$''. 

The potential corresponds to a $(p + 1)$-field-strength form ${\mathbf F}^{(p+1)} = {\rm d}\, {\mathbf C}^{(p+1)}$.
The Hodge dual form of ${\bf F}^{(p+1)}$, on the other hand, ${}^\star {\mathbf F}^{(D-p-1)} = {\rm d} \, {}^\star {\mathbf C}^{(D-p-2)}$, corresponds to a dual (``magnetic'') potential ${}^\star {\mathbf C}^{(D-p-2)}$, which couples to the ``dual'' extended object with world volume ${\mathcal W}_{D-p-3}$ (``magnetically'' charged $p$-brane): 
\be\label{mag}
\mu_{D-p-3} \, \int_{\mathcal W_{D-p-2}} \, {}^\star \mathbf C^{(D-p-2)}\, ,
\ee
with $\mu_{D-p-3}$ playing the role of the corresponding brane magnetic charge ``$g$''. In the concrete example of the type~II superstrings in critical dimension $D=10$, and $(p-1)$-branes, the $(p-1)$-brane charges obey the standard DQC \eqref{dirac} (in the SI system of units):
\begin{align}
\mu_{p-1} \, \mu_{6-(p-1)} = 2\pi n \hbar, \quad n \in \mathbb Z, 
\end{align}
compatible with electric--magnetic duality. 

Dyonic $p$-branes,  with both electric and magnetic charges, have also been constructed~\cite{Deser:1997se,Deser:1998vc,Sen:1999mg,Bertolini:1998mg}, and by considering their scattering, one may arrive at generalizations of the Schwinger-quantization conditions \eqref{dyon} (in the SI system of units )~\cite{Deser:1997se}:
\be\label{quantbrane}
e_1\, g_2 + (-1)^p \, e_2\, g_1 = 2\pi  n \hbar, \quad n \in \mathbb Z \,,
\ee
where here we used the notation $e_i (g_i)$, $i=1,2$, to denote the electric (magnetic) charges of the scattered dyonic $p$-branes. The reader should notice the $p$-dependent  sign between the two terms on the left-hand side of \eqref{quantbrane}. 

These considerations can be extended to compactified branes, which allows for phenomenologically realistic applications in four-dimensional (brane) Universes.  For instance, D3-branes wrapped around a compact extra-dimensional manifold with the topology of an orbifold $T_6/Z_3$, may also be charged under both electric and magnetic charges, and, from a four-dimensional viewpoint, they may appear as dyonic black holes~\cite{Bertolini:1998mg}. If such microscopic black holes are producible at colliders, in the case of low-string-mass scale theories, then they may lead to high ionization. 

In this respect, we also mention that magnetically charged  black holes, and their similarities to magnetic monopoles, have recently been discussed in Ref.~\refcite{Maldacena:2020skw}, where it was suggested that in some sense such black holes may be viewed as high-charge bound states of monopoles. The reader should also recall in this spirit, but from a different perspective, that the self-gravitating KR monopole solution of Ref.~\refcite{Mavromatos:2016mnj}, discussed in Sec.~\ref{ref:KRmon}, is linked to a Reissner-Nordstrom magnetically charged black hole. It is interesting to pursue further such connections between black holes and monopoles, which may also have consequences for Astrophysics and Cosmology. For phenomenological searches of such magnetic black holes, the reader is referred to a very recent work~\cite{Bai:2020spd}, where various methods for detecting them are presented. It is pointed out that magnetically charged black holes may possess a ``hairy'' electroweak-symmetric corona, which extends outside the event horizon, and allows for an extension of the Parker bound~\cite{Turner:1982ag,Adams:1993fj} on magnetic monopoles by several orders of magnitude using the large-scale coherent magnetic fields in Andromeda galaxy. According to Ref.~\refcite{Bai:2020spd}, this sets a mass-independent constraint on their cosmic abundance to be a fraction less than $4 \times 10^{-4}$  of that of dark matter. Moreover, one might hope that lensing techniques for searching for standard astrophysical black holes, e.g. those in Refs.~\refcite{Virbhadra:2008ws,Virbhadra:1999nm}, could be extended  to the search of the magnetically charged ones.

\subsection{Other monopole solutions beyond the Standard Model} 

Topologically stable magnetic monopoles of HP type, but with masses of order of the electroweak scale have recently been proposed in SM extensions with Higgs triplet fields~\cite{Hung:2020vuo}. The model had been previously proposed as an explanation of the SM neutrino masses, and involved non-sterile right-handed neutrinos with masses of order of the electroweak scale, as a consequence of their coupling to a {\it complex} triplet of Higgs scalars. There is a custodial symmetry in the model that ensures that the $W$ and $Z$-boson masses are connected as in the SM, $M_W = M_Z \, \cos\theta_W$. This custodial symmetry necessitates the introduction of an extra, yet {\it real}, Higgs triplet which provides the conditions met in the Georgi-Glashow model~\cite{Georgi:1974sy} for the existence of magnetic monopoles of HP type, with masses of the order of the electroweak scale. We note that the existence of scalar triplets, different from the SM Higgs field, also characterize the KR magnetic monopole solution~\cite{Mavromatos:2016mnj,Mavromatos:2018drr} discussed in Sec.~\ref{ref:KRmon}. It is a necessary ingredient of any model that admits a HP magnetic monopole, which is connected to the spectrum structure of the Georgi-Glashow model, as discussed in Sec.~\ref{sec:hp}. 

New magnetic monopole solutions inspired by an exactly solvable model for a Berry phase in parameter space have recently been proposed in Ref.~\refcite{Deguchi:2019jtz}. The model has the interesting feature of exhibiting a smooth topology {\it change} from a Dirac-like {\it monopole}, when seen far away from the monopole's position in space, to a {\it dipole} near the monopole position, and half a monopole (with magnetic charge equal to half of the fundamental Dirac charge) in the transitional region. The  configuration vanishes at the origin. The smooth topology change occurs when the DQC \eqref{dirac} is in operation when the system couples to electrons, which makes the Dirac string unobservable.

In addition, quite recently,  another type of monopoles, termed {\it emergent monopoles} has been proposed~\cite{Gera:2020fvo}, which arise as non-singular solutions of the gravitational field equations in the presence of torsion, which generate the ``magnetic'' charge. These monopoles are purely geometrical, distinct from the standard magnetic monopoles in the sense that, although there is a topological interpretation of the magnetic charge of these solutions, the latter is expressed in terms of a geometric invariant (Nieh-Yan) in the contorted gravity theory~\cite{Nieh:1981ww,Nieh:2007zz}, and thus the relevant quantization rule does not involve the electric charge. From a geometrical  point of view, these ``emergent'' monopoles, have a finite-size core, characterized by an invertible ``static'' metric (three-geometry embedded in four dimensions), with the timelike direction ``frozen'', which has no curvature singularity. The core region is surrounded by a RN magnetically charged spacetime. The spacetime singularity at the center of the monopole core is inaccessible, though, to a physical observer living in the exterior spacetime, since there are no radial geodesics connecting the inner core with the outer RN region. Thus the monopole core would not be directly observed in nature, although the magnetic (Coulomb-type) field would be indirectly observable through the spacetime curvature of the RN outer region. 

We note at this point that the RN spacetime surrounding the core is partly reminiscent of the structure of the self-gravitating KR monopoles of Refs.~\refcite{Mavromatos:2016mnj,Mavromatos:2018drr}, discussed in Sec.~\ref{ref:KRmon}. However in that work, the RN spacetime is characterized by a deficit, and there is a de Sitter spacetime regularization of the core. Moreover, the magnetic charge of the RN solution is associated with the ``axion charge'' of the KR gravitational axion, $b(x)$, which is ``dual'' to the four-dimensional KR field strength \eqref{eq:21}, and is a physical excitation of the string spectrum. Hence, the KR monopoles have different properties and observational signatures from the emergent monopole of Ref.~\refcite{Gera:2020fvo}. Nonetheless, a potential link between the KR and the emergent monopoles, which needs to be explored further in our opinion, lies on the fact that the topological quantization of the magnetic charge characterizing the RN spacetime structures of both types of monopoles, is linked, in the solution of Ref.~\refcite{Gera:2020fvo}, to the Nieh-Yan topological invariant~\cite{Nieh:1981ww,Chandia:1997hu,Chandia:1997jf,Nieh:2007zz} which characterizes geometries with torsion. It is well known~\cite{Polchinski:1998rq,Polchinski:1998rr} that the KR field strength \eqref{eq:21}, appearing in the string-inspired solution of Refs.~\refcite{Mavromatos:2016mnj,Mavromatos:2018drr}, plays the role of a totally antisymmetric component of a spacetime torsion, hence the topological quantization rule of the magnetic charge might also be applicable to such solutions.\footnote{The closed form of the KR torsion  \eqref{eq:21} would imply a Bianchi identity $\epsilon_{\mu\nu\rho\sigma} \partial^\sigma H^{\mu\nu\rho} =0 $ which would lead to a vanishing Nieh-Yan invariant~\cite{Bossingham:2018ivs}. However, as mentioned in Sec.~\ref{ref:KRmon}, in string theory~\cite{Polchinski:1998rq,Polchinski:1998rr} there are modifications to the Bianchi identity induced by gravitational and gauge Chern-Simons terms, which need to be explored further in connection with the aforementioned quantization rule, before definite conclusions are reached in this respect.}

On closing this section we mention that there are several studies of variants of (or related to) the Cho-Maison monopole within the extended electroweak theory, predicting relatively light monopoles, the production of which is in principle feasible at current and future colliders; for a partial list of references the reader is referred to Refs.~\refcite{Blaschke:2017pym,Benes:2019ext,Teh:2014xva,Teh:2013rpa,Pak:2013jaa}. Other theoretical models of monopoles, not necessarily of low mass, include brane/strings and supersymmetric models, Nambu-type string-like configurations in the minimal SM consisting of {\it confined} monopole--antimonopole pairs connected by flux strings of the $Z^0$ gauge boson field (``dumbbell' monopolium configurations)~\cite{Nambu:1977ag}, as well as evolutions of local and global monopole and cosmic string networks. Since we do not discuss them in our review, we refer the interested reader in the relevant literature for details.~\cite{Wen:1985qj,Sakai:2005sp,Tong:2005un,Weinberg:2006rq,Martins:2008zz,Lopez-Eiguren:2016jsy,Achucarro:2019blr,Achucarro:1999it,Saurabh:2019rrp}$^\text{ and references therein}$ 

In addition to monopoles with magnetic charges that are integer multiples of the Dirac charge, there are also milli-magnetically charged particle-like objects, in theories with two photons, involving kinetic mixing between the photons. Indeed, if one of the photons is a ``magnetic'' (dual) one, as in the Dirac-string-free formulations of a magnetic monopole~\cite{Cabibbo:1962td,Salam:1966bd,Zwanziger:1970hk},  then the mixing parameter in the kinetic terms of these two photons can suppress the magnetic charge to milli-charge fractions~\cite{Hook:2017vyc}, in an analogous way to the milli-electric-charge particles in standard theories with dark photons. Phenomenological searches and detection strategies for such milli-magnetic-charge objects are reported in  Refs.~\refcite{Hook:2017vyc,Chandra:2019dnf,Bellini:2020kub}. Last, but not least, we also mention a very recent work on 't Hooft-Polyakov magnetic monopole-like solutions in {\it non-Hermitian} non-Abelian gauge theories~\cite{Fring:2020xpi}, within a Parity-Time-reversal (PT)-symmetry framework~\cite{Bender:2005tb,Bender:2005hf}. 

\section{Monopole Phenomenology \label{sec:pheno}} 

In this section we shall discuss the basic phenomenological principles, which the production of monopole--antimonopole pairs at colliders and in the early Universe is based upon. We shall commence our discussion by the study of scattering of matter off monopoles, paying particular attention to subtleties associated with the construction of an effective field theory, which is a topic still far from being complete. This first part is relevant for a better understanding of the limitations of data analysis in collider searches for monopoles,  which are discussed in Section \ref{sc:colliders}.

\subsection{Monopole scattering off matter and effective field theories}\label{sc:scatter}

When considering scattering of a point particle carrying electric charge $e$ off a dyon carrying magnetic charge $g$ and electric charge $e_d$,  within a non-relativistic quantum mechanical framework,\footnote{A fully fledged quantum field theory for the monopoles is not yet available, although there are attempts. The first field theoretic attempt for a local field theory without the Dirac string has been considered in Ref.~\refcite{Zwanziger:1970hk}, using two electromagnetic potentials, but only one propagating on-shell degree of freedom via an appropriate constraint. However, its second quantization was not considered. We shall come back to it later in the section. Relativistic scattering off magnetic charges stemming from microscopic theories, has been considered, e.g. see Ref.~\refcite{Boulware:1976tv}, and compared with the scattering off structureless background Dirac monopoles, but the full quantum fluctuations of the monopole have not been taken properly into account. Moreover, a non-local field theory (dual electrodynamics) has been considered in its high energy scattering limit in Ref.~\refcite{Gamberg:1999hq}. Other formal field theoretic actions for the monopole exist, for instance the manifestly dual quantum field theory proposal for electric and magnetic charges of Ref.~\refcite{Lechner:1999ga}, in which the Dirac string becomes dynamical.  All such formal approaches are interesting and worthy of pursuing, however, in our opinion, they are not yet in a form that could be used for practical applications in discussing their interaction with matter, of interest to monopole searches.} as appropriate for small relative velocities $v$ of the scattered dyons, one arrives at the following differential cross section, for small scattering angle, sufficient for our purposes in this section~\cite{Schwinger:1976fr,Milton:2006cp}:
\be\label{ddcs}
\frac{d\sigma}{d\Omega} \simeq \left(\frac{1}{2\mu v}\right)^2 \, \Big[ (e g)^2 + \frac{(e\, e_d)^2}{v^2} \Big]\, \frac{1}{(\theta/2)^4}~.
\ee
Let us concentrate on the case of the magnetic monopole ($e_d=0$).  One observes that the cross section~\eqref{ddcs} in this case can be obtained from the Rutherford differential cross section 
\be\label{Rutherford1}
\left.\frac{d\sigma}{d\Omega}\right|_\text{Ruth}=\left(\frac{e^2}{2\mu v^2}\right)^2\frac{1}{(\theta/2)^4}
\ee
upon the replacement $e^2\to e g_\text{eff}$, where the effective monopole charge is~\cite{Milton:2006cp}
\be\label{geff1}
g_\text{eff}\equiv g  v \equiv g  \beta.
\ee
Upon invoking {\it electric--magnetic duality}, one therefore might expect that $g_\text{eff}$ defines an effective ``velocity-dependent'' magnetic charge that describes the behavior of a magnetic monopole in matter (or equivalently its production from the collision of SM matter particles, such as quarks or charged leptons, at colliders).  

Effective $U(1)$ gauge theories with the coupling~\eqref{geff1} of monopoles to ordinary photons are used in collider searches for extracting monopole mass limits from search data. In such cases, one employs Lorentz-invariant effective magnetic charges, using the center-of-mass velocity~\cite{Kurochkin:2006jr,Dougall:2007tt,Epele:2012jn,Baines:2018ltl}
\be\label{defbeta}
\beta = \sqrt{1 - \frac{4M^2}{s}},
\ee
where $M$ is the monopole mass and $s=(p_1+p_2)^2$ is the Mandelstam variable, with $p_i$, $i=1,2$, the momenta of the colliding particles; quarks or photons, in Drell-Yan (DY) or photon-fusion processes for monopole production, respectively, to be discussed in more detail in Sections~\ref{sc:composite} and~\ref{sc:cross}. The strong nature of the magnetic charge, as a result of the Dirac quantization \eqref{dirac}, invalidates any attempt to use such effective perturbative Feynman graph approach to interpret the data. 

The important feature of the ``velocity-dependent'' magnetic coupling \eqref{geff1}, \eqref{defbeta} lies exactly on its perturbative nature for small $v$, which could validate small coupling expansions of {\it near-threshold production} of magnetic monopole--antimonopole pairs ($\beta \ll 1$), thus partially allowing for placing monopole mass lower bounds in collider searches~\cite{Baines:2018ltl}, to be discussed further in Sec.~\ref{sc:scatter}. The result \eqref{geff1}, however, is disputed, since at present there is no rigorous justification for its validity,  given that the fundamental effective field theory underlying monopoles production and/or scattering of matter is not fully understood. 

Moreover, the use of a local and Lorentz $U(1)$ gauge invariant effective action to describe physical processes involving the monopole is in contradiction with a well-known paradox (termed ``Weinberg paradox'')~\cite{Weinberg:1965rz}, according to which a single electromagnetic-photon exchange between an electric and a magnetic current leads to a non-Lorentz-invariant result, as a consequence of the Dirac string. This paradox stems from the fact that magnetic monopoles, such as the HP or other configurations (like those studied above) arise as detailed solutions of the classical equations of motion of Lorentz- (and gauge-) invariant field theories. 

A modest attempt towards proving the existence of a velocity momentum effective magnetic charge, but also resolving the Weinberg paradox, has been undertaken in Ref.~\refcite{Alexandre:2019iub}, where an effective $U(1)_{\rm em} \times U(1)_{\rm dual}$ gauge field theory for monopole--photon coupling has been developed, where $U(1)_{\rm em}$ denotes the (weakly) coupled gauge group of electromagnetism, while $U(1)_{\rm dual}$ is a strongly coupled (``dual'') gauge group. One uses a Schwinger-Dyson (SD) approach for resumming the photon--monopole graphs in a gauge and Lorentz-invariant way, using techniques developed in Ref.~\refcite{Terning:2018udc}. The approach has its starting point in the two-potential formalism of Ref.~\refcite{Zwanziger:1970hk} for constructing a {\it local} effective action for the magnetic monopole, avoiding the use of Dirac strings.
 
The SD analysis of Ref.~\refcite{Alexandre:2019iub}, results in a dressed monopole--photon vertex, implying a renormalized coupling $Z (k) e_A$. The quantity $e_A$ denotes the bare coupling of the monopole to photons and  $Z(k)$ (with $k$ a momentum scale) is the wave-function renormalization, due to effects of the strongly coupled dual $U(1)_{\rm dual}$ gauge field:
\be\label{zea}
Z(k) \simeq \frac{Z_0}{k_0}(k-k_0)~, \quad 
Z_0=\frac{e_A^2+e_B^2}{8\pi^2+3(e_A^2+\tilde e_B^2)M_0/M}~, \quad \tilde e_B^2 =\frac{3\,e_B^2\, (e_A^2+e_B^2)}{3\,e_A^2+7\,e_B^2}\, ,
\ee
with $e_B$ the bare coupling of the dual $U(1)_{\rm dual}$ gauge field to the monopole. The quantities  $M_0, k_0$ are fixed mass scales. One may take $k_0=2M_0$, and interpret the mass scale $M_0$ as the monopole (rest) mass~\cite{Alexandre:2019iub}. The bare couplings $e_A$ and $e_B$ are renormalized ($R$) in a similar way, through $Z(k)$: $(e_A)_R = Z(k) e_A$, $(e_B)_R = Z(k) e_B$. The wave-function renormalization \eqref{zea} is purely non-perturbative and vanishes as the bare couplings go to zero, unlike the standard perturbative treatments.

\begin{figure}[ht]
 \centering
 \includegraphics[width=0.4\textwidth]{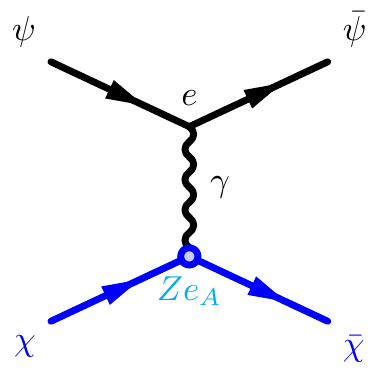} 
\caption{Typical scattering graph between an electron ($\psi$, black line) and a fermion-monopole ($\chi$, blue line), involving the exchange of a single electromagnetic photon ($\gamma$, wavy line). The blue blob denotes the dressed coupling $Ze_A$ due to quantum corrections induced by the strongly coupled dual photon.}
\label{fig:scat}
\end{figure}

On considering the single electromagnetic photon exchange  graph of Fig.~\ref{fig:scat} in this $U_{\rm em}(1) \times U(1)_{\rm dual}$ theory, one may compute the corresponding differential cross section in a frame where the monopole is initially at rest. The Feynman diagram of Fig.~\ref{fig:scat} makes sense if the dressed coupling of the monopole to the electromagnetic photon $\gamma$ is small, which necessitates $Z < 1$ \eqref{zea}, and can be achieved for $k \to k_0$. At small scattering angles, of interest to us for comparison with \eqref{ddcs} in the case $e_d=0$, the differential cross section is:
\be\label{ddscat}
\left.\frac{d \sigma}{d \Omega}\right|_{\rm LAB} \simeq \left(\frac{ Z \, e_A\, e }{2 \mu |\vec{v}|^2 \sin^2 \frac{\theta}{2}}\right)^2 \, \stackrel{\theta \ll 1}{\simeq}  \,
\left(\frac{ Z_0 \frac{k - k_0}{k_0} \, e_A\, e }{2 \mu |\vec{v}|^2 }\right)^2 \, \frac{1}{(\theta/2)^4}, 
\ee
where $\mu$ is the reduced mass of the electron--monopole system. 

Comparing \eqref{ddscat} with \eqref{ddcs} (in the magnetic monopole case ($e_d=0$)),   yields an expression for the effective coupling of the monopole to photons, which is identified with the effective magnetic charge \eqref{geff1}: \bea\label{geff4}
g_\text{eff} &=&Ze_A \simeq Z_0\, \frac{k-k_0}{k_0} \, e_A~,  \nonumber \\
\frac{|k - k_0|}{k_0} & \rightarrow & \frac{\sqrt{E^2 - 4M^2_0}}{2M_0} = \frac{E}{2M_0}\, \sqrt{1 - \frac{4M_0^2}{s}} \simeq \beta,
\eea
where we identified the scale $|k-k_0|$  with a proper (Lorentz invariant) center-of-mass momentum scale, consistently with the effective field theory approach for monopole--antimonopole pair production~\cite{Kurochkin:2006jr,Dougall:2007tt,Epele:2012jn,Baines:2018ltl};  $E^2=s=(p_1+p_2)^2$ is the relevant Mandelstam variable  (with $p_1$ ($p_2$) the incoming (outgoing) four momenta of the electron $\psi$ in the scattering process of Fig.~\ref{fig:scat}). The magnetic charge is identified with 
\be\label{magneticcharge}
g=Z_0 \, e_A~.
\ee
The DQC for the monopole \eqref{dirac} would then require 
\be\label{zquant}
Z_0 \, e_A =  \frac{n}{\alpha} \, e~, \quad  n \in \mathbb Z~,
\ee 
with $\alpha$ the fine structure constant of electromagnetism. Compatibility with the quantization condition \eqref{zquant} restricts the range of the model parameters~\cite{Alexandre:2019iub}. The above effective description can also describe production of monopole--antimonopole pairs from charged matter, taking into account that by rotating the graph of Fig.~\ref{fig:scat} by 90\textdegree\ counterclockwise. 

An important remark is in order at this stage, concerning the fact that the wave-function renormalization is less than one, $Z < 1$,  in the regime of interest. This would violate the unitarity bound~\cite{Peskin:1995ev} that requires the wave-function renormalization to be larger than one, should the monopole be an elementary particle asymptotic state. However, such a bound can be evaded for {\it composite} states, either of known type discussed in the monopole literature so far, and reviewed in previous sections of this review, or new, yet unknown, structured solutions to be discovered in theories beyond the Standard Model. Hence, for consistency, the effective field theory of Ref.~\refcite{Alexandre:2019iub}, should describe only composite monopoles, like the ones we have discussed in our previous section. 

\begin{figure}[ht]
  \centering
  \includegraphics[height=0.241\textwidth]{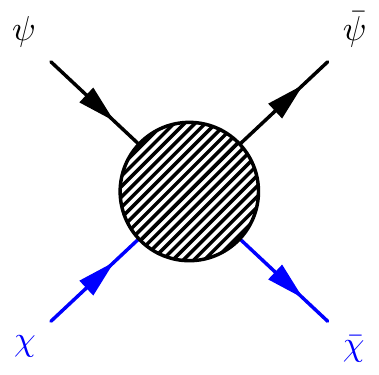} \hfill
  \includegraphics[height=0.241\textwidth]{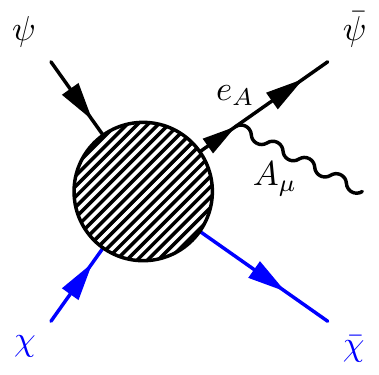} \hfill
  \includegraphics[height=0.241\textwidth]{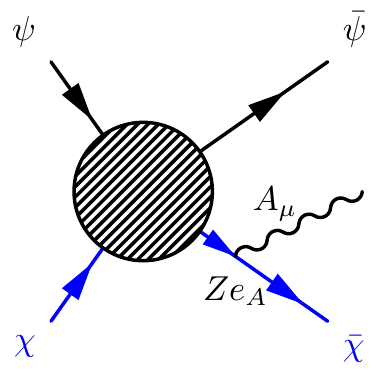} \hfill
  \includegraphics[height=0.241\textwidth]{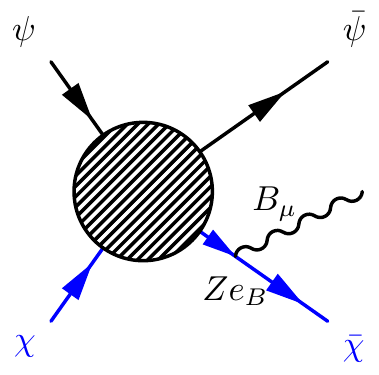} 
  \caption{Typical scattering graphs between a charged matter fermion $\psi$ (black lines) and a spin-\half monopole $\chi$ (blue lines) in a two-potential  $U(1) \times U(1)$ effective field theory. Wavy lines denote real (on-shell) soft electrodynamical photons $A_\mu$, or the dual gauge bosons $B_\mu$ of the strongly coupled $U(1)$, the latter associated only with the monopole/dyon. The gray blobs denote generic scattering processes, involving all fields. For slowly moving monopoles, $Z \ll 1$ ensuring perturbativity of all couplings, hence the depicted graphs denote the leading corrections in both $A^\mu$ and $B^\mu$ sectors. In such a case, resummation of the soft on-shell photons exponentiates the Dirac-string effects into phases of the pertinent scattering amplitudes~\cite{Terning:2018udc}, thus resolving the Weinberg paradox~\cite{Weinberg:1965rz}.}
\label{fig:IR}
\end{figure}

In the model of Ref.~\refcite{Alexandre:2019iub}, both the magnetic and electric charge couplings of the monopole to photon and dual photon are {\it perturbative} for {\it slowly moving} monopoles, due to the wave-function renormalization screening effects on the effective magnetic charge \eqref{magneticcharge}. For such perturbative couplings, the leading soft gauge field (photons and dual photons) emissions that affect generic scattering processes of electrons off fermion monopoles are depicted in Fig.~\ref{fig:IR}. The resummation of such processes can be performed as in Ref.~\refcite{Terning:2018udc}, leading to the  exponentiation of any (Lorentz-violating) Dirac-string dependent terms, that they now appear {\it only} in the phases of the amplitude~\cite{Terning:2018udc}, and thus do not affect the cross section. We also remark, however, that upon the DQC \eqref{dirac}, these effects disappear also from the phase of the scattering amplitude~\cite{Terning:2018udc}, thus resolving the ``Weinberg paradox''~\cite{Weinberg:1965rz}. 

This completes our discussion on the self-consistency of the use of the perturbative effective $U(1)$ gauge field theory approach to the (slow) monopole--matter scattering, as well as the (near-threshold) monopole--antimonopole pair-production from Drell-Yan or photon fusion processes. We shall discuss in some detail such collider production~\cite{Kurochkin:2006jr,Dougall:2007tt,Epele:2012jn,Baines:2018ltl} and monopole--matter scattering~\cite{Vento:2018sog,Vento:2019auh} processes in Sec.~\ref{sc:cross} and Sec.~\ref{sc:box}, respectively. 

On closing this section, we would like to mention that there are other relativistic effective field theory approaches to describe quantum effects of the interaction of magnetic monopoles with photons, which do not address the aforementioned issue of ``velocity-dependent'' magnetic charge. Such approaches can be roughly separated into two classes, depending on how the quantum effects renormalize the electric $e$ and magnetic  $g$ charges. In the first class, pioneered by Schwinger, the magnetic and electric charges are renormalized in the same way, given that the corresponding renormalized couplings are expressed in terms of a {\it common} wave-function renormalization~\cite{Schwinger:1966nj,Schwinger:1966zza,Schwinger:1966zzb,Deans:1981qs,Panagiotakopoulos:1982fp,Panagiotakopoulos:1982ne}. In this class of approaches, DQC \eqref{dirac} seems to depend on the renormalization scheme. From this point of view, the model of Ref.~\refcite{Alexandre:2019iub} belongs to this class of approaches, but, as already mentioned, the solution \eqref{zea} of the wave-function renormalization is novel and purely non-perturbative. The second class of approaches to renormalization of the magnetic charge~\cite{Coleman:1982cx,Calucci:1982wy,Calucci:1982fm,Goebel:1983we,Tolkachev:1992qx} argues in favor of the magnetic and electric charges wave-function renormalization functions, $Z_g$ and $Z_e$, respectively, satisfying $Z_g  Z_e =1$, which implies that the DQC \eqref{dirac} is independent of the renormalization point.  For a comprehensive discussion on, and further comparison between, these approaches we refer the interested reader to the literature~\cite{Shnir:2005xx,Blagojevic:1985sh} and references therein.

\subsection{Composite versus pointlike (Dirac) monopoles at colliders}\label{sc:composite}

Having discussed detailed models of light monopoles, based on spontaneously broken gauge theories, it is interesting to discuss their production at colliders. As we shall see below, a small mass is not the only criterion for production, given that composite monopoles with structure, as opposed to pointlike monopoles, envisaged by Dirac, exhibit extremely suppressed production rates at colliders. This was discussed in Ref.~\refcite{Drukier:1981fq} some time ago, in the context of the 't Hooft-Polyakov monopole~\cite{tHooft:1974kcl,Polyakov:1974ek}, but the arguments can be straightforwardly generalized to apply to the other structured monopoles we mentioned above. The main idea is that a composite monopole, like the ones we discussed in the previous sections (HP, GUT, CM and variants, KR monopoles) are finite extended objects, and as such can be represented as {\it coherent} quantum states of the quanta of the constituent fields. Below we review briefly those arguments.

\begin{figure}[ht]
\centering
\begin{minipage}{0.35\textwidth}%
     \subfloat[Pointlike monopoles\label{fig:ff-mm-DY}]{%
       \includegraphics[width=\textwidth]{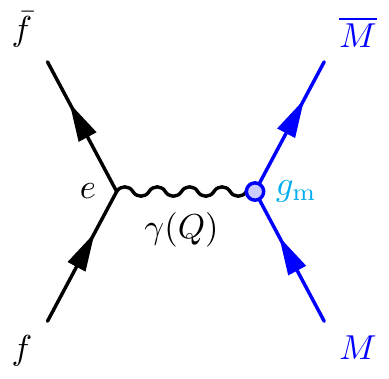}}
       \end{minipage}    
       \hspace{0.05\textwidth}
       \begin{minipage}{0.35\textwidth}%
     \subfloat[Composite monopoles\label{fig:ff-mmcomposite-DY}]{%
       \includegraphics[width=\textwidth]{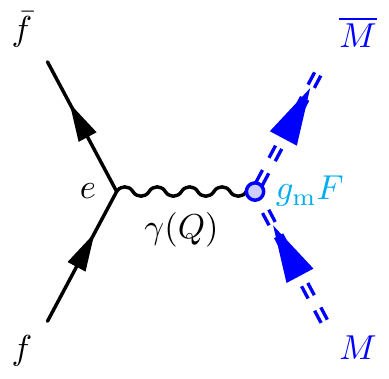}}
\end{minipage}       
     \subfloat[Decomposed monopoles\label{fig:ff-mm-decomposed}]{%
       \includegraphics[width=0.65\linewidth]{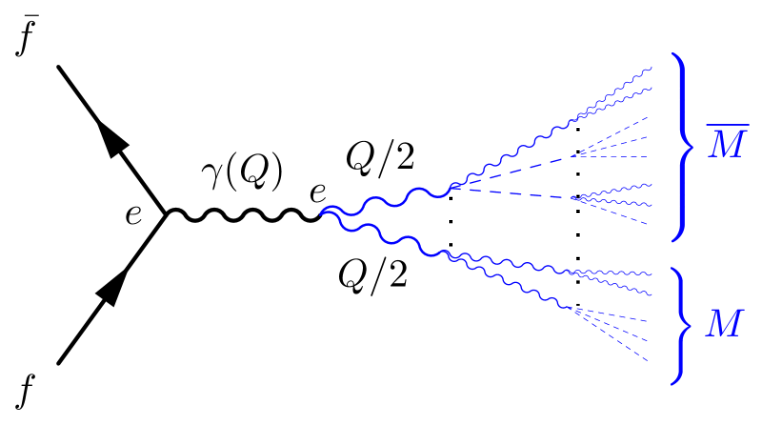}}
\caption{Cross sections for the production of pointlike (Dirac) monopoles at colliders, through the scattering of SM fermions (cf.\ Fig.~\protect\subref{fig:ff-mm-DY}), are very different from the corresponding processes involving production of composite monopoles (cf.\ Fig.~\protect\subref{fig:ff-mmcomposite-DY}). The latter processes involve the creation of coherent quantum states of the constituent particles (Higgs-like scalars, gauge bosons, etc.), which eventually result in the production of monopole--antimonopole pairs (cf.\ Fig.~\protect\subref{fig:ff-mm-decomposed}), only after a sufficient number of field quanta (of order $1/\alpha$, where $\alpha$ is the fine structure constant of the gauge group of the model that contains monopoles) is perturbatively generated. As a result, the corresponding production is extremely suppressed. \label{fig:dn}}
\end{figure}

As we have discussed above, a HP monopole comprises Higgs (``$\phi$'') and non-Abelian (``$W$''-)gauge boson field quanta. The authors of Ref.~\refcite{Drukier:1981fq} estimated that the average numbers of the quanta of those constituent fields, $\overline n_\phi$, $\overline n_W$, respectively, that constitute the magnetic monopole are {\it both} of order $1/\alpha$, where $\alpha$ is the fine structure constant of the gauge group in question. The basic assumption that lead them to such estimates was that the self-interaction scalar coupling was of order $\lambda \sim \alpha$. If a monopole occupies a volume $\delta V$, then the Higgs field has an average magnitude $\langle \phi\rangle  = \phi_0$ over that volume, where $\langle \dots \rangle $ denotes its vacuum expectation value. The authors of Ref.~\refcite{Drukier:1981fq} made an analogy of the Higgs field quanta inside the volume $\delta V$ with a collection of quantum harmonic oscillators in a coherent quantum state $|\xi \rangle $ with $\xi = \phi_0$. The average number of quanta in a coherent state of the harmonic oscillator with Hamiltonian $H_{\rm osc} = \frac{1}{2} k \xi^2 + \frac{1}{2}\, \mu \, {\dot \xi}^2$, where the overdot denotes time derivative, is $\overline n_\xi = \frac{1}{2}\frac{k\, \xi^2}{\hbar \, \omega}$, where the frequency of the harmonic oscillator is $\omega= (k/\mu)^{1/2}$. In the field theoretic analogue of the Higgs field, of interest here, one has a field-theoretic Hamiltonian density ${\mathcal H} = \frac{1}{2} \dot \phi^2 + \frac{1}{2} m_H^2 \phi^2$, where $m_H$ is the Higgs mass in the spontaneously broken phase of the gauge group. Thus, the analogy with the harmonic oscillator problem goes through upon making the correspondence $\mu \rightarrow 1, k \rightarrow m_H^2$ and $\omega \rightarrow m_H$, which implies that the Higgs-field quantum density in the monopole volume is 
 \be\label{hquant}
 \frac{dn_\phi}{dV}  = m_H\, \phi_0^2~.
 \ee
 In the HP prototype monopole~\cite{tHooft:1974kcl,Polyakov:1974ek}, as we have discussed, the monopole core radius is $R_m \sim  m_W^{-1}$, where $m_W$ is the mass of the $W$-gauge bosons in the spontaneously broken gauge symmetry phase, with $m_W \sim m_H =\lambda^{1/2} \, \phi_0$. hence, from \eqref{hquant}, we have that in a typical monopole volume $V_m=\frac{4}{3}\pi R_m^3$, the average number of Higgs quanta is ~\cite{Drukier:1981fq}
 \be\label{thq}
\overline n_\phi = V_m m_H\, \phi_0^2 = \frac{4\, \pi }{3} \, \frac{m_H}{q_e^3 \, \phi_0} \sim \frac{\lambda^{1/2} \, 4\pi}{q_e^3} \sim \frac{1}{\alpha}~,
\ee
where $q_e $ denotes the electric charge, satisfying the quantization rule \eqref{hpcharge}. The average number of gauge quanta is also estimated to be of the same order 
\be\label{thq2}
\overline n_W \sim \frac{1}{\alpha}~, 
\ee
since inside the monopole the magnitude of the gauge fields is or order $A_\mu \sim g_m/R_m \sim \frac{m_W}{q_e} =\phi_0$, with $g_m$ the magnetic charge satisfying the quantization rule \eqref{hpcharge}. In fact, the estimates \eqref{thq}, \eqref{thq2} are consistent with the interpretation of the large magnetic coupling of the monopole $g_m =\frac{1}{\alpha} \, q_e$ (cf.\ \eqref{dirac2}), which characterizes production processes involving the decays of an intermediate  photon to monopole--antimonopole pairs (see Fig.~\ref{fig:dn}), as a collective coupling of $1/\alpha$ bosons inside the monopole to a soft photon. For further arguments on the consistency of this picture, which is in agreement with generic expectations on the behavior and properties (including stability) of monopoles in the quantum theory, we refer the reader to Ref.~\refcite{Drukier:1981fq}.

We next use the above results to estimate the production cross section of a monopole--antimonopole pair at colliders. Typical processes are the ones depicted in Fig.~\ref{fig:dn}, involving annihilation of charged SM matter fermions (charged leptons or quarks) into intermediate photons, which then decay to monopole--antimonopole pairs ({\it Drell-Yan} processes). As we shall argue below, following Ref.~\refcite{Drukier:1981fq}, there is an extreme suppression of such processes if the monopoles are composite. For other processes on monopole production through {\it photon fusion} (see Sec.~\ref{sc:cross}, below), similar suppression arguments apply, so restricting our discussion to Drell-Yan production suffices. 

To this end, we consider first a typical tree-level Feynman diagram for the Drell-Yan production of pointlike (structureless, Dirac-type) monopole--antimonopole pairs, $M\overline M$, like the one depicted in Fig.~\ref{fig:ff-mm-DY}. The corresponding total cross section is easily estimated from this tree level graph:
\be\label{xsecdy}
\sigma^{(0)} ( e^+e^-\rightarrow M\overline M )\simeq \frac{4\pi}{3} \, \frac{\alpha_e \, \alpha_m}{Q^2} \sim \frac{1}{Q^2},
\ee
where $Q$ indicates the momentum of the intermediate photon, and $\alpha_e (\alpha_m) $ are the fine structure constants for the electric $q_e$ (magnetic $g_m$) coupling constants, satisfying the quantization rule \eqref{hpcharge}, which leads to the approximate equality on the right-hand side of \eqref{xsecdy}. The caveat here is that the strong magnetic coupling invalidates any calculation based on this graph, but this does not affect our arguments in this section; for more discussion see Sec.~\ref{sc:cross}

The production cross section of composite monopole--antimonopole pairs like the one depicted in Fig.~\ref{fig:ff-mmcomposite-DY} is expected to be suppressed by a form factor $|F^2|$, compared to \eqref{xsecdy}:
 \be\label{xseccomp}
\sigma (e^+e^-\rightarrow M\overline M)_{\rm comp} = |F^2|\, \sigma^{(0)} (e^+e^-\rightarrow M\overline M )~,
\ee
which we now proceed to estimate, following Ref.~\refcite{Drukier:1981fq}. 

To this end, we first note that the intermediate steps involved in the process (see Fig.~\ref{fig:ff-mm-decomposed}) can be handled using naive perturbation theory, especially if the intermediate processes involved are electroweak, which is most relevant for experimental monopole searches at current colliders. Since in view of our previous discussion, each monopole contains $2/\alpha$ scalar (Higgs-like) and gauge-boson quanta (\eqref{thq}, \eqref{thq2}), the latter are produced by at $2/\alpha$-order in the respective perturbation theory, and as a result the respective cross section will be given by the corresponding for pointlike monopole production, suppressed by the square of a form factor $F$:
\be\label{form}
F \simeq (q_e)^{2/\alpha},  \quad |F^2| \simeq (q_e)^{4/\alpha}. 
\ee
Expressing the $q_e$ (``electric charge'') coupling in terms of $\alpha$ as $q_e = (4\pi \alpha)^{1/2}$, and using that for the electroweak Drell-Yan processes of Fig.~\ref{fig:dn} one has in order of magnitude $\alpha \sim e^{-1}$, where $e\simeq2.72$ is the exponential  mathematical constant, we arrive from \eqref{form} at~\cite{Drukier:1981fq}
\be\label{formvalue}
|F|^2 \sim e^{-4/\alpha} \sim 10^{-250}~.
\ee
One arrives at similar suppression by applying more sophisticated arguments, as discussed in Ref.~\refcite{Drukier:1981fq}. Notice that the form of the suppression, with the inverse square coupling in  the exponent of the exponential in \eqref{formvalue} is reminiscent of the probability for tunnelling processes in quantum mechanics. This indicates that any attempt to produce composite monopoles in current colliders seems futile, and one has to resort to cosmic searches, where such composites monopoles have been produced in the early Universe. 

\subsection{Monopole vacuum production in strong external conditions}\label{sc:schwinger}

However, there may be an alternative, which can put collider searches for composite monopoles back on track. This is associated with unsuppressed monopole--antimonopole pair ($M \overline M$) production from the \emph{vacuum} \`a la Schwinger~\cite{Schwinger:1951nm}, but in strong external conditions, such as high temperature and/or strong magnetic fields. 
It was shown~\cite{Affleck:1981ag} (and also argued independently~\cite{Drukier:1981fq}) that such a mechanism will lead to unsuppressed $M\overline M$-pair production, with a rate of production per unit volume (in the presence of a magnetic field at zero temperature):
\be\label{schpair}
\frac{d N_{M \overline M}}{dV dt} \simeq g_m^2 \, B^2 \, \exp \left(-\frac{m_M^2}{g_m  B^2}\right)~,
\ee
where $m_M$ is the mass of the monopole $M$, $g_m$ is the magnetic coupling, $V$ denotes the spatial volume, and $B$ is the external magnetic field. 

Strong magnetic fields are required in order to lead to appreciable rates. Recently this issue was re-examined in more details in Ref.~\refcite{Gould:2017fve}, and actually the case was generalized to include thermal environments as well, corresponding to high temperatures and external magnetic fields, such as neutron stars or heavy-ion collisions at LHC~\cite{Gould:2017zwi}. The perturbative computations of Ref.~\refcite{Gould:2017fve} were valid as long as 
\be\label{range}
m_M^2 \gtrsim g_m^2\, B, \quad T \sqrt{2} \pi^{-3/4} \Big(\frac{g_m \, B^3}{m_M^2}\Big)^{1/4}~,
\ee
and the $M\overline M$-production rate per unit volume $\Gamma_T$ is estimated to be~\cite{Gould:2017fve}:
\begin{gather}
 \log(\Gamma_T) = -\frac{m_V^2}{g_m B} \left[ \tilde S(g_m, m_M, B, T) + 
{\mathcal O}\left( \frac{g_m B}{m_M^2}\, \log \left( \frac{g_m B}{m_M^2} \right) \right)\, \right],  \nonumber \\
 \tilde S(g_m, m_M, B, T) = 2\, \frac{g_m  B}{m_M  T}\, \left(1 - \sqrt{\frac{g_m^3 B}{4\pi m_M^2}}\right) .
\end{gather}
In heavy-ion environments, such as those of LHC (or the interior of neutron stars), where the above conditions can be met, the total cross section $\sigma_{M\overline M}$ for the thermal-and-magnetically-catalyzed $M\overline M$-production is expressed in terms of the inelastic differential heavy-ion (HI) collision cross section $d\sigma^{\rm inel}_{{\rm HI}}/db$ as~\cite{Gould:2017zwi}
\be
\sigma_{M\overline M} = \int  db \, \frac{d\sigma^{\rm inel}_{{\rm HI}}}{db}\, p(b), \quad p(b) = \int_{{\mathcal R}(b)} d^4x \, \Gamma_T\big(m_M, g_m, B(x; b), T(x; b)\big),
\ee
where $b$ is the impact parameter, and  ${\mathcal R}(b)$ denotes the spacetime region over which the fireball that is created during the HI collisions is extended. There are strong magnetic fields inside that fireball, as a result of the fast moving electric charges. 

In HI LHC collisions, considered in Ref.~\refcite{Gould:2017fve}, the magnetic field has been estimated~\cite{Gould:2017zwi} to be $B_{160~{\rm GeV}} = 0.0097~\gev^2$. The upper bound (UB) on $\sigma_{M\overline M} < \sigma^{\rm UB}_{M\overline M} = 1.9~{\rm nb}$ is taken into account in Ref.~\refcite{Gould:2017zwi} to determine the minimum allowed monopole mass for such a process to take place, via the inequality
\be
\log\left(\frac{\sigma_{\rm HI}^{\rm inel}}{\sigma^{\rm UB}_{M\overline M}}\right) \lesssim \frac{m_M^2}{g_m\, B}\,  \tilde S(g_m, m_M, B, T)~, \ee
from which the minimum allowed monopole mass is estimated to be~\cite{Gould:2017fve}:
\be\label{tsdlim}
m_M \gtrsim \left(2.0 + 2.6  \left( \frac{g_m}{\gd} \right)^{3/2}\right)~\gev,
\ee
with $\gd$ the fundamental Dirac charge \eqref{fdc}.

The above limit  might look much weaker than the one imposed in collider searches for magnetic monopoles (see Section~\ref{sc:colliders}), Nonetheless, it is based in a much more physically realistic situation,  given the aforementioned suppression of the composite-monopole production cross section, which invalidates the collider bounds for composite monopoles. Moreover, thermal production of monopoles is more or less independent of their compositeness. We shall discuss more about the relevant research strategies in the next section.

Before closing this section, we would like to mention that inhomogeneities of the magnetic field in HI collisions have been taken into account in Ref.~\refcite{Gould:2019myj}. Unfortunately, the computations of the associated cross sections for the monopole--antimonopole pair production are limited to small-size monopoles, smaller than the curvature of the worldline instanton employed in the calculations of Ref.~\refcite{Gould:2019myj}. This makes the study not suitable for realistic HI collisions at LHC. The role of the finite size of the monopoles on their thermal production has also been discussed in the context of HP monopoles~\cite{tHooft:1974kcl,Polyakov:1974ek} in the Georgi-Glashow $SU(2)$ model~\cite{Georgi:1974sy} in Ref.~\refcite{Ho:2019ads}, with the conclusion that the rate for thermal production of finite-size monopoles is higher than that of pointlike monopoles. Moreover, it was shown that the production of monopole--antimonopole pairs occurs as a consequence of classical instabilities of the magnetic fields, above a critical value of their intensity, which equals roughly the value above which the thermal Schwinger vacuum monopole--antimonopole pair production becomes unsuppressed~\cite{Gould:2017fve}. 

\subsection{Monopolium}\label{sc:monopolium}

A possible explanation for the lack of experimental confirmation of monopoles is Dirac's proposal~\cite{Dirac:1931kp,Dirac:1948um,Zeldovich:1978wj} that monopoles are not seen freely because they form a bound state called \emph{monopolium}~\cite{Hill:1982iq,Dubrovich:2002gp,Vento:2007vy} being confined by strong magnetic forces. This hypothetical state could be produced in the laboratory~\cite{Epele:2007ic,Epele:2008un,Epele:2012jn,Barrie:2016wxf,Reis:2017rvb,Fanchiotti:2017nkk} or may have been formed naturally in the early Universe. Such objects, though unstable, could have an interesting physical evolution in time, dependent upon their masses, their initial classical radii and their core structure. For GUT monopoles with mass of the order of $10^{16}~\gev$, the monopolium lifetime can range from days --- for an initial diameter of about a fermi --- up to many times the lifetime of the Universe, when their diameter is $\gtrsim 10~{\rm pm}~$\cite{Hill:1982iq}. While behaving as a classical system, they will radiate characteristic dipole radiation up to high energies. Hence a monopolium system provides a window on the physics of elementary processes up to the extremely high energy scale characterized by its mass and could, in principle, yield information about the physics between current collider energies and the grand unification scale.

The production of monopolium states in colliders and its subsequent decays to di- or multi-photon final states, as well as proposals for its detection, has been studied thoroughly in the literature~\cite{Epele:2007ic,Epele:2008un,Epele:2012jn,Barrie:2016wxf,Reis:2017rvb,Fanchiotti:2017nkk}. We stress here that monopolia can be relatively light, due to their strong binding, and can therefore serve as possible probes of heavy (yet not of GUT scale) monopoles in colliders\cite{Epele:2008un}. This aspect of the monopolium phenomenology is discussed in some detail in Sec.~\ref{sc:cross}, while the scattering of charged particles off monopolia~\cite{Vento:2018sog,Vento:2019auh} is briefly presented in Sec.~\ref{sc:scatter}.

\subsection{Monopoles and the dark sector of the Universe}\label{sc:dark}

The possibility that pointlike topological defects (PLTD), such as monopoles, may be viable dark matter candidates has been considered in Ref.~\refcite{Murayama:2009nj}. The originally proposed Kibble mechanism~\cite{Kibble:1976sj} for the production of such defects in the early Universe, which is associated with the  topology of  the coset space $G/H$ of the spontaneous  breaking of a symmetry group $G$ down to a subgroup $H  \subseteq G$, with  a (nontrivial) second homotopy group $\pi_2(G/H) \ne 0$, drastically underestimates the abundance of such primordial defects. Kibble mechanism actually provides a {\it lower} bound for the density of PLTD which is roughly one per horizon. This, in turn, would imply the following estimate for the density $n_{\rm D}$ of such defects per entropy density $s$ of the Universe: $n_{\rm D}/s \sim (T_c^3/M_{\rm P})^3$, where $T_c$ is the critical temperature for the phase transition in the early Universe associated with their production, which means that only phase transitions close to the GUT scale would produce non-negligible abundances. However, Zurek~\cite{Zurek:1985qw} has subsequently provided a different estimate for the production  of PLTD, predicting substantially increased abundances in the early Universe, such that even phase transitions above a few \tev energies might produce interesting (and even dangerous!) abundance of such defects. 

In fact in Ref.~\refcite{Murayama:2009nj}, it was pointed out that, if such defects, including magnetic monopoles arising from the breaking of appropriate gauge symmetry groups,  are produced non thermally in the early Universe, via the Kibble-Zurek extended mechanism, then they could play the role of  cosmological dark matter candidates with masses in the range  1--$10^9$~\gev.  Indeed, since their correlation length $\xi$ scales with the temperature as: $\xi \sim \xi_0\, | 1-T/T_c|^{-\nu}$, near the critical temperature regime, where $\nu$ is the critical exponent of the second-order phase transition assumed in Ref.~\refcite{Murayama:2009nj}, then the estimated density at production becomes much larger than the one in the original Kibble mechanism, $n_{\rm D}/s\Big|_{T=T_c}  \simeq 0.006 \Big(30\, T_c/M_{\rm Pl}\Big)^{3\nu/(1 + \nu)}$. With $\nu$ close to~\cite{Murayama:2009nj} $\nu \sim 2/3$, this predicts an abundance that can be several orders of magnitude larger than the one by the original Kibble mechanism, implying the potential role of  the PLTD as dark matter. After their production the magnetic monopoles are stable, and their number can only attenuate via monopole--antimonopole annihilation, which has been properly taken into account in Ref.~\refcite{Murayama:2009nj}, leading to a monopole density $n_M \sim 7.9 \times 10^{-22}  (T_c/(1~\tev))$. The latter can be used to  discuss dark matter phenomenology for a range of $T_c$, which leads to the aforementioned mass range for acceptable dark-mater candidates.

The above arguments on significant production of magnetic monopoles in the early Universe appear generic, provided the above assumptions of second order phase transitions and critical behavior are valid. In this sense, the above described type of dark matter phenomenology may include several types of magnetic monopoles produced in GUT gauge theories, including relatively low-mass ones~\cite{Kephart:2017esj}, global monopoles~\cite{Barriola:1989hx}, associated with the spontaneous breaking of $O(3)$ internal symmetry, as well as their string-inspired magnetic monopole extensions~\cite{Mavromatos:2016mnj,Mavromatos:2018drr} discussed in Sec.~\ref{ref:KRmon}. In the latter case, though, one should carefully study the associated phase transitions, as the cosmology of the string universe might lead to different results, depending on the model used. 

In other scenarios involving monopoles, namely primordial black holes and/or domain wall defects in the early Universe, the predictions for significant cosmic abundance of monopoles are rather pessimistic. For instance~\cite{Stojkovic:2004hz}, under some conditions, primordial black holes, produced in the early Universe, can accrete magnetic monopoles before the latter dominate the energy density of the Universe. The fast evaporation of these small black holes, then, leads to the conversion of most of the monopole energy density into radiation. Estimates of the range of parameters for which such a solution is possible, under very conservative assumptions, taking into account the charged nature of the black holes after the monopole capture, have been performed~\cite{Stojkovic:2004hz}, concluding that the black hole mass must be less than $10^9$~g. The black hole abundance is also constrained so as not to violate other observational constraints. In other scenarios, involving domain walls in the early Universe, assumed to be formed through some phase transition, the latter can remove (``sweep away'',``unwind'') the magnetic monopoles~\cite{Dvali:1997sa,Pogosian:1999zi,Pogosian:2000xv,Alexander:2000yx}, through appropriate interactions among these cosmic defects, in the context of specific theoretical models. Moreover, it was argued~\cite{Stojkovic:2005zh} that primordial black holes, that move relatively fast compared to the domain walls, perforate the latter, literary piercing a hole in  them, and thus change their topology. The holes expand with the Universe expansion, and this eventually destroys the domain walls, provided a sufficient number of such holes exists. In this way, the authors of Ref.~\refcite{Stojkovic:2005zh} argue that one obtains a solution of both the domain wall and monopole problems without fine tuning.  

In both of the above scenarios, a small number of monopoles is still expected to be within the cosmic horizon, unlike in the standard scenarios in which inflation wipes out completely GUT scale monopoles. This small allowed window in the above models, might be relevant for cosmic searchers of magnetic monopoles, to be discussed in Sec.~\ref{sc:cosmics}.

In some extra-dimensional models for the Universe, with non-simply connected compact manifolds, cosmic string loops, produced at the end of the inflationary phase~\cite{Dvali:1998pa,Burgess:2001fx}, may wrap around nontrivial cycles of the extra-dimensional manifold, thus becoming topologically stable~\cite{Dvali:2003zj}. Such objects can behave as ``monopole matter'' during the radiation era of the Universe, and they may pose a monopole problem, unless their density is appropriately diluted, depending on parameters of the model. Their cosmic time evolution has been studied in Ref.~\refcite{Avgoustidis:2005vm}, in the context of brane-inflation Universe models. Generically it was found that to avoid domination of their density before the matter--radiation transition in the Universe, the wrapped cosmic-string loops should be extremely light, with linear mass density  ${\rm G} \mu < 10^{-18}$ (with ${\rm G} = M_{\rm P}^{-2}$ the four-dimensional Newton's gravitational constant). However, there are regions in the parameter space of the models, which allow for heavier cosmic string loops ${\rm G} \mu \sim 10^{-14}$, which make them phenomenologically consistent dark-matter candidates. 

Moreover, as already mentioned in the introduction of the review, upon considering SSB in the two-potential formalism for Dirac strings\cite{Cabibbo:1962td,Salam:1966bd,Zwanziger:1970hk}, Singleton~\cite{Singleton:1995cc,Singleton:2011ru}  concluded that a {\it massive} ``magnetic (dual) photon'' emerges from the formalism, under some proper phase transitions in the Universe. This could play the role of a dark matter component (a specific kind of dark photon), and in fact there have already been dedicated experimental laboratory searches for such magnetic photons~\cite{Lakes:2004rc}, stemming from their property to easily penetrate metallic matter, with negative results. However, cosmic searches for such massive, dark, SSB magnetic photons might be worthy of pursuing.

Another association of magnetic monopoles with the dark sector concerns models in which the monopoles live only in the hidden sector of a beyond-the-Standard-Model theory, which could involve extra spacelike dimensions~\cite{Zhang:2019ona,Daido:2019tbm,Sato:2018nqy,Baek:2013dwa,Sousa:2009is,Rahaman:2006kw,Terning:2018lsv}. They could have interesting implications for the observable sector as dark-matter candidates through appropriate portal interactions that connect the dark and visible sectors in such models. Moreover, dark sector monopoles may be used as a toy laboratory for the study of scattering of matter off them, given that in certain models the coupling of the dark-sector magnetic monopole to matter or radiation may be assumed perturbative, in contrast to the case of the observable sector magnetic monopoles, whose coupling to ordinary matter and radiation is strong, as a consequence of the DQC~\eqref{dirac}. 
 
Indeed, one such model was presented in Ref.~\refcite{Terning:2018udc}, in which the dark-sector magnetic monopoles couple perturbartively to the observable sector photons via mixing of the respective Maxwell field strengths. In this way, the authors of Ref.~\refcite{Terning:2018udc} have managed to demonstrate, via resummation of soft-photon emission graphs, that any Lorentz-violating effects of the (dark) magnetic monopole due to the Dirac string can be absorbed in the phase of the matter--monopole scattering amplitude, leaving the cross section Lorentz- (and gauge-) invariant. Upon the imposition of the DQC~\eqref{dirac}, such effects cancel out in the phase, so the amplitude itself is Lorentz invariant. In this way, the model provides a resolution, within this perturbative dark-monopole framework, of the Weinberg paradox~\cite{Weinberg:1965rz}, associated with the violation of Lorentz invariance  in a single-photon-exchange graph between and electric and a magnetic current, due to the Dirac string.

Such interesting avenues for research are worth pursuing in the future, as they may provide useful information that would not only further our understanding of the properties of magnetic monopoles, but also assist our hunt for dark matter, and more generally, our quest for understanding the dark sector of the Universe, including dark energy.

\section{Monopole Detection Techniques}\label{sc:techniques}

Monopole detection techniques rely on the various types of their interactions with matter, which are briefly discussed in this section.  It is worth noting that the same technique may be applied to monopoles of different origin, i.e.\ for cosmic rays or when produced in colliders. However, the sensitivity --- and thus the detector optimization --- may vary and depend, e.g.\ on the monopole velocity. 

\subsection{Ionization and excitation}\label{sc:gas}

Charged particles carrying electromagnetic or magnetic charge will deposit some amount of energy through the ionization process and excitation of atoms when they traverse matter. To calculate the energy loss of monopoles passing through matter, the energy transfer of the monopole to the surrounding medium is considered. There are three primary ways via which the energy can be dissipated and their importance depends on the monopole velocity $\beta$  and the medium~\cite{Giacomelli:2003yu}.
\begin{compactdesc}
\item[Ionization] The energy transferred leads to the production of free electrons. This is given by the Bethe-Bloch formula discussed below and is the dominant energy-loss mechanism for fast monopoles $(\beta>0.05)$ moving in gaseous detectors.
\item[Atomic excitation] The energy is transferred to atoms of higher energy states. It starts to dominate the energy loss at slower monopole speeds, i.e.\ for $10^{-3}\lesssim\beta\lesssim10^{-2}$.
\item[Elastic collisions with atoms] The energy loss is due to atoms (nuclei) recoiling through the monopole coupling to atomic or nuclear magnetic moment. It becomes important for even slower monopoles  $(\beta\lesssim10^{-3})$ and different energy-loss calculations have been performed for diamagnetic and paramagnetic materials.
\end{compactdesc}
 
The energy loss per path length traveled due to ionization, \dedx, for a particle with electric charge $ze$ is well described by the Bethe-Bloch formula~\cite{Tanabashi:2018oca}
\be
-\dedx = z^2 \frac{KZ}{A} \frac{1}{\beta^2} \left[ \frac{1}{2} \ln \left( \frac{2m_{\rm e}\beta^2\gamma^2}{I^2} \right)
 - \beta^2 \right], \label{eq:bethe-bloch}
\ee
\noindent where $K=4\pi N_{\rm A} r_{\rm e}^2 m_{\rm e} c^2$, with $Z$ $(A)$ the atomic number (mass) of the medium and $I$ its excitation energy; $m_{\rm e}$ $(r_{\rm e})$ the electron mass (radius); and $N_{\rm A}$ the Avogadro number. The ionization energy loss for electric charges can be adapted for a magnetic charge $n\gd$ by replacing $ze \to n\gd\beta$~\cite{Cecchini:2016vrw}. The stopping power then becomes
\be
-\dedx = (n\gd/e)^2 \frac{KZ}{A} \left[ \frac{1}{2} \ln \left( \frac{2m_{\rm e}\beta^2\gamma^2}{I^2} \right)
 - \beta^2 \right]
\label{eq:bethe2}
\ee
For instance, a relativistic monopole with charge \gd\ loses energy as a nucleus with $z \simeq 69$, or $(\gd/e)^2 = 68.5^2 \simeq 4,\!700$ times more than an electron. This makes even a singly charged monopole a highly ionizing particle (HIP).  

Through the ionization process and excitation of atoms, liquid or plastic scintillators, gas detectors, and nuclear track detectors (NTDs) are sensitive to monopoles. 
\begin{compactdesc}
\item[Scintillators] Atom excitation induced by slow monopoles of $\beta\gtrsim 10^{-4}$ passing through a scintillation counter invokes a light yield much larger than that of a minimum ionizing particle. The light yield is saturated for velocities  $10^{-3}\lesssim\beta\lesssim10^{-1}$, whilst it increases again for  $\beta > 0.1$ due to secondary emission of $\delta$-rays~\cite{Patrizii:2015uea}.
\item[Gaseous detectors] Drift and streamer tubes have a lower cost compared to scintillators. At velocities $\beta\gtrsim 10^{-3}$ the high ionization provides a good handle to distinguish monopoles from minimum ionizing particles such as muons. For slower monopoles and for gases such as H and He, the Drell~\cite{Drell:1982zy} and Penning effects can be exploited: a magnetic monopole leaves the atoms in a metastable excited state and the monopole can be detected by the radiation produced by the subsequent return of electrons to their ground state.
\item[Nuclear track detectors] When traversing NTD panels, highly ionizing particles, such as magnetic or high electric charges, damage the material  at the level of polymeric bonds within a cylindrical region extending to a few tens of nanometers around the particle trajectory, forming the so-called \emph{latent track}~\cite{NIKEZIC200451}. The latter is related to the restricted energy loss, which is the fraction of the total energy loss localized in this cylindrical region. When the NTD sheets are chemically etched after exposure, the latent tracks are revealed as cone etch pits, which can be identified after proper scanning of the NTDs. Concerning the material used, plastic CR-39 is the most sensitive having a threshold of $z/\beta \simeq 5$~ \cite{Cecchini:1995rw,Cecchini:2016vrw,Patrizii:2015uea}. The Makrofol\textregistered\ and Lexan\texttrademark\ polycarbonates have a higher threshold of $z/\beta \simeq 50$, hence they are sensitive only to extremely slow monopoles and to very high electric charges. The efficiency of these detectors also depends on the incidence angle; the steeper the angle the lower the thresholds. NTDs are calibrated by exposure to heavy-ion beam fragments~\cite{Cecchini:1995rw,Patrizii:2010jla}.
\end{compactdesc}

One disadvantage of these detection techniques is that it is in general difficult to disentangle magnetic from high electric charges solely from the energy deposit information. So a careful review of the radiation environment of the setup is necessary in order to exclude the possibility that the candidate event is due to background sources, e.g.\ from heavy ions. However, it is worth stressing here that, due to the different energy loss dependence on $\beta$~\eqref{eq:bethe-bloch},\eqref{eq:bethe2}, the etch-pit cone shapes of subsequent layers of NTDs may provide a means to discriminate between magnetic and electric charge. 

As an alternative to NTDs, the use of solid state breakdown counters (SSBC) has been proposed~\cite{Ostrovskiy:2014hfa}. The SSBC exhibits high $dE/dx$ thresholds, convenience of electronic registration and simplicity of fabrication, operation, and signal extraction, however more research and development is required to render it an attractive means for monopole detection. 

\subsection{Induction}\label{sc:squid}

This detection technique is based on the long-range electromagnetic interaction of a monopole with the microscopic state of a superconducting loop and it is directly sensitive to the magnetic charge $g$.  The magnetic flux of a monopole passing through the loop is given by $4\pi g = hc/e$, where $h$ in the Planck constant. A superconducting loop compares this flux with the elementary flux quantum $\phi_0 = hc/2e$, where the factor two arises from the electrons appearing as Cooper pairs. The induced persistent electric current $\Delta i$ in a coil with $N$ turns and inductance $L$ is given by the formula
\be
\Delta i = 4 \pi N g / L.
\label{eq:squid}
\ee

The major background for these experiments are small changes in Earth's magnetic field, therefore shielding of the ambient field is required with extreme caution, leading to high costs for detectors with broad surveillance areas. As a consequence, this technique is no longer used to search directly for monopoles of cosmic origin, yet it is still widely used in searches for monopoles bound in matter, as we shall discuss in Sections~\ref{sc:bound}, \ref{sc:coll-direct} and \ref{sc:moedal}.

\subsection{Cherenkov light}\label{sc:cherenkov}

When traversing a medium such as water or ice, relativistic monopoles would lose some of their energy to Cherenkov radiation. When the monopole speed exceeds the group velocity of light in that particular medium, photons are emitted from excited atoms in the medium. Electrically charged particles also give rise to Cherenkov radiation, yet the number of photons emitted is then much smaller. In water and ice, having a refractive index $n_{\rm r} \simeq 1.33$, a monopole with one Dirac charge generates $(\gd n_{\rm r}/e)^2 \simeq 8,300$ more photons than a particle with one electric unit charge traveling with the same speed. The radiation can only be produced by particles with speeds above a threshold of $\beta_{\rm thr} = 1/n_{\rm r} \simeq 0.75$. The photons are emitted coherently under a fixed angle $\cos\theta_{\rm C} = 1/\beta n_{\rm r}$, which for water or ice is $\theta_{\rm C} \simeq 41.2^\circ $ for relativistic monopoles. As presented in Sec.~\ref{sc:direct} below, this phenomenon is the principal technique used in neutrino telescopes to look for monopoles.

\subsection{Catalysis of nucleon decay}\label{sc:callan}

As explained in detail in Section~\ref{sc:gut}, it has been proposed that the boson in the core of a GUT monopole may cause nucleons to decay by performing transitions between quarks and leptons, as predicted by the \emph{Callan-Rubakov mechanism}~\cite{Rubakov:1981rg,Rubakov:1983sy,Callan:1982ac}. Such processes, such as $uud \to e^+ \bar{d}d$ and $udd \to e^+ \bar{u}d$, violate the baryon-number conservation. The process cross section, $\sigma_0$, is of the same order as that of the strong interactions while the branching fractions of the aforesaid transitions exceed 90\%. 

The decay products, being much lighter than their parents, are highly relativistic, e.g.\ pions subsequently decay into neutrinos. Besides this two-stage neutrino production, the possibility of utilzing \emph{direct} proton decay to \emph{monochromatic} neutrinos from the Sun to detect this effect has been proposed in Ref.~\refcite{Houston:2018rvz}. Therefore, along the trajectory of a catalyzing monopole in medium, outbursts of Cherenkov radiation would occur. The catalysis cross section depends on the monopole--nucleon relative velocity as $\sigma_{\rm cat}=(\sigma_0/\beta)F(\beta)$, where $F(\beta)$ is a correction factor relevant for speeds below a threshold. Hence, nucleon-decay catalysis allows detection of extremely massive sub-relativistic monopoles, e.g.\ in neutrino telescopes (see Sec.~\ref{sc:nucl-decay}).

\section{Searches for Monopoles of Cosmic Origin}\label{sc:cosmics}

If magnetic monopoles of cosmic origin do exist, they must have been formed shortly after the Big Bang, presumably as topological defects arising when the Universe expanded and cooled. The existence of the galactic magnetic field $B \simeq 3~{\rm \mu G}$ would accelerate such monopoles, thus draining energy from the magnetic field. In order for the galactic field to sustain, its dissipation must not exceed its regeneration. This requirement implies that an upper flux limit should be respected, the so-called \emph{Parker bound}~\cite{Turner:1982ag}
\be
\Phi \lesssim 10^{-15}~{\rm cm^{-2} s^{-1} sr^{-1}}. 
\label{eq:parker}
\ee
The (tighter) \emph{extended Parker bound} takes into account the survival of a small galactic seed field and lowers the flux bound to~\cite{Adams:1993fj}
\be
\Phi \lesssim 10^{-16}(M/10^{17}~\gev)~{\rm cm^{-2} s^{-1} sr^{-1}},
\label{eq:ext-parker}
\ee
where $M$ is the monopole mass and a magnetic charge of 1\gd is assumed.

Blas Cabrera and collaborators at Stanford University set up an experiment with a four-turn, \mbox{5-cm}-diameter loop, with its axis vertically oriented, connected to the superconducting input coil of a SQUID (superconducting quantum interference device) magnetometer~\cite{Cabrera:1982gz}. He reported a single candidate event during the 151~days of his experiment operation on February $14^{\rm th}$ 1982. Although the event had the right flux-step size for a Dirac monopole, another plausible explanation could be a mechanically induced offset. Despite further improvements of the experimental setup to suppress possible background sources, this result was not confirmed. If this candidate event is considered to be spurious, these data set an upper limit of $\rm 6.1\times 10^{-10}~cm^{-2} s^{-1} sr^{-1}$, which is much larger than the Parker bound. The best bound given by an induction detector on cosmic monopoles was obtained later by the same group: 90\% confidence level (CL) limit on monopole flux of $\rm  7.2\times 10^{-10}~cm^{-2} s^{-1} sr^{-1}$~\cite{Huber:1990an}.

\begin{figure}[ht]
	\includegraphics[width=0.46\textwidth]{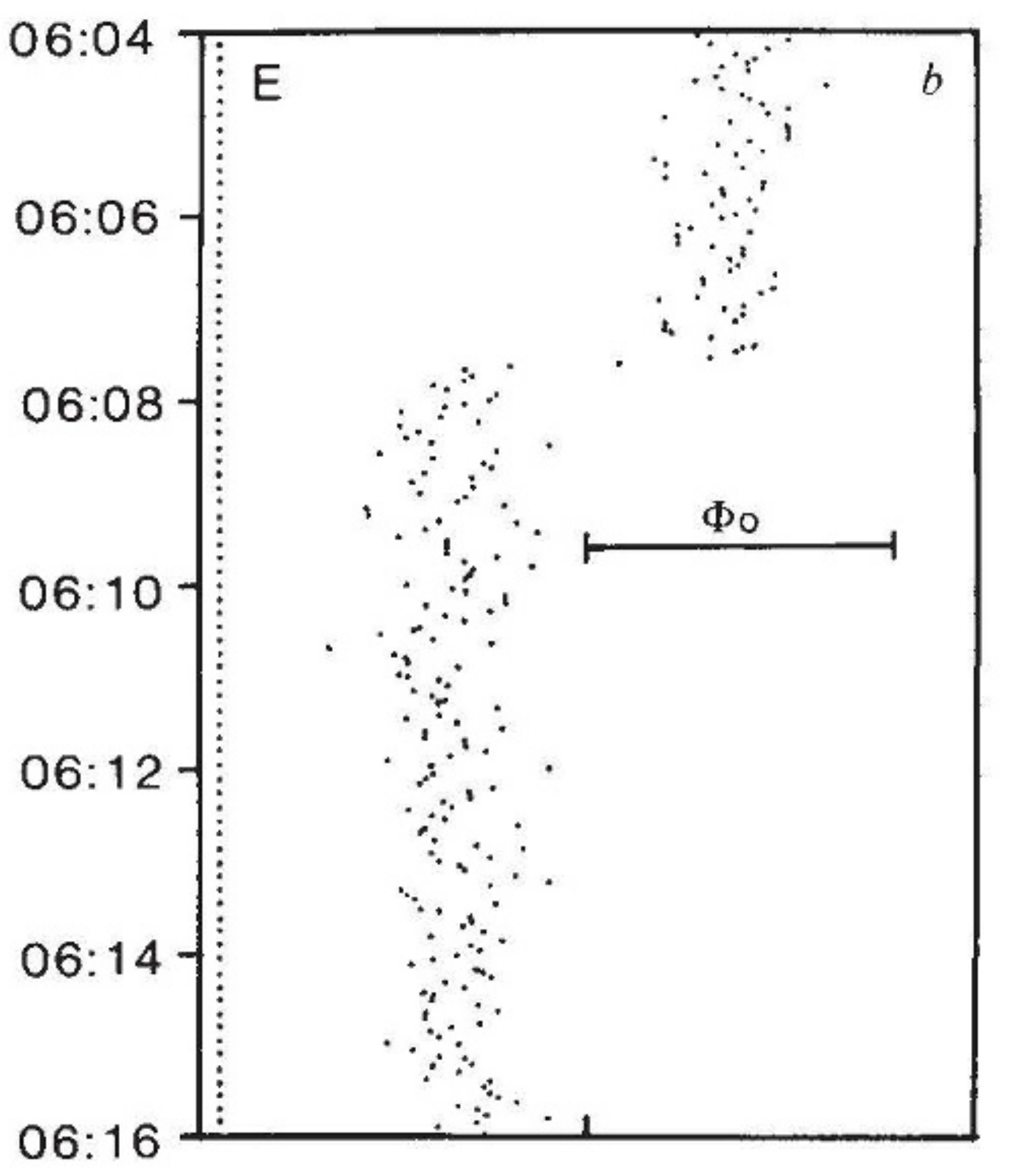}\hfill
    	\includegraphics[width=0.52\textwidth]{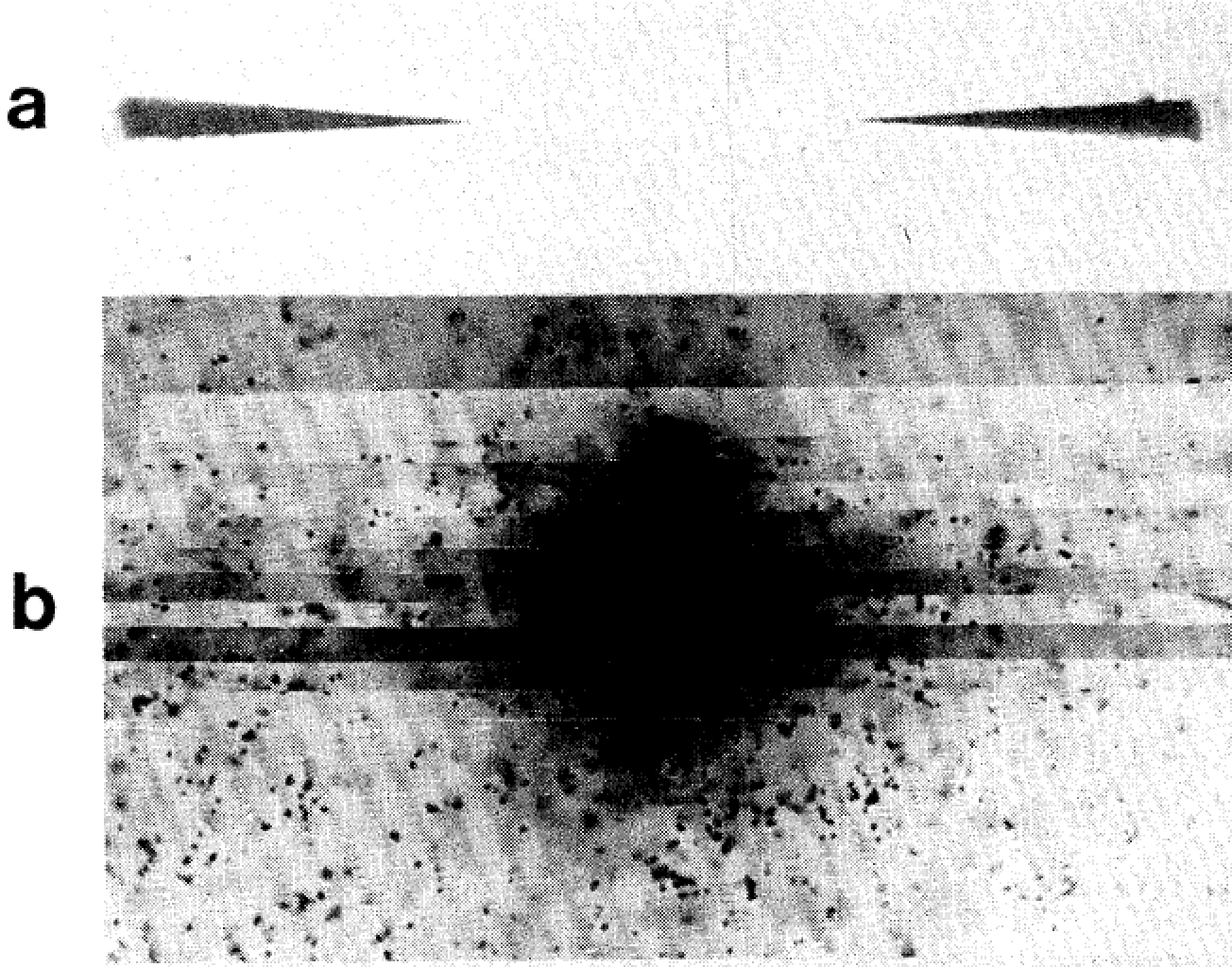}
  \caption{{\bf Left:} Section of the detailed high-frequency record, showing event-160 recorded at 07:06 (BST) on August $11^{\rm th}$ 1985 at Imperial College. Time is shown in minutes and seconds. The bar indicates the offset corresponding to a change of a flux quantum of superconductivity, $\Phi_0$. From Ref.~\protect\refcite{Caplin:1986kw}. {\bf Right:} Photomicrographs of the anomalous track in the balloon-borne experiment: (a) a Lexan\texttrademark\ sheet, viewed edge-on; and (b) the emulsion sheet, viewed nearly vertically. From Ref.~\protect\refcite{Price:1975zt}.}
\label{fig:events}
\end{figure}

A few years after the Stanford event observation, another unusual event was recorded this time at Imperial College London in a setup with two superconducting loops~\cite{Caplin:1986kw}. The event, which occurred on August $11^{\rm th}$ 1985, is shown in Fig.~\ref{fig:events} (left). It was observed during an operational period of total exposure$\times$area corresponding to 400 times that of Cabrera's. Possible explanations such as mechanical shock, residual magnetic flux or cosmic rays have been ruled out by the research team and it remains unexplained as of today. 

Another monopole candidate had been seen earlier in 1973 at Berkeley by a team led by P.~Buford~Price, in a balloon-borne Lexan\texttrademark\  emulsion stack~\cite{Price:1975zt}. The NTD and emulsion sheets   had recorded an unusual track corresponding to a HIP moving downward at near-relativistic speed, as shown in Fig.~\ref{fig:events} (right). The track seemed consistent with a magnetic monopole of charge 2\gd, velocity $0.5c$ and a mass of $\gtrsim 200~\gev$. However, comprehensive studies indicated that the track was probably due to the double fragmentation of a platinum nucleus~\cite{Alvarez:1975gm,Fowler:1975db}.

Events compatible with the passage of a magnetic monopole, as the aforementioned cases, showcase an interesting feature of monopole searches. Due to their unique feature, the magnetic charge, their existence is expected to be inferred from the observation of singular events with practically no obvious background sources involved. This is in contrast to modern-era particle-physics experiments where signals of New Physics are expected to manifest themselves as excesses over accumulated background distributions, the latter being estimated frequently through data-driven methods. On the downside, these unique monopole-like events may be due to background sources too rare to be accounted for or studied with real data. This different paradigm of monopole searches is by no means a defect; it rather underlines the great care with which such candidate events should be interpreted.   

Monopoles of cosmic origin can be detected by exploiting any of the techniques outlined in Sec.~\ref{sc:techniques}. The search may involve old material where monopoles may be either trapped or may have left tracks, as discussed in Sections~\ref{sc:bound} and~\ref{sc:mica}, respectively. However, most of the searches, described in Sec.~\ref{sc:direct}, seek monopoles in-flight interacting with detectors covering a wide spectrum of monopole masses and velocities. Lastly, searches for effects of monopole-catalyzed decays, e.g.\ proton decay, are discussed in Sec.~\ref{sc:nucl-decay}. 

\subsection{Monopoles bound in matter}\label{sc:bound}

During the last decades the induction technique has been mostly deployed in searches for monopoles \emph{bound in matter}, such as lunar rocks~\cite{Eberhard:1971re,Ross:1973it}, meteorites~\cite{Kovalik:1986zz,Jeon:1995rf}, seawater~\cite{Kovalik:1986zz}, iron ores~\cite{Ebisu:1986dw} and ferromanganese nodules~\cite{Kovalik:1986zz}, by passing samples through superconducting loops. With this method, a stringent upper limit on the monopoles per nucleon ratio of $\sim10^{-29}$ has been obtained~\cite{Kovalik:1986zz,Jeon:1995rf}. Moreover, as we shall see in Sec.~\ref{sc:colliders}, SQUIDs have been used in experiments to look for monopoles produced in high-energy collisions and trapped in Al and/or Be volumes.

Another widely used method is the \emph{extraction} technique, which involves the application of strong ($\geq 5~{\rm T}$) magnetic field to samples, sufficient to dislodge the bound monopoles, accelerate them to an appropriate velocity and identify them via the high $dE/dx$ loss in a scintillator or an NTD array. It has been used in the past to search for monopoles in seawater, air and ocean bottom samples~\cite{Fleischer:1969mj,Fleischer:1970zy,Kolm:1971xb,Carrigan:1975bk}, as well as material exposed in high-energy collisions, as we shall see in Sec.~\ref{sc:coll-direct}. A detailed account of searches for monopoles bound in matter is provided in Ref.~\refcite{Burdin:2014xma}.

\subsection{Traces in matter from past passages of monopoles}\label{sc:mica}

Another sort of experiment relies on ionization to look for traces of traversing monopoles in ancient ($4.6\times10^8~{\rm yr}$) mica. The analysis involves a search for defects in the molecular structure of the material caused by the propagation of a monopole in a similar way as for plastic NTDs. The observed absence of monopole tracks in the mica detector placed an upper limit of $10^{-17}$ to $\rm 10^{-16}~cm^{-2} s^{-1} sr^{-1}$ on the flux of GUT monopoles having velocity $\beta \sim 3\times10^{-4} - 1.5\times10^{-3}$~\cite{Price:1983ax}. This was the first direct search for monopoles with adequate sensitivity to detect a flux as small as the Parker flux limit.

\subsection{Direct cosmic searches}\label{sc:direct}

\begin{figure}[ht]
  \centering
  \includegraphics[width=0.9\linewidth]{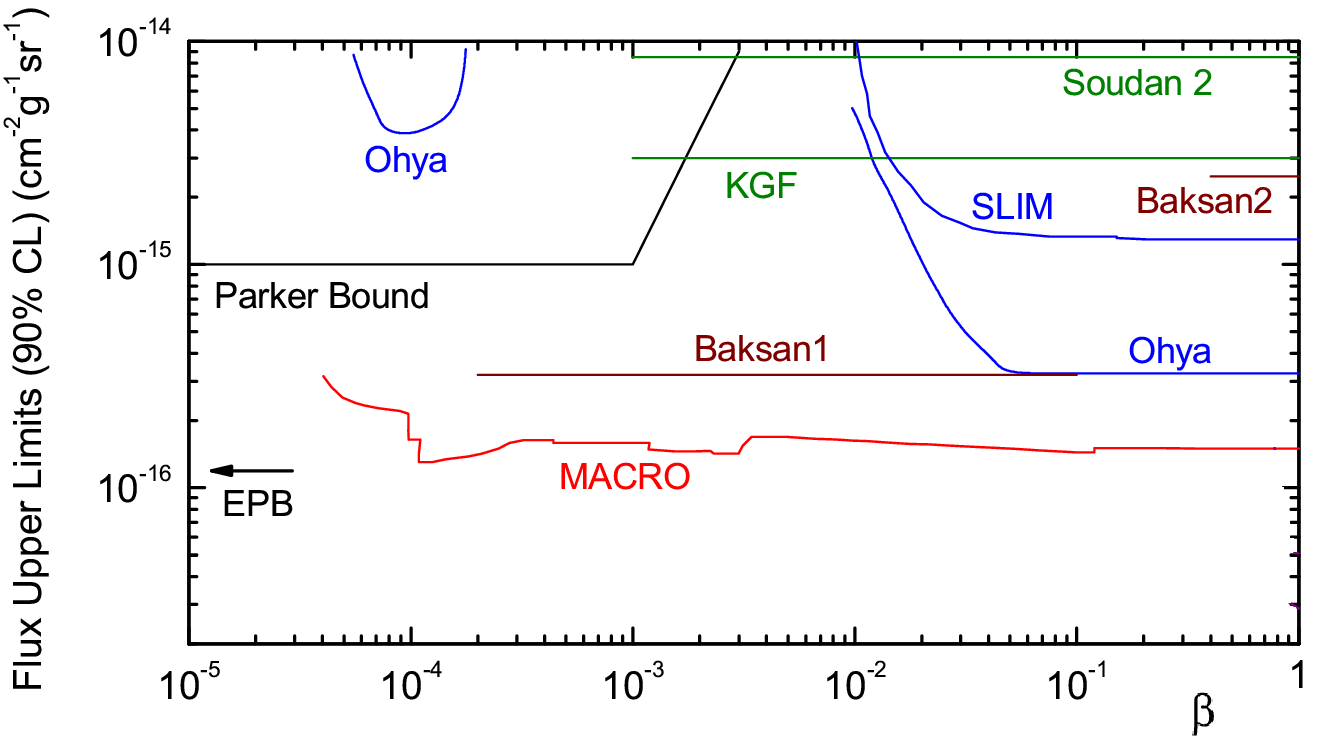}
  \caption{90\%-CL upper limits versus velocity $\beta$ for a flux of cosmic GUT monopoles with magnetic charge of $g = \gd$. The Parker bound is given in~\eqref{eq:parker}, whereas the extended Parker bound (EPB) refers to~\eqref{eq:ext-parker}. Results from the following experiments are shown: Ohya~\cite{Orito:1990ny}, Baksan~\cite{Novoseltsev:2006mw}, 
Soudan~2~\cite{Thron:1992ri}, Kolar Gold Field (KGF)~\cite{Krishnaswamy:1984fu}, SLIM~\cite{Balestra:2008ps} and MACRO~\cite{Ambrosio:2002qq}. From Ref.~\protect\refcite{Spurio:2019oaq}.}
\label{fig:cosmic}
\end{figure}

Searches for non-relativistic monopoles~\cite{Patrizii:2015uea,Patrizii:2019eud} have been performed on underground, surface and balloon-borne experiments targeting GUT monopoles spanning masses of 100--$10^4~\tev$ with a velocity range of $10^{-5} < \beta < 1$. When considering downgoing monopoles, the main background sources are cosmic-ray muons and natural radioactivity, which can be partly rejected by two classes of detection methods, also applicable to heavy stable electrically charged particles. The first method, suitable for very slow monopoles, is based on a time-of-flight measurement or a wide signal in a thick detector plate. The second method is complementary to the first and provides sensitivity for less massive and faster monopoles and involves anomalously high ionization energy loss. As of to date, there is no experimental evidence for cosmic magnetic monopoles, only bounds on their flux as a function of mass and velocity. The present limits for slow-moving monopoles are summarized in Fig.~\ref{fig:cosmic}~\cite{Spurio:2019oaq}. 

MACRO~\cite{Ambrosio:2002mb} was a large underground detector operated in the Gran Sasso laboratory during the 1990s at a depth of more than 3,100~m of water equivalent. It provided the best limits for super-heavy GUT monopoles up to date with a sensitivity that covers most of the phase space in Fig.~\ref{fig:cosmic}, mostly thanks to the redundancy and complementarity of the various detector components it was comprising: liquid scintilation counters; limited streamer tubes; and NTDs~\cite{Ahlen:1993mg,Ambrosio:1997cd,Ambrosio:2001ai,Ambrosio:2002qq}. However, due to its underground location, it was not sensitive to lower-energy monopoles, which are blocked by the Earth. As shown in Fig.~\ref{fig:cosmic}, the upper limit to the local monopole flux set is $\rm 1.4\times 10^{-16}~cm^{-2} s^{-1} sr^{-1}$, i.e.\ well below the Parker bound in almost all the $\beta$ range for GUT monopoles~\cite{Ambrosio:2002qq}. 

Besides some (early) sea-level NTD experiments~\cite{Fleischer:1971bd,Bartlett:1981hy,Doke:1983if}, underground experiments include the Ohya stone quarries near Tokyo, which used a $2,\!000~{\rm m}^2$ array of CR-39 NTDs. As shown in Fig.~\ref{fig:cosmic}, it placed an upper flux limit of $\rm 3.2\times 10^{-16}~cm^{-2} s^{-1} sr^{-1}$ in monopoles in the velocity range of $4\times 10^{-5} < \beta <1$~\cite{Orito:1990ny}. Another telescope, deployed at Baksan~\cite{Novoseltsev:2006mw} in Russia, used liquid scintillation counters to probe both slow {\it (Baksan1)} and fast {\it (Baksan2)} monopoles. Soudan~2~\cite{Thron:1992ri} in the United States, on the other hand, was a large fine-grained tracking calorimeter composed of long drift tubes. Similar techniques were used earlier by other nuclear-decay detectors, namely the tracking calorimeter, Kolar Gold Fields (KGF)~\cite{Krishnaswamy:1984fu} deployed in India and the Mont-Blanc Nucleon Stability Experiment~\cite{Battistoni:1983ka}, which used plastic streamer tubes; both detectors placed looser flux bounds than MACRO, Ohya and Baksan.

The SLIM detector, on the other hand, installed at high altitude at the Mt Chacaltaya laboratory in Bolivia with an elevation of $5,\!400~{\rm m}$, probed a region for intermediate-mass monopoles ($10^5 \lesssim M \lesssim 10^{12}~\gev$), well below the GUT scale, which do not have enough energy to penetrate the entire atmosphere. The SLIM NTDs array covered an area $427~{\rm m}^2$ that after four years of exposure no signal of magnetic monopoles was observed and set the limits~\cite{Balestra:2008ps} shown in Fig.~\ref{fig:cosmic} for single magnetic charge. This detector was also sensitive to monopoles of charge 2\gd in the range $4\times 10^{-5} < \beta < 1$.

NO$\nu$A is a long-baseline neutrino experiment studying neutrino oscillations in the Fermilab NuMI beam. The far detector, located at the surface, consists of liquid scintillator cells read out at both ends by avalanche photodiodes. This detector has the potential to cover a still unexplored phase-space region of intermediate-mass slow monopoles due to its location on the surface and large surface area. To this effect, a dedicated data-driven trigger for slow monopoles has been designed. After a three-month exposure, upper flux limits in monopoles of mass $10^7$--$10^{19}~\gev$ and $10^{-3} \lesssim \beta \lesssim 0.2$ have been placed~\cite{Wang:2015ery}.

\begin{figure}[ht]
  \centering
  \includegraphics[width=0.95\linewidth]{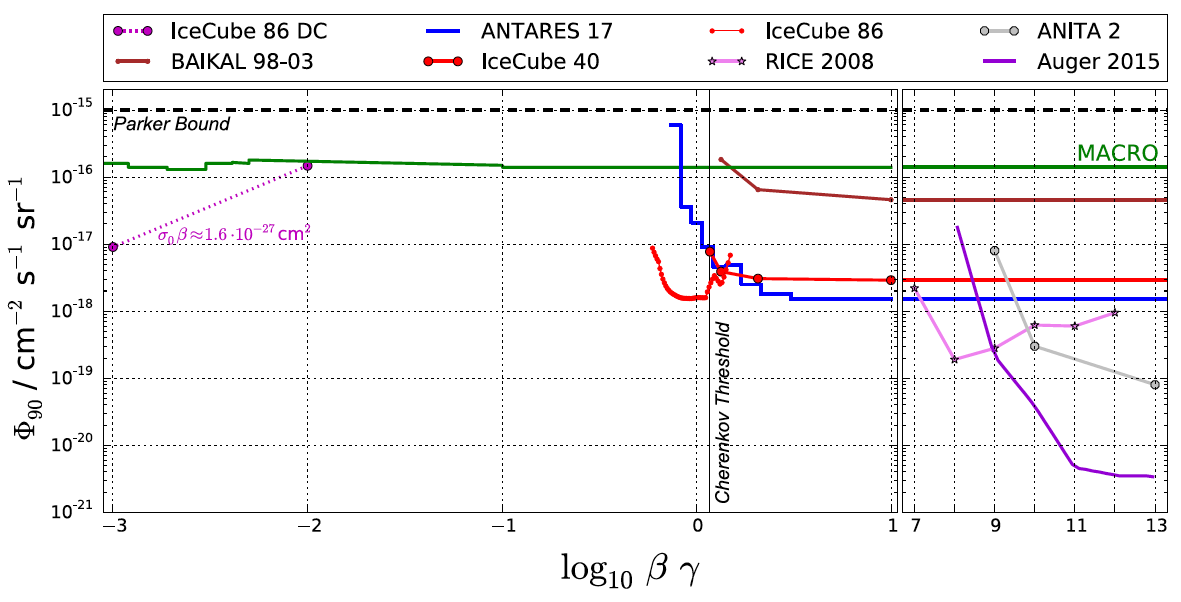}
  \caption{90\%-CL upper limits versus velocity $\beta$ for a flux of very energetic cosmic GUT monopoles with magnetic charge of $g = \gd$. The {\it IceCube 86 DC} nucleon-decay analysis is based on the IceCube DeepCore~\cite{Aartsen:2014awd}. Results from MACRO~\cite{Ambrosio:2002qq},  ANTARES~\cite{AdrianMartinez:2011xr}, RICE~\cite{Hogan:2008sx}, BAIKAL~\cite{Antipin:2007zz}, ANITA~\cite{Detrixhe:2010xi}, Auger~\cite{Aab:2016poe} are superimposed on IceCube limits from direct {\it (IceCube 40)}~\cite{Abbasi:2012eda} and indirect {\it (IceCube 86)}~\cite{Aartsen:2015exf} Cherenkov light~\cite{Aartsen:2015exf}. The Parker bound is given in~\eqref{eq:parker}. From Ref.~\protect\refcite{Pollmann:2018ihz}. }
\label{fig:pollmann}
\end{figure}

Relativistic monopoles can be sought through the emittance of Cherenkov radiation, when traveling through a homogeneous and transparent medium such as ice or water, which can be detected by neutrino telescopes, which feature arrays or strings of photomultiplier tubes. Neutrino telescopes such as Baikal~\cite{Antipin:2007zz}, AMANDA~\cite{Abbasi:2010zz}, ANTARES~\cite{AdrianMartinez:2011xr,Albert:2017fud}, IceCube~\cite{Abbasi:2012eda,Aartsen:2015exf} were/are sensitive to the huge quantity of visible Cherenkov light emitted by a monopole with $\beta > 0.75$ (\emph{direct} Cherenkov). Additional light is produced by Cherenkov radiation from $\delta$-ray electrons along the monopole path for velocities down to $\beta = 0.625$  (\emph{indirect} Cherenkov). Furthermore, luminescence may be induced by molecular excitation of the medium for monopole velocities of $\beta > 0.01$~\cite{ObertackePollmann:2016uvi,Pollmann:2015nmo}. The second phase of the Antarctic Muon And Neutrino Detector Array (AMANDA-II), the predecessor of IceCube, set an upper limit on $\beta=1$ monopoles flux of $\rm 3.8 \times 10^{-17}~cm^{-2} s^{-1} sr^{-1}$~\cite{Abbasi:2010zz,Niessen:2001jn}. Recent results for relativistic monopoles from MACRO~\cite{Ambrosio:2002qq}, ANTARES~\cite{Albert:2017fud} and IceCube~\cite{Aartsen:2015exf,Aartsen:2014awd} are depicted in Fig.~\ref{fig:pollmann}. As is the case for neutrinos, a large background from cosmic muons inhibits searches for downgoing candidates; upgoing monopoles having traversed the Earth before reaching the detector are probed instead. It is worth noting that GUT supermassive monopoles are unlikely to reach (nearly) relativistic velocities.

The flux of ultra-relativistic monopoles has been constrained by the Pierre Auger Observatory, which was sensitive to monopoles with Lorentz factor values \mbox{$\gamma \sim 10^9$--$10^{12}$}, leading to flux limits in the range $\rm 2.5 \times 10^{-21} - 10^{-15}~cm^{-2} s^{-1} sr^{-1}$~\cite{Aab:2016poe}. Two other experiments exploited the radio-wave pulses from the interactions of a primary particle with ice to search for monopoles. The Radio Ice Cherenkov Experiment (RICE), consisting of radio antennas buried in the Antarctic ice, set a flux upper limit of $\rm 10^{-18}~cm^{-2} s^{-1} sr^{-1}$ at 95\% CL for intermediate-mass monopoles with $10^7 < \gamma < 10^{12}$ and a total energy of $10^{16}~\gev$~\cite{Hogan:2008sx}. The ANITA-II balloon-borne radio interferometer, on the other hand, set a 90\%-CL flux upper limit on the order of $\rm 10^{-19}~cm^{-2} s^{-1} sr^{-1}$ for a Lorentz factor $\gamma > 10^{10}$ at a total energy of $10^{16}~\gev$~\cite{Detrixhe:2010xi}.

\subsection{Searches through catalysis of nucleon decay}\label{sc:nucl-decay}

Signals of a monopole-induced decay of a nucleon, as predicted by the Callan-Rubakov mechanism~\cite{Rubakov:1981rg,Rubakov:1983sy,Callan:1982ac} and discussed earlier in Sections~\ref{sc:gut} and~\ref{sc:callan}, have been sought, which are sensitive to the assumed value of the catalyzed-decay cross section $\sigma_{\text{cat}}$. These signals are ideal for probing ultra-heavy sub-relativistic monopoles. Searches have been made with the Soudan~\cite{Bartelt:1986cv} and MACRO~\cite{Ambrosio:2002qu} experiments, using tracking detectors. Searches at the Irvine-Michigan-Brookhaven detector (IMB)~\cite{BeckerSzendy:1994wb}, the underwater Lake Baikal experiment~\cite{Balkanov:1997da} and the IceCube experiment~\cite{Aartsen:2014awd} which exploit the Cherenkov effect have also been performed. The resulting $\beta$-dependent flux limits from these experiments typically lie in the range $10^{-18}$--$\rm 10^{-14}~cm^{-2} s^{-1} sr^{-1}$, as shown in Fig.~\ref{fig:pollmann}. A search for low-energy neutrinos, assumed to be produced from induced proton decay in the Sun, was made at Super-Kamiokande~\cite{Ueno:2012md}. A model- and $\beta$-dependent of limit of $\rm 6.3\times10^{-24}(0.001\beta)^2~cm^{-2} s^{-1} sr^{-1}$ was obtained.

\section{Searches in Colliders}\label{sc:colliders}

Present and proposed future accelerators feature a center-of-mass energy of ${\mathcal O}(10~\tev)$, thus it is virtually impossible to search for GUT monopoles in these machines. Nevertheless, searches have been carried out to detect direct or indirect signals of lower-mass monopoles. Searches have been performed at hadron--hadron, electron--positron and lepton--hadron experiments, mostly directly using scintillation counters, gas chambers and NTDs, taking advantage of the monopole high-ionization power. Other analyses focus on exposed material for trapped monopoles or peculiar magnetic-charge trajectories. In addition, virtual-monopole processes enhancing production rates of certain final states have also been considered as indirect probes for monopoles.

The last few years, the monopole searches interest has been shifted to the LHC, which is the largest and highest-energy particle collider to-date~\cite{Evans:2008zzb}. It was built at CERN between 1998 and 2008 in the existing Large Electron--Positron (LEP) tunnel of 27~km in circumference beneath the France--Switzerland border near Geneva. The LHC primarily collides proton beams, yet heavy-ions have been collided since its operation startup in 2010, too. Since then, it has collided protons at the record center-of-mass energy of 7, 8 and 13~\tev during two running periods: Run~1 (2010--2012) and Run~2 (2015--2018).

\subsection{Direct monopole production mechanisms \label{sc:cross}}
 
\begin{figure}[ht]
 \centering
 \subfloat[Drell-Yan production\label{fig:dy}]{%
       \includegraphics[width=0.3\textwidth]{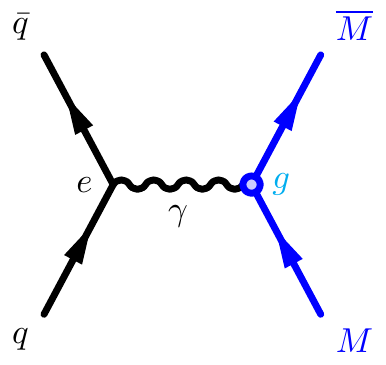}}
       \hspace{0.07\textwidth}  
     \subfloat[Photon-fusion production\label{fig:pf}]{%
       \includegraphics[width=0.3\textwidth]{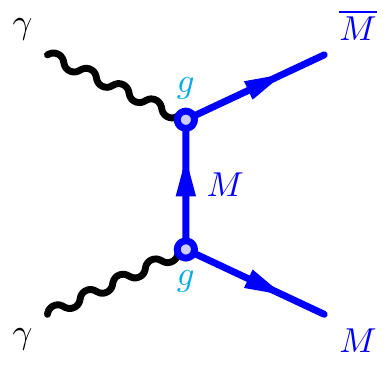}}
  \caption{Monopole $M$ pair production diagrams in hadron colliders: Drell-Yan~\protect\subref{fig:dy} and via photon fusion~\protect\subref{fig:pf}.}
\label{fig:direct}
\end{figure}

Direct monopole pair production in colliders can proceed via two processes: a Drell-Yan-like (DY) process in photon $s$-channel intermediation (see Fig.~\ref{fig:dy}) and a photon-fusion $t$-channel diagram~\cite{Kurochkin:2006jr,Dougall:2007tt,Baines:2018ltl} (see Fig.~\ref{fig:pf}). For both mechanisms, duality arguments justify an effective $\beta$-dependent magnetic charge in monopole-matter scattering processes, with $\beta$ defined in~\eqref{defbeta} and discussed in Sec.~\ref{sc:scatter}, which might also characterize monopole production~\cite{Baines:2018ltl}. 

An important cautionary remark, already mentioned in Sec.~\ref{sc:composite}, is reiterated here. In both cases, the monopole pair couples to the photon via a coupling that depends on $\gd$ and hence it is ${\cal O}(10)$. This large monopole--photon coupling invalidates any perturbative treatment of the cross-section calculation and hence any result based on it is \emph{only indicative} and used merely to facilitate comparisons between experiments. On the contrary, the upper bounds placed on production cross sections are solid and can be safely relied upon~\cite{Tanabashi:2018oca}.

This situation may be resolved if thermal Schwinger production of monopoles in heavy-ion collisions is considered~\cite{Schwinger:1951nm}. This mechanism becomes effective in the presence of strong electromagnetic fields and does not rely on perturbation theory, therefore it overcomes these limitations~\cite{Gould:2017zwi,Gould:2017fve,Gould:2019myj,Ho:2019ads}. This mechanism has been treated in detail in Sec.~\ref{sc:schwinger}.

Another possibility is the photon-fusion production for fermionic and vector boson monopoles with a $\beta$-dependent coupling. A magnetic-moment term proportional to a phenomenological parameter $\kappa$ is added to the effective Lagrangians for spins \half and~1~\cite{Baines:2018ltl}. As argued in that work, by applying the patchwise covering of space around the monopole (see Fig.~\ref{fig:wy} and relevant discussion in Sec.~\ref{sec:history}), the DQC is not affected by the introduction of this parameter. Unitarity of the effective field theory for vector (spin-1) monopoles requires the value $\kappa = 1$. The lack of unitarity and renormalizability for an arbitrary  $\kappa \ne 1$ value is not an issue, given that the microscopic high-energy (ultraviolet) completion of the models considered above is unknown. Thus unitarity may be restored in such models by high-energy modes of New Physics in the ultraviolet energy regime.

The possibility to use the parameter $\kappa$ in conjunction with the monopole velocity $\beta$ to achieve a perturbative treatment of the monopole--photon coupling was introduced in Ref.~\refcite{Baines:2018ltl}. Indeed, by limiting the discussion to very slow ($\beta \ll 1$) monopoles, the perturbativity is guaranteed, however, at the expense of a vanishing cross section in DY production. Nonetheless it turns out that the \emph{photon-fusion} cross section remains finite \emph{and} the coupling is perturbative at the formal limits $\kappa\to\infty$ and $\beta\to 0$ when the following condition is met:
\be g\kappa\beta^2 < 1,
\label{eq:pert}\ee
provided one is using a $\beta$-dependent magnetic charge~\cite{Milton:2006cp,Schwinger:1976fr}. An effective field theory for the monopole--photon interaction~\cite{Alexandre:2019iub} supporting this latter point of view has been discussed in Sec.~\ref{sc:scatter}. Such a treatment opens up the possibility to interpret the cross-section bounds set in collider experiments~\cite{Lee:2018pag,Alimena:2019zri} in a proper way, thus yielding sensible monopole-mass limits. 

\subsection{Virtual monopoles and monopolia}\label{sc:box}

\begin{figure}[ht]
 \centering
     \subfloat[Monopole box diagram\label{fig:box}]{%
      \includegraphics[width=0.36\textwidth]{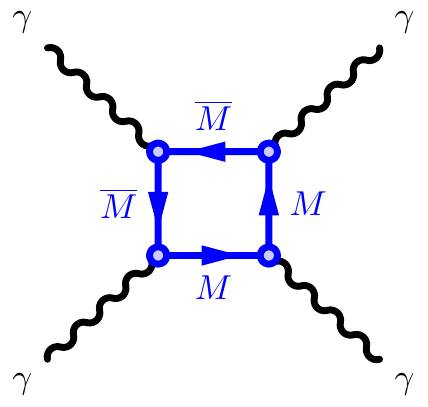}}
       \hspace{0.07\textwidth}  
     \subfloat[Monopolium production\label{fig:mono}]{%
       \includegraphics[width=0.32\textwidth]{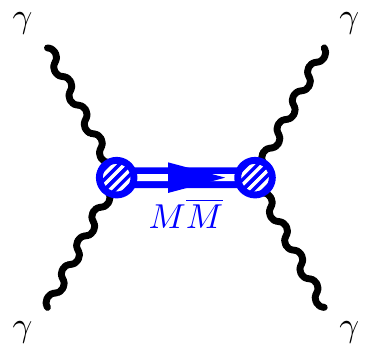}}
  \caption{Monopole-related diphoton production: Light-by-light scattering through a monopole loop~\protect\subref{fig:box} and monopolium $M\overline{M}$ production via $\gamma$-fusion~\protect\subref{fig:mono}.}
\label{fig:indirect}
\end{figure}

Virtual monopoles have been suggested to mediate processes giving rise to multi-photon final states via the ``box diagram'' shown in Fig.~\ref{fig:box}~\cite{DeRujula:1994nf,Ginzburg:1982fk}. Photon-based searches have been carried out by D0~\cite{Abbott:1998mw} at the Tevatron and L3~\cite{Acciarri:1994gb} at LEP. The D0 analysis led to spin-dependent lower mass limits of between 610 and 1580~\gev, whilst L3 reported a lower mass limit of 510~\gev. However, the uncertainties of the cross-section calculations used to derived these limits are difficult to estimate~\cite{Gamberg:1998xf,Milton:2006cp}. 

Indirect searches for monopoles can also proceed by seeking the monopolium~\cite{Hill:1982iq,Dubrovich:2002gp,Vento:2007vy}, a bound state of a monopole and an antimonopole, previously discussed in Sec.~\ref{sc:monopolium}. Most of the monopolium studies follow the low-energy effective theory of Ginzburg and Schiller~\cite{Ginzburg:1998vb,Ginzburg:1999ej}. This theory is based on the standard electroweak theory where the monopole is coupled to the photon and weak bosons assuming that its mass is much larger than the $Z$ mass and that the monopole interacts with the fundamental fields of the $SU(2) \otimes U(1)$ theory before symmetry breaking. This object can be produced via photon fusion, as shown in Fig.~\ref{fig:mono}, in $e^+e^-$ annihilation~\cite{Epele:2007ic,Reis:2017rvb} and in high-energy proton--(anti)proton collisions~\cite{Ginzburg:1998vb,Epele:2008un,Epele:2012jn}, with some specific predictions given in Refs.~\refcite{Barrie:2016wxf,Epele:2016wps}. 

\begin{figure}[htb]
\centering
\includegraphics[width=0.5\linewidth]{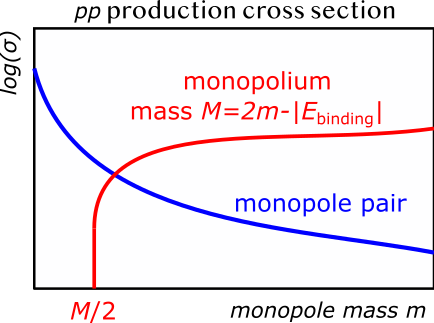} 
\caption{Indicative $pp$ cross sections for monopolium (red curve) and monopole-pair (blue curve) production from photon fusion as a function of the monopole mass $m$. The monopolium has a mass $M=2m-|E_{\text{binding}}|$, where $E_{\text{binding}}$ is the (varying) monopole--antimonopole binding energy. The monopolium is more abundantly produced than the monopole pair for the same monopole mass. Details on cross-section calculation can be found in Ref.~\protect\refcite{Epele:2008un}. \label{fig:monopolium}}
\end{figure}

Monopolium is a neutral state, hence it is difficult to detect directly at a collider detector, however its decay into two photons (see Fig.~\ref{fig:mono}) would give a rather clear signal in the ATLAS and CMS detectors~\cite{Epele:2012jn,Epele:2016wps}. Moreover, monopole--antimonopole annihilation and a lightly bound monopolium may lead to multi-photon events (four and more photons in the final state), while for a strongly bound monopolium --- although diphoton events are dominant --- four- and six-photon event production is also sizable~\cite{Fanchiotti:2017nkk}. The monopolium has lower mass than the monopole--antimonopole pair and, depending on the binding energy $E_{\text{binding}}$, may feature a much larger production cross section in LHC proton--proton ($pp$) collisions for the same monopole mass, as shown in Fig.~\ref{fig:monopolium}.

At LHC indirect searches for monopoles and monopolia can be performed by the ATLAS and CMS experiments with diphoton and multi-photon events. These searches can use forward-proton tagging to probe lower masses in photon-intiated central exclusive production~\cite{Harland-Lang:2018iur}. The latter may have increased sensitivity in ultraperipheral $\gamma\gamma$ collisions (UPCs) in heavy-ion collisions~\cite{Bruce:2018yzs}. The direct observation of light-by-light scattering at the LHC~\cite{dEnterria:2013zqi} by ATLAS~\cite{Aaboud:2017bwk} and CMS~\cite{Sirunyan:2018fhl} in Pb--Pb UPCs opens up the possibility to explore monopole-pair and monopolium diphoton signatures.

The existence of both monopole pairs and monopolia can be probed via the scattering of charged particles off them~\cite{Kazama:1976fm,Vento:2018sog,Vento:2019auh}. If they are produced in proton collisions at the LHC, the beam particles interact with the effective magnetic dipole they represent and they are deflected in off-forward directions. Monopole--antimonopole pairs lead to a sizable effect and thus the effect is suitable for detection in ATLAS, CMS and LHCb~\cite{Vento:2019auh}. Other speculative suggestions for the detection of monopolium is its breaking up in the medium into highly ionizing dyons, which subsequently can be detected in MoEDAL, or through its decay via photon emission which would produce a peculiar trajectory in the medium, should the decaying states are also magnetic multipoles~\cite{Acharya:2014nyr}.

\subsection{Past direct searches}\label{sc:coll-direct}

Several collider experiments utilized a variety of techniques to search for magnetic monopoles in the past. In most of these analyses, results were expressed as upper limits on production cross sections versus the monopole mass, under the ansatz that the kinematics of monopole--antimonopole pair production was determined by the Drell-Yan production. Others proceeded to set mass limits, assuming a (leading-order) DY total cross section. As thoroughly discussed in Sections~\ref{sec:pheno} and~\ref{sc:cross}, this process, as well as the photon fusion, cannot be used to calculate the rate and kinematic properties of produced monopoles, since perturbative field theory is not applicable. The set bounds are therefore only indicative and serve as means for analysis optimization and, to a lesser degree, for comparison purposes between different experiments. This issue may be sorted out by the thermal production in heavy-ion collisions or by  considering events at a kinematic limits in photon-fusion production, as outlined in Sec.~\ref{sc:cross}. 

The most recent direct searches for monopoles at the Tevatron were carried out by the CDF~\cite{Abulencia:2005hb} and the E882~\cite{Kalbfleisch:2000iz,Kalbfleisch:2003yt} experiments. The CDF collaboration used a time-of-flight system with a dedicated trigger requiring large light pulses in the scintillators~\cite{Mulhearn:2004kw} and an offline selection requiring large \dedx tracks not curving in the plane perpendicular to the magnetic field. With 35.7~\ipb of $p\bar{p}$ collisions at $\sqrt{s}=1.96~\tev$, this analysis yielded a monopole production cross-section limit of 0.2~pb for monopole masses in the range 100--700~\gev and a mass limit of 360~\gev  for the DY process~\cite{Abulencia:2005hb}.

The E882 experiment, also known as the ``Oklahoma experiment'', employed the induction technique to search for stopped monopoles in discarded material, such as a Be beam pipe and other Pb and Al parts of the CDF and D0 detectors exposed to $\sim 172~\ipb$ (D0) and $\sim 180~\ipb$ (CDF) proton--antiproton collisions~\cite{Kalbfleisch:2000iz,Kalbfleisch:2003yt}. Upper cross-section limits of 0.6, 0.2, 0.07 and 0.02~pb were obtained for magnetic charges of \gd, 2\gd, 3\gd and 6\gd assuming a uniform monopole--antimonopole pair production. These bounds are translated into 265, 355, 410 and 375~\gev lower mass limits for a DY-like cross-section calculation~\cite{Kalbfleisch:2003yt}. An interesting study was performed by assuming three different angular distribution $d\sigma / d\cos\theta$ scenarios for the monopole production: constant, $1+\cos^2\theta$ and $1-\cos^2\theta$. This variation leads to a $\pm 15~\gev$ spread in the obtained mass limits.  

Earlier searches~\cite{Price:1987py,Price:1990in,Bertani:1990tq} at the Tevatron used NTDs, such as polycarbonate stacks, CR-39 plastic and BP-1 glass sheets, and were based on comparatively modest amounts of integrated luminosity at $\sqrt{s}=1.8~\tev$ $p\bar{p}$ collisions. An upper limit on the production of magnetic monopoles with mass lower than 850~\gev was set at a cross section of 200~pb at 95\% CL~\cite{Bertani:1990tq}.  Lower-energy hadron--hadron experiments have employed a variety of search techniques including plastic track detectors at the CERN ISR~\cite{Hoffmann:1978mp} and $\rm Sp\bar{p}S$~\cite{Aubert:1982zi} and the extraction method to search for trapped monopoles (see also Sec.~\ref{sc:bound}) in proton beam dumps of 300-\gev~\cite{Carrigan:1973mw} and 400-\gev~\cite{Carrigan:1974un} at Fermilab and the CERN ISR~\cite{Carrigan:1977ku}.

Regarding $e^+e^-$ colliders, the only LEP-2 search was made by OPAL based on 62.7~\ipb of data collected on average at $\sqrt{s} = 206.3~\gev$~\cite{Abbiendi:2007ab}. By searching for two back-to-back particles with an anomalously high \dedx in the tracking chambers, an average upper limit of 0.05~pb was acquired on the monopole-pair-production cross section in the mass range 45--102~\gev. Earlier at LEP-1, NTDs deployed around the interaction point, allowing probing high charges for masses up to \mbox{$\sim45~\gev$.} Specifically, the L6-MODAL experiment~\cite{Pinfold:1993mq} set limits for monopoles with charges in the range 0.9--3.6\gd, while a previous search by the MODAL experiment was sensitive to magnetic charges as low as $0.1\gd$~\cite{Kinoshita:1992wd}. The deployment of NTDs around the beam interaction point was also used at $e^+e^-$ facilities of lower collision energies such as PEP~\cite{Kinoshita:1982mv,Fryberger:1983fa} at SLAC, PETRA~\cite{Musset:1983ii} at DESY and the TRISTAN ring~\cite{Kinoshita:1988cn,Kinoshita:1989cb} at KEK. Other detection techniques such as non-helical trajectories have been performed at $e^+e^-$ colliders with the CLEO~\cite{Gentile:1986sf} detector at Cornell and the TASSO~\cite{Braunschweig:1988uc} detector at PETRA.

Up to now the only search for monopole production in lepton--hadron scattering used the induction method on the Al beam pipe used by the H1 experiment at HERA exposed to $e^+p$ collisions at $\sqrt{s} = 300~\gev$~\cite{Aktas:2004qd}. Parts of the beam pipe were scanned by a SQUID magnetometer with a sensitivity as low as 0.1\gd and no monopoles were found. With an integrated luminosity of 62~\ipb, upper limits on the monopole pair production cross section were set for magnetic charges in the range 1--6\gd for masses up to $\sim 140~\gev$.

Nowadays, searches for monopoles produced at the highest available energies in hadron--hadron collisions are being carried out in $pp$ collisions at the LHC~\cite{Evans:2008zzb} by the ATLAS~\cite{Aad:2008zzm} and MoEDAL~\cite{Pinfold:2009oia} experiments utilizing different detection techniques, as elaborated in Sec.~\ref{sc:atlas} and Sec.~\ref{sc:moedal}, respectively.

\subsection{Searching for monopoles with ATLAS}\label{sc:atlas}

ATLAS (A Toroidal LHC ApparatuS)~\cite{Aad:2008zzm} is the largest, general-purpose particle detector experiment at the LHC at CERN, designed to measure the broadest possible range of SM properties and signals of New Physics. In 2012, together with the CMS experiment, the ATLAS Collaboration announced the discovery of a particle consistent with the Higgs boson~\cite{Aad:2015zhl}. The ATLAS detector consists of a series of concentric cylinders around the interaction point where the proton (or heavy-ion) beams from the LHC collide. It is divided into four major parts, namely the Inner Detector, the calorimeters (electromagnetic and hadronic), the Muon Spectrometer and the magnet systems. The Inner Detector consists of silicon detectors and the Transition Radiation Tracker (TRT)~\cite{Akesson:2001su,Aaboud:2017odu}, which is a combination of a straw tracker and a transition radiation detector.

In ATLAS, searches for magnetic monopoles have been performed on 7~\tev~\cite{Aad:2012qi}, on 8~\tev~\cite{Aad:2015kta} and recently on 13~\tev~\cite{Aad:2019pfm} data using the TRT sensitivity to high-ionization signals. The 13~\tev analysis used 34.4~\ifb of data recorded during 2015--2016 and relied on a dedicated trigger for highly ionizing particles, which made use of two \dedx\ variables: (i) the fraction of TRT hits passing a predefined high threshold (HT), $f_{\rm HT}$, and (ii) the number of hits in the TRT passing the HT, $N_{\rm HT}$. The discriminating variables used in this search were the energy dispersion in the electromagnetic calorimeter~\cite{Aad:2010ai}, $w$, and the $f_{\rm HT}$. The energy dispersion measures the fraction of the cluster energy contained in the most energetic cells of a cluster in each of the layers of the electromagnetic calorimeter, in order to identify the characteristic ``pencil-shape'' energy deposit of the signal. Figure~\ref{fig:2d} shows the 2D distribution of $w$ and $f_{\rm HT}$ for data and a hypothetical magnetic-charge signal of 1\gd~\cite{Aad:2019pfm}. It is evident that the variables $w$ and $f_{\rm HT}$ are strongly discriminating and thus monopoles can be sought after in signal region~A with a low expectation of background events. This background was estimated  in a data-driven way and validated in the regions~B, C and D. 

\begin{figure}[ht]
\centering
	\subfloat[Discriminating variables\label{fig:2d}]{%
    	\includegraphics[width=0.54\textwidth]{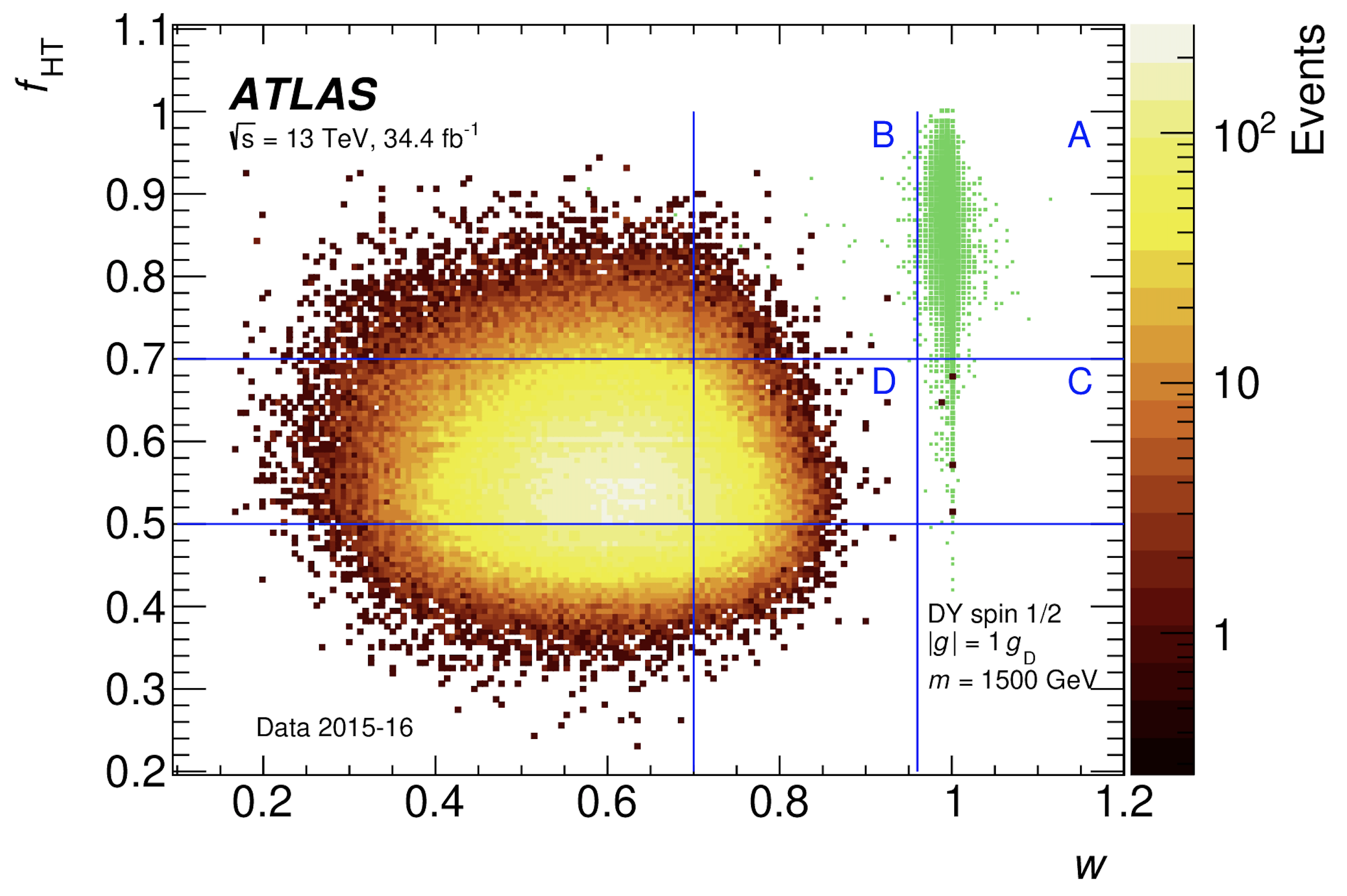}}\hfill
	\subfloat[Monopole limits\label{fig:limit}]{%
    	\includegraphics[width=0.44\textwidth]{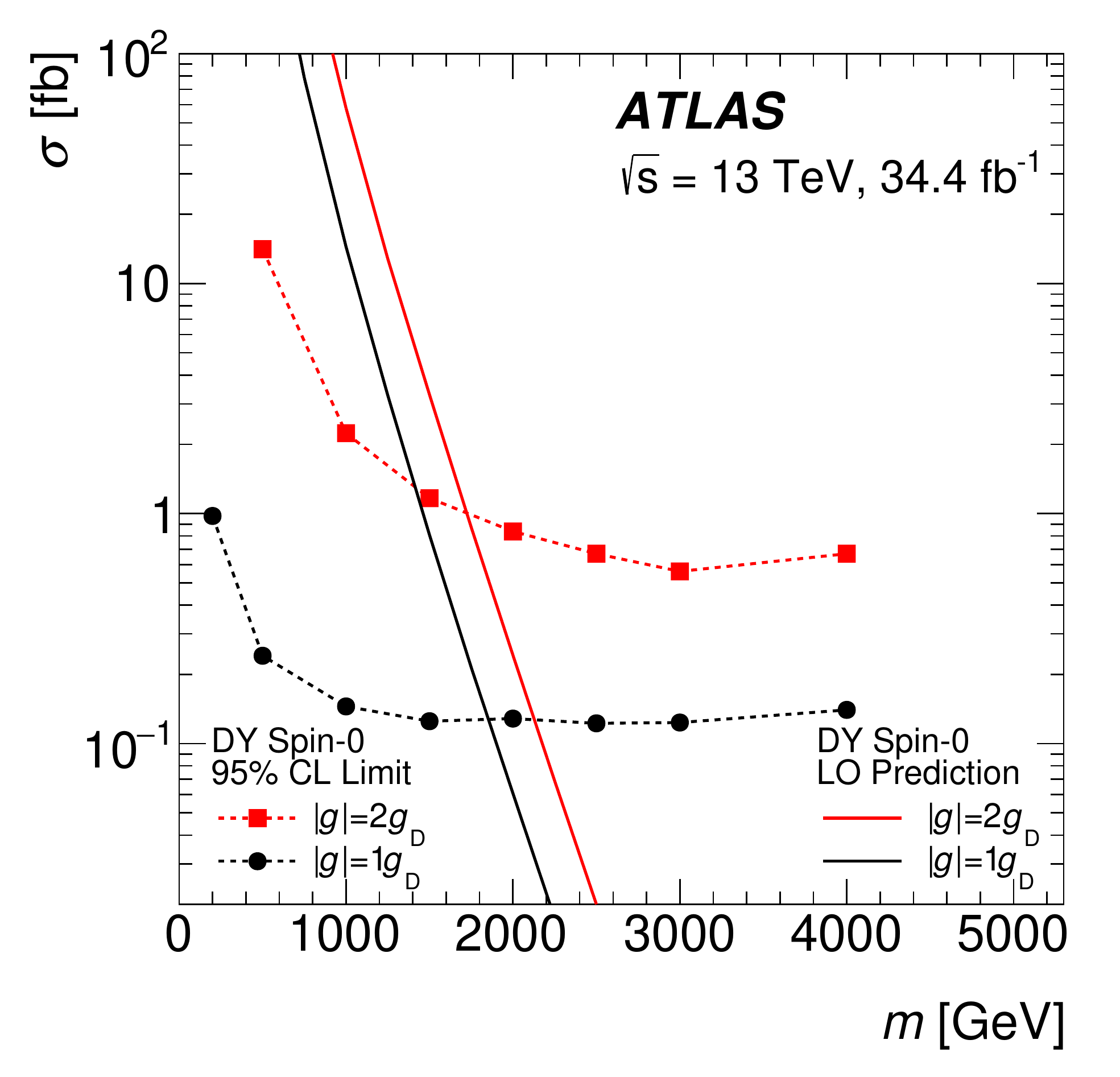}}
  \caption{ATLAS monopole analysis at 13~\tev. \protect\subref{fig:2d} Two-dimensional distribution of the two discriminating variables $f_{\rm HT}$ versus $w$ in data (color scale) and a typical HIP signal (green squares). The regions for defining the signal (A) and for the background estimate and background validation (B, D and C) are indicated. \protect\subref{fig:limit}  95\% cross-section upper limits (dashed lines) and theoretical cross sections (solid lines) for magnetic monopoles with spin~0 and various magnetic charges. From~\protect\refcite{Aad:2019pfm}.}
\label{fig:atlas}
\end{figure}

No excess of data events was observed and this search result was interpreted assuming the Drell-Yan production process with modified electromagnetic couplings, as seen in Fig.~\ref{fig:limits} for the case of a scalar monopole~\cite{Aad:2019pfm}. The analysis was sensitive to magnetic charges of $1\gd\leq|g|\leq2\gd$ and set limits for spin-0 and spin-\half monopoles. The search excluded monopoles with a magnetic charge of 1\gd (2\gd) up to masses of 2125~\gev (2370~\gev) for a spin-\half hypothesis of the particle. These limits are the most stringent bounds placed by an LHC experiment on magnetic charges of $1\gd\leq|g|\leq2\gd$ to-date. In a previous 8~\tev analysis~\cite{Aad:2015kta}, a model-independent upper limit on the production cross section of 0.5~fb was obtained for signal particles with magnetic charge in the range $0.5\gd\leq|g|\leq2\gd$.

\subsection{The MoEDAL experiment}\label{sc:moedal}

MoEDAL (Monopole and Exotics Detector at the LHC)~\cite{Pinfold:2009oia} is designed to search for manifestations of New Physics through highly ionizing particles in a manner complementary to ATLAS and CMS~\cite{DeRoeck:2011aa}. Its main motivation is to pursue the quest for magnetic monopoles at LHC energies, nonetheless, the detector is also designed to search for any massive, long-lived, slow-moving particles~\cite{Fairbairn:2006gg,Burdin:2014xma} with single or multiple electric charges arising in many scenarios of physics beyond the SM~\cite{Acharya:2014nyr}, such as supersymmetry~\cite{Sakurai:2019bac,Felea:2020cvf,Acharya:2020uwc} and D-particles~\cite{Ellis:2009vq,Shiu:2003ta,Mavromatos:2010nk,Elghozi:2015jka}. 

\subsubsection{The MoEDAL detector}\label{sc:moedal-det}

The MoEDAL detector~\cite{Pinfold:2009oia,Acharya:2014nyr} is deployed around the intersection region at Point~8 of the LHC in the LHCb experiment Vertex Locator cavern. It is a unique and largely passive LHC detector comprising four subdetector systems. 

\begin{description}
\item[Nuclear track detectors] The main subdetector system is made of a large array of CR-39,  Makrofol\textregistered\ and Lexan\texttrademark\ NTD stacks surrounding the intersection area. The analysis procedure outlined in Sec.~\ref{sc:gas} is followed for each plastic sheet and then each one is scanned looking for aligned etch pits in multiple sheets. The MoEDAL NTDs have a threshold of $z/\beta\sim5$, where $z$ is the charge and $\beta=v/c$ the velocity of the incident particle. 

Another type of (relatively high-threshold) NTD installed is the Very High Charge Catcher ($z/\beta\sim50$). It consists of two flexible low-mass stacks of Makrofol\textregistered, deployed in the LHCb acceptance between RICH1 and the Trigger Tracker. It is the only NTD (partly) covering the forward region, adding only $\sim0.5\%$ to the LHCb material budget while enhancing considerably the overall geometrical coverage of MoEDAL.

\item[Magnetic trappers] A unique, for an LHC experiment, feature of the MoEDAL detector is the use of magnetic monopole trappers (MMTs) to capture magnetically charged HIPs. The aluminium absorbers of MMTs are subject to an analysis looking for magnetic monopoles or dyons at the SQUID magnetometer facility at ETH Zurich~\cite{DeRoeck:2012wua}, following the induction technique described in Sec.~\ref{sc:squid}. The advantage of this method is that it is fast and allows for a virtually infinite number of measurements for any sample showing signal-like behavior.

\item[TimePix radiation monitors] The only non-passive MoEDAL subdetector is an array of MediPix pixel devices distributed throughout the MoEDAL cavern, forming a real-time radiation monitoring system of beam-related backgrounds, such as spallation products. The operation in time-over-threshold mode allows a 3D mapping of the charge spreading in the volume of the silicon sensor, thus differentiating between various particles species from mixed radiation fields and measuring their energy deposition.

\end{description}

\subsubsection{Searches for monopoles and dyons in MoEDAL}\label{sc:lightsearch}

As explained in Sec.~\ref{sc:squid}, the high charge of a monopole, expected to be at least one Dirac charge $\gd = 68.5 e$~\eqref{dirac}, implies a strong magnetic dipole moment, which in turn may result in a strong binding of the monopole with the $^{27}_{13}\mathrm{Al}$ nuclei of the MoEDAL MMTs. In such a case, the presence of a monopole trapped in an aluminum bar of an MMT would be detected through the existence of a persistent current, defined as the difference between the currents in the SQUID of a magnetometer before and after the passage of the bar through the sensing coil. 

The MoEDAL experiment published its first physics analysis paper in 2016 based on the MMT data taken at the collision energy of 8~\tev during 2012 in Run~1~\cite{MoEDAL:2016jlb}. Model-independent cross-section limits were obtained in fiducial regions of monopole energy and direction for $1\gd \leq |g| \leq 6\gd$. Since then it has released more MMT results from exposures to 13~\tev $pp$ collisions in LHC  Run~2~\cite{Acharya:2016ukt,Acharya:2017cio,Acharya:2019vtb}.

In its most recent search for monopoles, MoEDAL used data taken at a center-of-mass energy of 13~\tev with a delivered integrated luminosity of 4.0~\ifb~\cite{Acharya:2019vtb}. In this search, the photon-fusion monopole production mechanism, characterized by much higher cross-section than DY at LHC energies~\cite{Baines:2018ltl}, was considered for the first time at LHC when interpreting such results. Different monopole--photon couplings were assumed --- both $\beta$-independent and $\beta$-dependent ---, different spins of monopole (spin~0, \half and~1) and both Drell-Yan and photon-fusion production mechanisms. This interpretation used the results of a detailed phenomenological study which compared Drell-Yan and photon-fusion mechanisms for both assumptions of the photon--monopole coupling~\cite{Baines:2018ltl}.

In this analysis, the full MoEDAL trapping detector, consisting of 794~kg of aluminium samples installed in the forward and lateral regions, was analyzed by searching for induced persistent currents after passage through the SQUID magnetometer at ETH Zurich. The measurements were compatible with the absence of monopoles and therefore magnetic charges equal to or above the Dirac charge were excluded in all samples~\cite{Acharya:2019vtb}.  Cross-section upper limits as low as 11~fb were set, improving previous limits of 40~fb also set by MoEDAL~\cite{Acharya:2017cio}. Mass limits in the range 1500--3750~\gev were set for magnetic charges up to $5\gd$ for monopoles of spins~0, \half and~1 --- the strongest to-date at a collider experiment~\cite{Tanabashi:2018oca} for charges ranging from 3\gd to 5\gd. 

In Fig.~\ref{fig:monopole}, cross-section upper limits are shown for spin-\half monopoles, for $\beta$-dependent coupling and for various magnetic charges, together with the leading-order calculations for the combined photon-fusion and Drell-Yan mechanisms. The weaker limits for $|g|= \gd$, when compared to higher charges, are mostly due to loss of acceptance from monopoles punching through the trapping volume. For higher charges, monopoles ranging out before reaching the trapping volume decrease the acceptance for DY monopoles with increasing charge. Under the assumption of simultaneous photon-fusion and Drell-Yan production, mass limits are derived for $1\gd\leq|g|\leq5\gd$ with MoEDAL~\cite{Acharya:2019vtb}, complementing previous results  from ATLAS~\cite{Aad:2019pfm} presented in Sec.~\ref{sc:atlas}, as shown in Fig.~\ref{fig:limits}. 

\begin{figure}[ht]
\centering
	\subfloat[Monopole limits\label{fig:monopole}]{%
    	\includegraphics[width=0.45\textwidth]{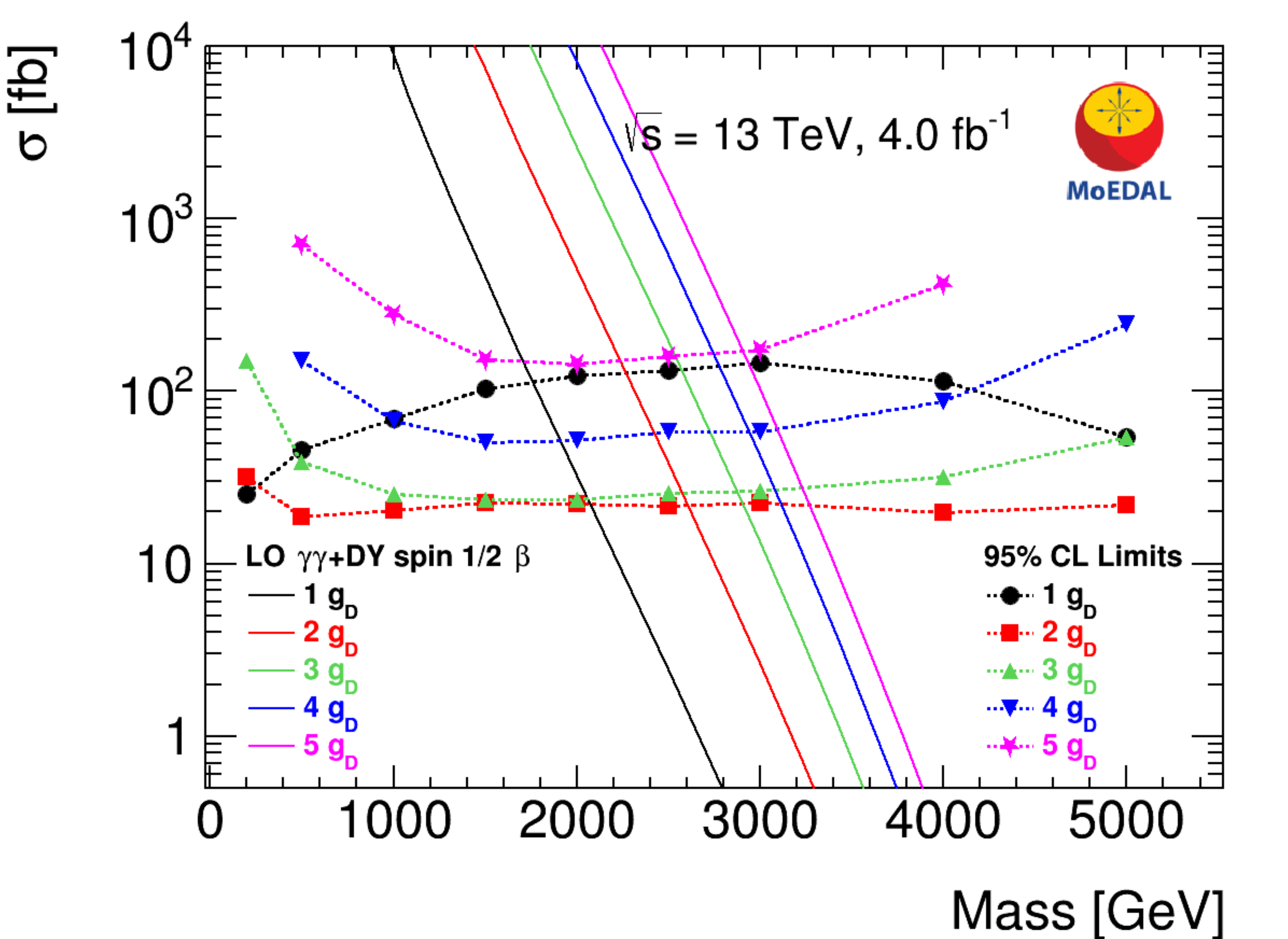}}\hfill
	\subfloat[Dyon limits\label{fig:dyon}]{%
    	\includegraphics[width=0.54\textwidth]{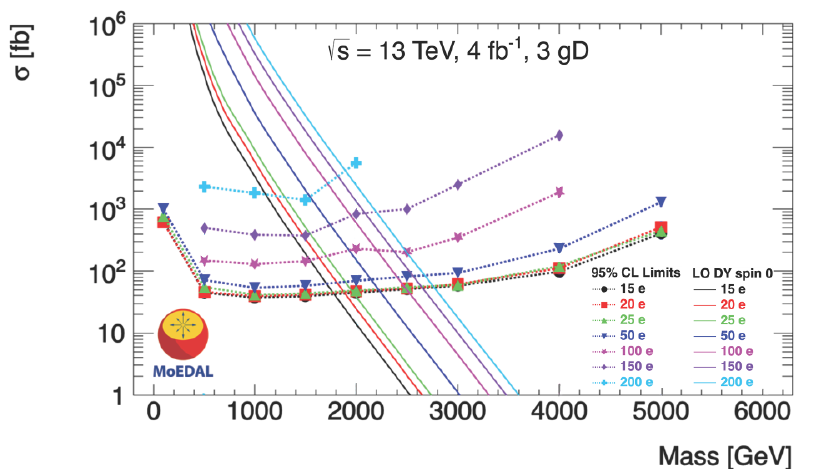}}
  \caption{MoEDAL 95\% upper limits on production cross sections (dashed lines) and theoretical cross-section calculations at leading order (solid lines). \protect\subref{fig:monopole} Limits for fermionic magnetic monopoles and $\beta$-dependent coupling, where the various line colors correspond to different magnetic charges. From Ref.~\protect\refcite{Acharya:2019vtb}. \protect\subref{fig:dyon} Limits for scalar dyons of magnetic charge 3\gd and $\beta$-independent coupling, where the various line colours correspond to different electric charges ranging from $15e$ to $200e$. From Ref.~\protect\refcite{Acharya:2020bed}. }
\label{fig:moedal}
\end{figure}

\begin{figure}[ht]
   \centering
   \includegraphics[width=0.85\linewidth]{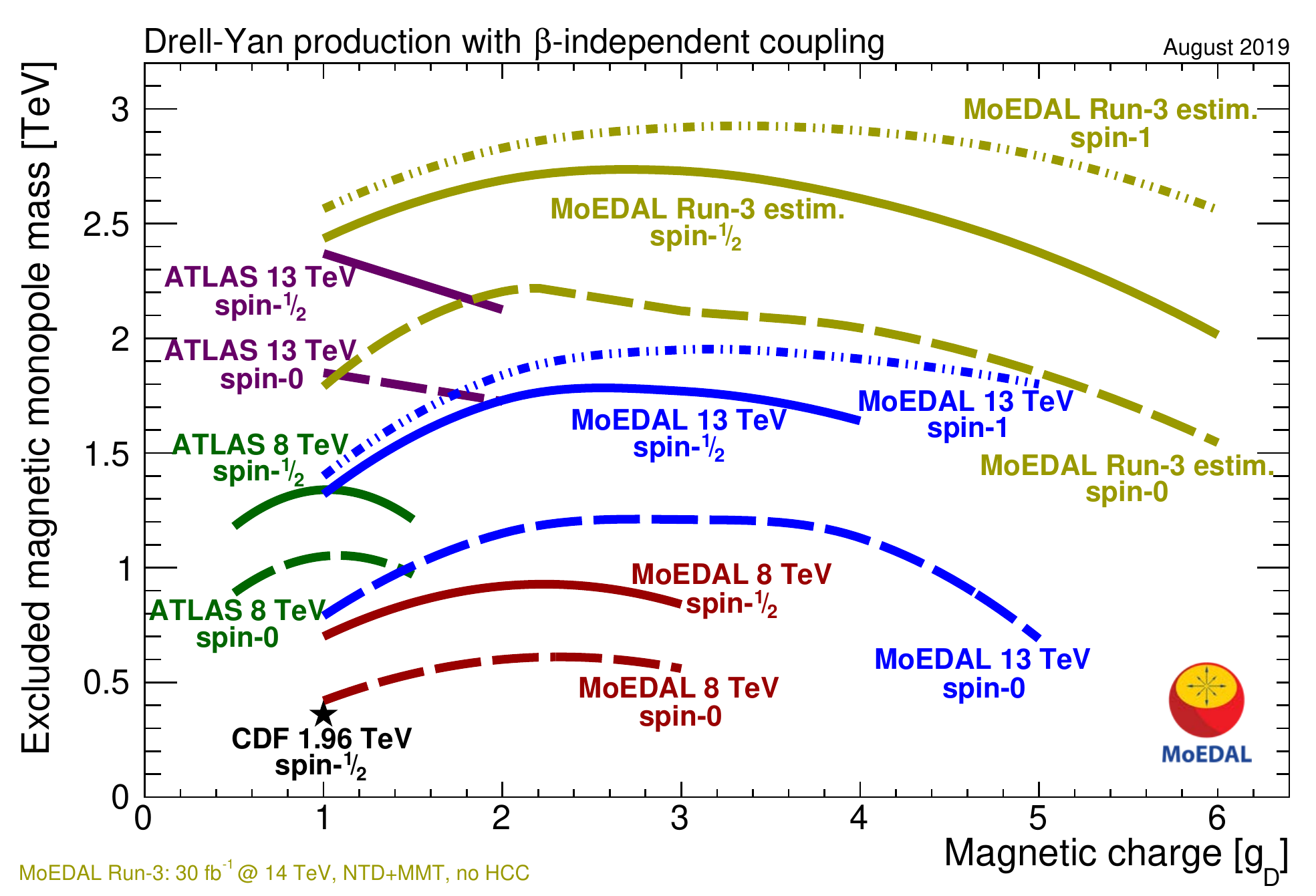}
   \caption{Magnetic monopole mass limits from CDF~\cite{Abulencia:2005hb}, ATLAS~\cite{Aad:2015kta,Aad:2019pfm} and MoEDAL searches~\cite{MoEDAL:2016jlb,Acharya:2017cio,Acharya:2019vtb} as a function of magnetic charge for various spins, assuming Drell-Yan pair-production mechanism and a $beta$-independent coupling. The MoEDAL projection for LHC Run~3 assuming a 30~\ifb integrated luminosity and combined NTD and MMT data is superimposed.}
\label{fig:limits} 
\end{figure}

Furthermore, these MMT scanning results, corresponding to 2015--2017 exposure, were interpreted recently in terms of \emph{dyon production}~\cite{Acharya:2020bed}; the very first search for dyons in a collider experiment. The direct DY production of dyon--antidyon pairs was considered and cross-section upper limits as low as 30~fb were placed. Mass limits in the range 830--3180~\gev were set for dyons with magnetic charge up to 6\gd, for electric charge from $1e$ to $200e$ and for spins~0, \half and~1. Cross-section limits for scalar dyons of magnetic charge 3\gd and electric charges ranging from $15e$ to $200e$ are shown in Fig.~\ref{fig:dyon}. The total dyon energy loss is the sum of the energy losses due to the electric and the magnetic  charge. 

An overview of the monopole mass limits for DY production and $\beta$-independent coupling set by CDF~\cite{Abulencia:2005hb}, ATLAS~\cite{Aad:2015kta,Aad:2019pfm} and MoEDAL~\cite{MoEDAL:2016jlb,Acharya:2017cio,Acharya:2019vtb} is given in the graph of Fig.~\ref{fig:limits} for various magnetic charges. The ATLAS bounds are better that the MoEDAL ones for $|g| \leq 2\gd$ due to the higher luminosity delivered in ATLAS\footnote{A factor of $\sim 20$ less luminosity was delivered to MoEDAL than ATLAS during LHC Run~2 due to the operation requirements of the LHCb experiment. This factor is planned to be reduced to $\sim 10$ in the upcoming Run~3.} and the loss of acceptance in MoEDAL for small magnetic charges. On the other hand, higher charges are difficult to be probed in ATLAS due to the limitations of the electromagnetic-calorimeter-based level-1 trigger deployed for such searches. A comparison of the upper limits on monopole production cross sections set by other colliders with those set by MoEDAL is presented in Refs.~\refcite{Tanabashi:2018oca,Rajantie:2016paj,pdg}. 

\section{Summary and Outlook}\label{sc:outlook}

The existence of magnetic monopoles, if confirmed experimentally,  would modify our understanding of Electromagnetism, rendering the Maxwell equations fully symmetric. The Dirac electric-charge quantization condition is a beautiful consequence of the existence of monopoles and hence it represents an extremely appealing motivation for studying and looking for isolated magnetic charges. 

A wide range of theoretical microscopic models of magnetic monopoles, ranging from string/brane and GUT theories to (extensions of) the Standard Model, have been proposed. Some relatively recent developments support the existence of stable monopole with masses of the order of the electroweak scale. The classic monopole solution of 't Hooft-Polyakov exists in GUT models; some variants of this construction based on the Georgi-Glashow model with real Higgs triplet fields, occur in some string-inspired models with a Kalb-Ramond antisymmetric tensor field, or in some extensions of the Standard Model, with right-handed neutrinos. The Cho-Maison electroweak monopole and its finite energy variants in theories beyond the Standard Model, such as the Born-Infeld theory, have been considered. Other scenarios predict dyonic D-brane solutions and the resulting dyonic black holes under appropriate compactifications, as well as some new objects that exhibit a smooth topology-change interpolation between a monopole-like solution (asymptotically far away from the object's position in space) and a dipole (near the origin). The interest in magnetic monopole searches, both at colliders and in the Cosmos, has been revived in the past few years also due to the construction of the aforementioned theoretical models which predict relatively light monopoles, with masses of order of the electroweak scale, thus accessible, in principle, at colliders.

Important ongoing developments in the phenomenolgy of the magnetic monopoles have addressed the building of effective field theories describing monopole interactions with matter, in particular the production of monopole--antimonopole pairs at colliders or from the vacuum \`a la Schwinger. The feasibility of the production of such monopoles depends on their composite or fundamental (Dirac) nature. The results may be very different; for instance, the composite monopole production is expected to be strongly suppressed at colliders, based on some arguments, but this is not the case if they are thermally produced from the vacuum in the presence of high temperatures and external magnetic fields, a situation met, for instance, in neutron stars and heavy-ion collisions. Monopoles can potentially play a role as dark matter candidates through their connection with (hidden) dark sectors. The formation of a bound state of monopoles (monopolium) in high-energy collisions may be more abundant than a monopole pair of the same mass, offering an additional handle in the hunt for monopoles. 

Numerous monopoles searches have beed carried out for decades by utilizing diverse detection techniques in both observational facilities and experiments in colliders. The CERN LHC, being the most powerful collider to-date, leads this effort via two complementary detection approaches applied by the ATLAS and MoEDAL experiments. ATLAS is more sensitive in low magnetic charges, while MoEDAL has set the stringent bounds in high charges. As shown in Fig.~\ref{fig:limits}, the projected MoEDAL reach for Run~3 at $\sqrt{s}=14~\tev$, combining low-threshold NTDs and MMTs is as high as $\sim 3~\tev$ for an integrated luminosity of 30~\ifb, if only Drell-Yan production is assumed. The large monopole--photon coupling makes such cross-section calculations unreliable, yet a proposal involving the thermal Schwinger production of monopoles in heavy-ion collisions may overcome this limitation~\cite{Ho:2019ads}. Moreover, the introduction of a magnetic-moment phenomenological parameter $\kappa$ combined with a $\beta$-dependent coupling may lead to a perturbative treatment of the cross-section calculation in spin-\half and spin-1 monopoles~\cite{Baines:2018ltl}.

The possibility of analyzing decommissioned parts of the LHC beam-pipe system at the ATLAS, CMS and LHCb/MoEDAL sites using the induction technique to search for trapped magnetic monopoles has been proposed~\cite{beampipe}. In this context, the MoEDAL experiment may serve as a formal platform for coordinating machining, scanning and analysis work, in collaboration with interested ATLAS, CMS and LHCb members. The Run~1 CMS beam pipe has already been handed over to MoEDAL~\cite{beampipe-cc} and the analysis is underway. 

Furthermore, magnetic monopoles can be probed at the LHC indirectly in events with multiple photons in the final state~\cite{Epele:2012jn}. Magnetic monopoles can enhance the production of such events by participating in a box diagram. In addition, the production and subsequent decay of a monopolium would also yield multiphoton events~\cite{Fanchiotti:2017nkk}. 

In the cosmic front, neutrino telescopes, such as ANTARES and IceCube currently, and KM3NeT~\cite{LeBreton:2019lpq} and IceCube-Gen2~\cite{Blaufuss:2017wzl} in the future, pursue the search for GUT monopoles. Besides them, neutrino-oscillation experiments such as the currently running NO$\nu$A at Fermilab and the Iron CALorimeter (ICAL) detector, to be built at the proposed India-based Neutrino Observatory (INO) facility~\cite{Dash:2014fba}, are expected to be sensitive to magnetic monopoles in the sub-relativistic range.\footnote{The detector sensitivity for a live-time of 10~yr is $\rm\sim 1.6 \times 10^{-16}~cm^{-2} s^{-1} sr^{-1}$ for particles carrying magnetic charge in the mass range from $10^7$--$10^{17}~\gev$ with velocities  $\beta \sim 0.001$--$0.7$.~\cite{Dash:2014fba}}

In addition to neutrino experiments, a large array of $\sim 10,\!000~{\rm m}^2$ of plastic \mbox{CR-39} NTDs deployed at high altitude, the \emph{Cosmic-MoEDAL} has been proposed by James Pinfold~\cite{Pinfold:2019zwp}. Following the same approach as the SLIM experiment, Cosmic-MoEDAL would be able to take the search for cosmic monopoles with velocities $\beta\gtrsim 0.05$ from the \tev scale to the GUT scale for monopole fluxes well below the Parker bound.  A possible site for the detector installation includes Mt Chacaltaya in Bolivia at an altitude of 5,400~m. 

Magnetic monopoles continue to fascinate in the fields of Particle Physics and Cosmology. Exciting developments are being made in theoretical scenarios, while experimental advancements allow their exploration using diverse detection approaches. Future prospects keep looking promising for this elusive particle.

\section*{Acknowledgments}

We thank Prof.\ E.~Vagenas for the invitation to write this review.  NEM was supported in part by the STFC (UK) research grants ST/P000258/1 and 
ST/T000759/1. He also acknowledges a scientific associateship (``\emph{Doctor Vinculado}'') at IFIC-CSIC-University of Valencia, Spain. VAM acknowledges support by the Generalitat Valenciana through MoEDAL-supporting agreements and the Excellence Project PROMETEO-II/2017/033, by Spanish and European funds under the project PGC2018-094856-B-I00 (MCIU/AEI/FEDER, EU), and by a 2017 Leonardo Grant for Researchers and Cultural Creators, BBVA Foundation. The authors acknowledge partial support by the EU through the COST Association Action CA18108 {\it ``Quantum Gravity Phenomenology in the Multimessenger Approach (QG-MM)''}.

\bibliographystyle{ws-mpla-vaso}
\bibliography{MM_review_monopoles.bib}

\begin{thebibliography}{100}

\bibitem{Dirac:1931kp}
P.~A.~M. Dirac, {\em Proc. Roy. Soc. Lond. A} {\bf 133}, 60  (1931).

\bibitem{Dirac:1948um}
P.~A.~M. Dirac, {\em Phys. Rev.} {\bf 74}, 817  (1948).

\bibitem{tHooft:1974kcl}
G.~'t~Hooft, {\em Nucl. Phys. B} {\bf 79}, 276  (1974), [,291(1974)].

\bibitem{Polyakov:1974ek}
A.~M. Polyakov, {\em JETP Lett.} {\bf 20}, 194  (1974), [,300(1974)].

\bibitem{Guth:1980zm}
A.~H. Guth, {\em Phys. Rev. D} {\bf 23}, 347  (1981), [{\it Adv. Ser.
  Astrophys. Cosmol.} {\bf 3}, 139 (1987)].

\bibitem{Schwinger:1969ib}
J.~S. Schwinger, {\em Science} {\bf 165}, 757  (1969).

\bibitem{Vento:2013jua}
V.~Vento and V.~Sari~Mantovani  (2013),
  \href{http://arxiv.org/abs/1306.4213}{{\sffamily arXiv:1306.4213 [hep-ph]}}.

\bibitem{Rossi:1982fq}
P.~Rossi, {\em Phys. Rept.} {\bf 86}, 317  (1982).

\bibitem{Weinberg:2006rq}
E.~J. Weinberg and P.~Yi, {\em Phys. Rept.} {\bf 438}, 65  (2007),
  \href{http://arxiv.org/abs/hep-th/0609055}{{\sffamily arXiv:hep-th/0609055
  [hep-th]}}.

\bibitem{Wen:1985qj}
X.-G. Wen and E.~Witten, {\em Nucl. Phys. B} {\bf 261}, 651  (1985).

\bibitem{Fairbairn:2006gg}
M.~Fairbairn, A.~C. Kraan, D.~A. Milstead, T.~Sjostrand, P.~Z. Skands and
  T.~Sloan, {\em Phys. Rept.} {\bf 438}, 1  (2007),
  \href{http://arxiv.org/abs/hep-ph/0611040}{{\sffamily arXiv:hep-ph/0611040
  [hep-ph]}}.

\bibitem{Giacomelli:2003yu}
G.~Giacomelli and L.~Patrizii, {\em ICTP Lect. Notes Ser.} {\bf 14}, 121
  (2003), \href{http://arxiv.org/abs/hep-ex/0302011}{{\sffamily
  arXiv:hep-ex/0302011 [hep-ex]}}.

\bibitem{Burdin:2014xma}
S.~Burdin, M.~Fairbairn, P.~Mermod, D.~Milstead, J.~Pinfold, T.~Sloan and
  W.~Taylor, {\em Phys. Rept.} {\bf 582}, 1  (2015),
  \href{http://arxiv.org/abs/1410.1374}{{\sffamily arXiv:1410.1374 [hep-ph]}}.

\bibitem{Patrizii:2015uea}
L.~Patrizii and M.~Spurio, {\em Ann. Rev. Nucl. Part. Sci.} {\bf 65}, 279
  (2015), \href{http://arxiv.org/abs/1510.07125}{{\sffamily arXiv:1510.07125
  [hep-ex]}}.

\bibitem{Cho:1996qd}
Y.~M. Cho and D.~Maison, {\em Phys. Lett. B} {\bf 391}, 360  (1997),
  \href{http://arxiv.org/abs/hep-th/9601028}{{\sffamily arXiv:hep-th/9601028
  [hep-th]}}.

\bibitem{Cho:2013vba}
Y.~M. Cho, K.~Kim and J.~H. Yoon, {\em Eur. Phys. J. C} {\bf 75},  ~67  (2015),
  \href{http://arxiv.org/abs/1305.1699}{{\sffamily arXiv:1305.1699 [hep-ph]}}.

\bibitem{Ellis:2016glu}
J.~Ellis, N.~E. Mavromatos and T.~You, {\em Phys. Lett. B} {\bf 756}, 29
  (2016), \href{http://arxiv.org/abs/1602.01745}{{\sffamily arXiv:1602.01745
  [hep-ph]}}.

\bibitem{Kephart:2017esj}
T.~W. Kephart, G.~K. Leontaris and Q.~Shafi, {\em JHEP} {\bf 10},   176
  (2017), \href{http://arxiv.org/abs/1707.08067}{{\sffamily arXiv:1707.08067
  [hep-ph]}}.

\bibitem{Arunasalam:2017eyu}
S.~Arunasalam and A.~Kobakhidze, {\em Eur. Phys. J. C} {\bf 77},   444  (2017),
  \href{http://arxiv.org/abs/1702.04068}{{\sffamily arXiv:1702.04068
  [hep-ph]}}.

\bibitem{Ellis:2017edi}
J.~Ellis, N.~E. Mavromatos and T.~You, {\em Phys. Rev. Lett.} {\bf 118},
  261802  (2017), \href{http://arxiv.org/abs/1703.08450}{{\sffamily
  arXiv:1703.08450 [hep-ph]}}.

\bibitem{Arai:2018uoy}
M.~Arai, F.~Blaschke, M.~Eto and N.~Sakai, {\em PTEP} {\bf 2018},   083B04
  (2018), \href{http://arxiv.org/abs/1802.06649}{{\sffamily arXiv:1802.06649
  [hep-ph]}}.

\bibitem{Mavromatos:2016mnj}
N.~E. Mavromatos and S.~Sarkar, {\em Phys. Rev. D} {\bf 95},   104025  (2017),
  \href{http://arxiv.org/abs/1607.01315}{{\sffamily arXiv:1607.01315
  [hep-th]}}.

\bibitem{Mavromatos:2018drr}
N.~E. Mavromatos and S.~Sarkar, {\em Phys. Rev. D} {\bf 97},   125010  (2018),
  \href{http://arxiv.org/abs/1804.01702}{{\sffamily arXiv:1804.01702
  [hep-th]}}.

\bibitem{Acharya:2014nyr}
MoEDAL Collaboration, B.~Acharya {\em et~al.}, {\em Int. J. Mod. Phys. A} {\bf
  29},   1430050  (2014), \href{http://arxiv.org/abs/1405.7662}{{\sffamily
  arXiv:1405.7662 [hep-ph]}}.

\bibitem{Gould:2017fve}
O.~Gould and A.~Rajantie, {\em Phys. Rev. D} {\bf 96},   076002  (2017),
  \href{http://arxiv.org/abs/1704.04801}{{\sffamily arXiv:1704.04801
  [hep-th]}}.

\bibitem{Gould:2017zwi}
O.~Gould and A.~Rajantie, {\em Phys. Rev. Lett.} {\bf 119},   241601  (2017),
  \href{http://arxiv.org/abs/1705.07052}{{\sffamily arXiv:1705.07052
  [hep-ph]}}.

\bibitem{Shnir:2005xx}
{\relax Ya}.~M. Shnir, {\em {Magnetic monopoles }}Text and Monographs in
  Physics (Springer, Berlin/Heidelberg, 2005).

\bibitem{Milton:2006cp}
K.~A. Milton, {\em Rept. Prog. Phys.} {\bf 69}, 1637  (2006),
  \href{http://arxiv.org/abs/hep-ex/0602040}{{\sffamily arXiv:hep-ex/0602040
  [hep-ex]}}.

\bibitem{Rajantie:2012xh}
A.~Rajantie, {\em Contemp. Phys.} {\bf 53}, 195  (2012),
  \href{http://arxiv.org/abs/1204.3077}{{\sffamily arXiv:1204.3077 [hep-th]}}.

\bibitem{Evans:2008zzb}
L.~Evans and P.~Bryant, {\em JINST} {\bf 3},   S08001  (2008).

\bibitem{Aad:2008zzm}
ATLAS Collaboration, G.~Aad {\em et~al.}, {\em JINST} {\bf 3},   S08003
  (2008).

\bibitem{Pinfold:2009oia}
MoEDAL Collaboration, J.~Pinfold {\em et~al.}, {\em {Technical Design Report of
  the MoEDAL Experiment}},
  $\!$~{\href{http://cds.cern.ch/record/1181486}{CERN-LHCC-2009-006}},
  MoEDAL-TDR-001  (2009).

\bibitem{Curie:1894}
P.~Curie, {\em J.Phys. Theor. Appl.} {\bf 3(1)}, 415 (Jan 1894).

\bibitem{hp1896crh}
H.~Poincar\'e, {\em Comptes rendus hebdomadaires de l'Acad\'emie des sciences
  de Paris} {\bf 123}, 530  (1896).

\bibitem{birkeland1}
K.~Birkeland, {\em Elektroteknisk Tidsskrift, Kristiania} {\bf 9}, 104  (1896).

\bibitem{birkeland2}
K.~Birkeland, {\em Elec. Rev.} {\bf 38}, 752, 782  (1896).

\bibitem{thomson1904}
J.~J. Thomson, {\em The London, Edinburgh, and Dublin Philosophical Magazine
  and Journal of Science} {\bf 8}, 331  (1904).

\bibitem{Castelnovo:2007qi}
C.~Castelnovo, R.~Moessner and S.~L. Sondhi, {\em Nature} {\bf 451}, 42
  (2008), \href{http://arxiv.org/abs/0710.5515}{{\sffamily arXiv:0710.5515
  [cond-mat.str-el]}}.

\bibitem{Bender:2014xva}
C.~M. Bender, M.~DeKieviet and K.~A. Milton  (2014),
  \href{http://arxiv.org/abs/1408.4051}{{\sffamily arXiv:1408.4051 [hep-th]}}.

\bibitem{Weinberg:1965rz}
S.~Weinberg, {\em Phys. Rev.} {\bf 138}, B988  (1965).

\bibitem{Cabibbo:1962td}
N.~Cabibbo and E.~Ferrari, {\em Nuovo Cim.} {\bf 23}, 1147  (1962).

\bibitem{Salam:1966bd}
A.~Salam, {\em Phys. Lett.} {\bf 22}, 683  (1966).

\bibitem{Zwanziger:1970hk}
D.~Zwanziger, {\em Phys. Rev. D} {\bf 3},   880  (1971).

\bibitem{Singleton:1995cc}
D.~Singleton, {\em Int. J. Theor. Phys.} {\bf 35}, 2419  (1996),
  \href{http://arxiv.org/abs/hep-th/9509157}{{\sffamily arXiv:hep-th/9509157}}.

\bibitem{Singleton:2011ru}
D.~Singleton, {\em Am. J. Phys.} {\bf 64}, 452  (1996),
  \href{http://arxiv.org/abs/1106.1505}{{\sffamily arXiv:1106.1505 [hep-th]}}.

\bibitem{Wu:1975es}
T.~T. Wu and C.~N. Yang, {\em Phys. Rev. D} {\bf 12}, 3845  (1975).

\bibitem{Wu:1976ge}
T.~T. Wu and C.~N. Yang, {\em Nucl. Phys. B} {\bf 107},   365  (1976).

\bibitem{Wu:1976qk}
T.~T. Wu and C.~N. Yang, {\em Phys. Rev. D} {\bf 14}, 437  (1976).

\bibitem{Lanyi:1977kp}
G.~Lanyi and R.~Pappas, {\em Phys. Lett. B} {\bf 68}, 436  (1977).

\bibitem{Constantinidis:2016wjo}
C.~Constantinidis, L.~Ferreira and G.~Luchini, {\em J. Phys. A} {\bf 52},
  155202  (2019), \href{http://arxiv.org/abs/1611.07041}{{\sffamily
  arXiv:1611.07041 [hep-th]}}.

\bibitem{Ferreira:2012aj}
L.~Ferreira and G.~Luchini, {\em Phys. Rev. D} {\bf 86},   085039  (2012),
  \href{http://arxiv.org/abs/1205.2088}{{\sffamily arXiv:1205.2088 [hep-th]}}.

\bibitem{Ferreira:2011ed}
L.~Ferreira and G.~Luchini, {\em Nucl. Phys. B} {\bf 858}, 336  (2012),
  \href{http://arxiv.org/abs/1109.2606}{{\sffamily arXiv:1109.2606 [hep-th]}}.

\bibitem{Mazur:1990ak}
P.~O. Mazur and J.~Papavassiliou, {\em Phys. Rev. D} {\bf 44}, 1317  (1991).

\bibitem{Ren:1993hr}
H.~Ren, {\em Phys. Lett. B} {\bf 325}, 149  (1994),
  \href{http://arxiv.org/abs/hep-th/9312074}{{\sffamily arXiv:hep-th/9312074
  [hep-th]}}.

\bibitem{BezerradeMello:1996si}
E.~R. Bezerra~de Mello and C.~Furtado, {\em Phys. Rev. D} {\bf 56}, 1345
  (1997).

\bibitem{RoderiguesSobreira:1998tb}
A.~A. Roderigues~Sobreira and E.~R. Bezerra~de Mello, {\em Grav. Cosmol.} {\bf
  5}, 177  (1999), \href{http://arxiv.org/abs/hep-th/9809212}{{\sffamily
  arXiv:hep-th/9809212 [hep-th]}}.

\bibitem{Lousto:1991rh}
C.~O. Lousto, {\em Class. Quant. Grav.} {\bf 9}, 2417  (1992).

\bibitem{Mavromatos:2017qeb}
N.~E. Mavromatos and J.~Papavassiliou, {\em Eur. Phys. J. C} {\bf 78},  ~68
  (2018), \href{http://arxiv.org/abs/1712.03395}{{\sffamily arXiv:1712.03395
  [hep-ph]}}.

\bibitem{Georgi:1974sy}
H.~Georgi and S.~Glashow, {\em Phys. Rev. Lett.} {\bf 32}, 438  (1974).

\bibitem{Halpern:1978ik}
M.~B. Halpern, {\em Phys. Rev. D} {\bf 19},   517  (1979).

\bibitem{Rubakov:1981rg}
V.~A. Rubakov, {\em JETP Lett.} {\bf 33}, 644  (1981), [Pisma Zh. Eksp. Teor.
  Fiz.33,658(1981)].

\bibitem{Rubakov:1983sy}
V.~A. Rubakov and M.~S. Serebryakov, {\em Nucl. Phys. B} {\bf 218}, 240
  (1983).

\bibitem{Callan:1982ac}
C.~G. Callan, Jr., {\em Nucl. Phys. B} {\bf 212}, 391  (1983).

\bibitem{Preskill:1984re}
J.~Preskill, {\it {Monopoles in 1983}}, in {\em {MONOPOLE '83. Proceedings,
  NATO Advanced Research Workshop, Ann Arbor, USA, Oct 6-9, 1983}\/},  {\em
  NATO Sci. Ser. B} {\bf 111} (1984).

\bibitem{Dokos:1979vu}
C.~P. Dokos and T.~N. Tomaras, {\em Phys. Rev. D} {\bf 21},   2940  (1980).

\bibitem{Blaer:1981ps}
A.~S. Blaer, N.~H. Christ and J.-F. Tang, {\em Phys. Rev. Lett.} {\bf 47},
  1364  (1981).

\bibitem{Wilczek:1981du}
F.~Wilczek, {\em Phys. Rev. Lett.} {\bf 48}, 1144  (1982).

\bibitem{Witten:1979ey}
E.~Witten, {\em Phys. Lett. B} {\bf 86}, 283  (1979).

\bibitem{grattoni}
{Grattoni, Christopher}, {\it {The Riemann Sphere as a Stereographic
  Projection}}
  \url{http://demonstrations.wolfram.com/TheRiemannSphereAsAStereographicProjection/},
   (2015), {Wolfram Demonstrations Project}.

\bibitem{Yang:2001bb}
Y.~Yang, {\em {Solitons in field theory and nonlinear analysis}} (Springer, New
  York, USA, 2001).

\bibitem{Cho:2019vzo}
Y.~Cho, {\em Phil. Trans. Roy. Soc. Lond. A} {\bf 377},   20190038  (2019).

\bibitem{TheATLASandCMSCollaborations:2015bln}
ATLAS and CMS Collaborations, {\em {Measurements of the Higgs boson production
  and decay rates and constraints on its couplings from a combined ATLAS and
  CMS analysis of the LHC pp collision data at $\sqrt{s}$ = 7 and 8~TeV}},
  $\!$~{\href{http://cds.cern.ch/record/2052552}{ATLAS-CONF-2015-044}}  (2015).

\bibitem{Ellis:2014jta}
J.~Ellis, V.~Sanz and T.~You, {\em JHEP} {\bf 03},   157  (2015),
  \href{http://arxiv.org/abs/1410.7703}{{\sffamily arXiv:1410.7703 [hep-ph]}}.

\bibitem{Ellis:2014dva}
J.~Ellis, V.~Sanz and T.~You, {\em JHEP} {\bf 07},   036  (2014),
  \href{http://arxiv.org/abs/1404.3667}{{\sffamily arXiv:1404.3667 [hep-ph]}}.

\bibitem{Cho:2016npz}
Y.~M. Cho, K.~Kimm and J.~H. Yoon, {\em Phys. Lett. B} {\bf 761}, 203  (2016),
  \href{http://arxiv.org/abs/1605.08129}{{\sffamily arXiv:1605.08129
  [hep-th]}}.

\bibitem{Mavromatos:2018kcd}
N.~E. Mavromatos and S.~Sarkar, {\em Universe} {\bf 5},  ~8  (2018),
  \href{http://arxiv.org/abs/1812.00495}{{\sffamily arXiv:1812.00495
  [hep-ph]}}.

\bibitem{Aaboud:2017bwk}
ATLAS Collaboration, M.~Aaboud {\em et~al.}, {\em Nature Phys.} {\bf 13}, 852
  (2017), \href{http://arxiv.org/abs/1702.01625}{{\sffamily arXiv:1702.01625
  [hep-ex]}}.

\bibitem{Sirunyan:2018fhl}
CMS Collaboration, A.~M. Sirunyan {\em et~al.}, {\em Phys. Lett. B} {\bf 797},
   134826  (2019), \href{http://arxiv.org/abs/1810.04602}{{\sffamily
  arXiv:1810.04602 [hep-ex]}}.

\bibitem{dEnterria:2013zqi}
D.~d'Enterria and G.~G. da~Silveira, {\em Phys.\ Rev.\ Lett.} {\bf 111},
  080405  (2013), \href{http://arxiv.org/abs/1305.7142}{{\sffamily
  arXiv:1305.7142 [hep-ph]}}, [Erratum: {\it Phys.\ Rev.\ Lett.} {\bf 116},
  129901 (2016)].

\bibitem{dEnterria:2016sca}
D.~d'Enterria, {\it {Physics at the FCC-ee}}, in {\em {17th Lomonosov
  Conference on Elementary Particle Physics}\/},  (2017).
\newblock pp. 182--191.
\newblock \href{http://arxiv.org/abs/1602.05043}{{\sffamily arXiv:1602.05043
  [hep-ex]}}.

\bibitem{Mangano:2017tke}
{\em CERN Yellow Rep. Monogr.} {\bf 3}  (2017),
  \href{http://arxiv.org/abs/1710.06353}{{\sffamily arXiv:1710.06353
  [hep-ph]}}.

\bibitem{Zarnecki:2020ics}
A.~F. Zarnecki, {\it {On the physics potential of ILC and CLIC}}, in {\em {19th
  Hellenic School and Workshops on Elementary Particle Physics and Gravity}\/},
   (2020).
\newblock \href{http://arxiv.org/abs/2004.14628}{{\sffamily arXiv:2004.14628
  [hep-ph]}}.

\bibitem{Polchinski:1998rq}
J.~Polchinski, {\em {String theory. Vol. 1: An introduction to the bosonic
  string }}Cambridge Monographs on Mathematical Physics (Cambridge University
  Press, 2007).

\bibitem{Polchinski:1998rr}
J.~Polchinski, {\em {String theory. Vol. 2: Superstring theory and beyond
  }}Cambridge Monographs on Mathematical Physics (Cambridge University Press,
  2007).

\bibitem{Das:2000ph}
P.~Das, P.~Jain and S.~Mukherji, {\em Int. J. Mod. Phys. A} {\bf 16}, 4011
  (2001), \href{http://arxiv.org/abs/hep-ph/0011279}{{\sffamily
  arXiv:hep-ph/0011279 [hep-ph]}}.

\bibitem{Kar:2000ct}
S.~Kar, P.~Majumdar, S.~SenGupta and A.~Sinha, {\em Eur. Phys. J. C} {\bf 23},
  357  (2002), \href{http://arxiv.org/abs/gr-qc/0006097}{{\sffamily
  arXiv:gr-qc/0006097 [gr-qc]}}.

\bibitem{Maity:2005ah}
D.~Maity, S.~SenGupta and S.~Sur, {\em Phys. Rev. D} {\bf 72},   066012
  (2005), \href{http://arxiv.org/abs/hep-th/0507210}{{\sffamily
  arXiv:hep-th/0507210 [hep-th]}}.

\bibitem{Alexandre:2008ew}
J.~Alexandre, N.~E. Mavromatos and D.~Tanner, {\em Phys. Rev. D} {\bf 78},
  066001  (2008), \href{http://arxiv.org/abs/0804.2353}{{\sffamily
  arXiv:0804.2353 [hep-th]}}.

\bibitem{Alexandre:2007hs}
J.~Alexandre, N.~E. Mavromatos and D.~Tanner, {\em New J. Phys.} {\bf 10},
  033033  (2008), \href{http://arxiv.org/abs/0708.1154}{{\sffamily
  arXiv:0708.1154 [hep-th]}}.

\bibitem{Barriola:1989hx}
M.~Barriola and A.~Vilenkin, {\em Phys. Rev. Lett.} {\bf 63},   341  (1989).

\bibitem{Harari:1990cz}
D.~Harari and C.~Lousto, {\em Phys. Rev. D} {\bf 42}, 2626  (1990).

\bibitem{Lopez-Eiguren:2017dmc}
A.~Lopez-Eiguren, J.~Lizarraga, M.~Hindmarsh and J.~Urrestilla, {\em JCAP} {\bf
  1707},   026  (2017), \href{http://arxiv.org/abs/1705.04154}{{\sffamily
  arXiv:1705.04154 [astro-ph.CO]}}.

\bibitem{Polchinski:1996na}
J.~Polchinski, {\it {Tasi lectures on D-branes}}, in {\em {Theoretical Advanced
  Study Institute in Elementary Particle Physics (TASI 96): Fields, Strings,
  and Duality}\/},  (1996).
\newblock pp. 293--356.
\newblock \href{http://arxiv.org/abs/hep-th/9611050}{{\sffamily
  arXiv:hep-th/9611050}}.

\bibitem{Li:2009tt}
T.~Li, N.~E. Mavromatos, D.~V. Nanopoulos and D.~Xie, {\em Phys. Lett. B} {\bf
  679}, 407  (2009), \href{http://arxiv.org/abs/0903.1303}{{\sffamily
  arXiv:0903.1303 [hep-th]}}.

\bibitem{Shiu:2003ta}
G.~Shiu and L.-T. Wang, {\em Phys. Rev. D} {\bf 69},   126007  (2004),
  \href{http://arxiv.org/abs/hep-ph/0311228}{{\sffamily arXiv:hep-ph/0311228
  [hep-ph]}}.

\bibitem{Ellis:2009vq}
J.~Ellis, N.~E. Mavromatos and D.~V. Nanopoulos, {\em Int. J. Mod. Phys. A}
  {\bf 26}, 2243  (2011), \href{http://arxiv.org/abs/0912.3428}{{\sffamily
  arXiv:0912.3428 [astro-ph.CO]}}.

\bibitem{Mavromatos:2010nk}
N.~E. Mavromatos, V.~A. Mitsou, S.~Sarkar and A.~Vergou, {\em Eur. Phys. J. C}
  {\bf 72},   1956  (2012), \href{http://arxiv.org/abs/1012.4094}{{\sffamily
  arXiv:1012.4094 [hep-ph]}}.

\bibitem{Mavromatos:2012ha}
N.~E. Mavromatos, M.~Sakellariadou and M.~F. Yusaf, {\em JCAP} {\bf 03},   015
  (2013), \href{http://arxiv.org/abs/1211.1726}{{\sffamily arXiv:1211.1726
  [hep-th]}}.

\bibitem{Elghozi:2015jka}
T.~Elghozi, N.~E. Mavromatos, M.~Sakellariadou and M.~F. Yusaf, {\em JCAP} {\bf
  1602},   060  (2016), \href{http://arxiv.org/abs/1512.03331}{{\sffamily
  arXiv:1512.03331 [hep-th]}}.

\bibitem{Kogan:1996zv}
I.~I. Kogan, N.~E. Mavromatos and J.~F. Wheater, {\em Phys. Lett. B} {\bf 387},
  483  (1996), \href{http://arxiv.org/abs/hep-th/9606102}{{\sffamily
  arXiv:hep-th/9606102}}.

\bibitem{Ellis:1999jf}
J.~R. Ellis, N.~Mavromatos and D.~V. Nanopoulos, {\em Phys. Rev. D} {\bf 61},
  027503  (2000), \href{http://arxiv.org/abs/gr-qc/9906029}{{\sffamily
  arXiv:gr-qc/9906029}}.

\bibitem{Mavromatos:2009pp}
N.~E. Mavromatos, {\em Found. Phys.} {\bf 40}, 917  (2010),
  \href{http://arxiv.org/abs/0906.2712}{{\sffamily arXiv:0906.2712 [hep-th]}}.

\bibitem{Bachas:1998rg}
C.~P. Bachas, {\it {Lectures on D-branes}}, in {\em {A Newton Institute
  Euroconference on Duality and Supersymmetric Theories}\/},  (1998).
\newblock pp. 414--473.
\newblock \href{http://arxiv.org/abs/hep-th/9806199}{{\sffamily
  arXiv:hep-th/9806199}}.

\bibitem{Deser:1997se}
S.~Deser, A.~Gomberoff, M.~Henneaux and C.~Teitelboim, {\em Nucl. Phys. B} {\bf
  520}, 179  (1998), \href{http://arxiv.org/abs/hep-th/9712189}{{\sffamily
  arXiv:hep-th/9712189}}.

\bibitem{Deser:1998vc}
S.~Deser, M.~Henneaux and A.~Schwimmer, {\em Phys. Lett. B} {\bf 428}, 284
  (1998), \href{http://arxiv.org/abs/hep-th/9803106}{{\sffamily
  arXiv:hep-th/9803106}}.

\bibitem{Sen:1999mg}
A.~Sen, {\it {NonBPS states and Branes in string theory}}, in {\em {Advanced
  School on Supersymmetry in the Theories of Fields, Strings and Branes}\/},
  (1999).
\newblock pp. 187--234.
\newblock \href{http://arxiv.org/abs/hep-th/9904207}{{\sffamily
  arXiv:hep-th/9904207}}.

\bibitem{Bertolini:1998mg}
M.~Bertolini, R.~Iengo and C.~A. Scrucca, {\em Nucl. Phys. B} {\bf 522}, 193
  (1998), \href{http://arxiv.org/abs/hep-th/9801110}{{\sffamily
  arXiv:hep-th/9801110}}.

\bibitem{Maldacena:2020skw}
J.~Maldacena  (2020), \href{http://arxiv.org/abs/2004.06084}{{\sffamily
  arXiv:2004.06084 [hep-th]}}.

\bibitem{Bai:2020spd}
Y.~Bai, J.~Berger, M.~Korwar and N.~Orlofsky (7 2020),
  \href{http://arxiv.org/abs/2007.03703}{{\sffamily arXiv:2007.03703
  [hep-ph]}}.

\bibitem{Turner:1982ag}
M.~S. Turner, E.~N. Parker and T.~J. Bogdan, {\em Phys. Rev. D} {\bf 26},
  1296  (1982).

\bibitem{Adams:1993fj}
F.~C. Adams, M.~Fatuzzo, K.~Freese, G.~Tarle, R.~Watkins and M.~S. Turner, {\em
  Phys. Rev. Lett.} {\bf 70}, 2511  (1993).

\bibitem{Virbhadra:2008ws}
K.~Virbhadra, {\em Phys. Rev. D} {\bf 79},   083004  (2009),
  \href{http://arxiv.org/abs/0810.2109}{{\sffamily arXiv:0810.2109 [gr-qc]}}.

\bibitem{Virbhadra:1999nm}
K.~Virbhadra and G.~F. Ellis, {\em Phys. Rev. D} {\bf 62},   084003  (2000),
  \href{http://arxiv.org/abs/astro-ph/9904193}{{\sffamily
  arXiv:astro-ph/9904193}}.

\bibitem{Hung:2020vuo}
P.~Hung  (2020), \href{http://arxiv.org/abs/2003.02794}{{\sffamily
  arXiv:2003.02794 [hep-ph]}}.

\bibitem{Deguchi:2019jtz}
S.~Deguchi and K.~Fujikawa, {\em Phys. Lett. B} {\bf 802},   135210  (2020),
  \href{http://arxiv.org/abs/1909.10916}{{\sffamily arXiv:1909.10916
  [hep-th]}}.

\bibitem{Gera:2020fvo}
S.~Gera and S.~Sengupta  (2020),
  \href{http://arxiv.org/abs/2004.13083}{{\sffamily arXiv:2004.13083 [gr-qc]}}.

\bibitem{Nieh:1981ww}
H.~Nieh and M.~Yan, {\em J. Math. Phys.} {\bf 23},   373  (1982).

\bibitem{Nieh:2007zz}
H.~Nieh and C.~Yang, {\em Int. J. Mod. Phys. A} {\bf 22}, 5237  (2007).

\bibitem{Chandia:1997hu}
O.~Chandia and J.~Zanelli, {\em Phys. Rev. D} {\bf 55},   7580  (1997),
  \href{http://arxiv.org/abs/hep-th/9702025}{{\sffamily arXiv:hep-th/9702025}}.

\bibitem{Chandia:1997jf}
O.~Chandia and J.~Zanelli, {\em AIP Conf. Proc.} {\bf 419}, 251  (1998),
  \href{http://arxiv.org/abs/hep-th/9708138}{{\sffamily arXiv:hep-th/9708138}}.

\bibitem{Bossingham:2018ivs}
T.~Bossingham, N.~E. Mavromatos and S.~Sarkar, {\em Eur. Phys. J. C} {\bf 79},
  ~50  (2019), \href{http://arxiv.org/abs/1810.13384}{{\sffamily
  arXiv:1810.13384 [hep-ph]}}.

\bibitem{Blaschke:2017pym}
F.~Blaschke and P.~{Bene\v s}, {\em PTEP} {\bf 2018},   073B03  (2018),
  \href{http://arxiv.org/abs/1711.04842}{{\sffamily arXiv:1711.04842
  [hep-th]}}.

\bibitem{Benes:2019ext}
P.~{Bene\v s} and F.~Blaschke, {\em J. Phys. Conf. Ser.} {\bf 1416},   012004
  (2019), \href{http://arxiv.org/abs/1912.09822}{{\sffamily arXiv:1912.09822
  [hep-ph]}}.

\bibitem{Teh:2014xva}
R.~Teh, B.-L. Ng and K.-M. Wong, {\em Annals Phys.} {\bf 354}, 489  (2015),
  \href{http://arxiv.org/abs/1406.0978}{{\sffamily arXiv:1406.0978 [hep-th]}}.

\bibitem{Teh:2013rpa}
R.~Teh, B.-L. Ng and K.-M. Wong, {\em Int. J. Mod. Phys. A} {\bf 28},   1350144
   (2013).

\bibitem{Pak:2013jaa}
D.~G. Pak, P.~M. Zhang and L.~P. Zou, {\em Int. J. Mod. Phys. A} {\bf 30},
  1550164  (2015), \href{http://arxiv.org/abs/1311.7567}{{\sffamily
  arXiv:1311.7567 [hep-th]}}.

\bibitem{Nambu:1977ag}
Y.~Nambu, {\em Nucl. Phys. B} {\bf 130},   505  (1977), [,329(1977)].

\bibitem{Sakai:2005sp}
N.~Sakai and D.~Tong, {\em JHEP} {\bf 03},   019  (2005),
  \href{http://arxiv.org/abs/hep-th/0501207}{{\sffamily arXiv:hep-th/0501207}}.

\bibitem{Tong:2005un}
D.~Tong, {\it {TASI lectures on solitons: Instantons, monopoles, vortices and
  kinks}}, in {\em {Theoretical Advanced Study Institute in Elementary Particle
  Physics}: {Many Dimensions of String Theory}\/},  (2005).
\newblock \href{http://arxiv.org/abs/hep-th/0509216}{{\sffamily
  arXiv:hep-th/0509216}}.

\bibitem{Martins:2008zz}
C.~Martins and A.~Achucarro, {\em Phys. Rev. D} {\bf 78},   083541  (2008).

\bibitem{Lopez-Eiguren:2016jsy}
A.~Lopez-Eiguren, J.~Urrestilla and A.~{Ach\'{u}carro}, {\em JCAP} {\bf 01},
  020  (2017), \href{http://arxiv.org/abs/1611.09628}{{\sffamily
  arXiv:1611.09628 [hep-ph]}}, [Erratum: {\it JCAP} {\bf 06}, E01 (2017)].

\bibitem{Achucarro:2019blr}
A.~Ach\'{u}carro, A.~Avgoustidis, A.~{L\'{o}pez}-Eiguren, C.~Martins and
  J.~Urrestilla, {\em Phil. Trans. Roy. Soc. Lond. A} {\bf 377},   0004
  (2019), \href{http://arxiv.org/abs/1912.12069}{{\sffamily arXiv:1912.12069
  [astro-ph.CO]}}.

\bibitem{Achucarro:1999it}
A.~Achucarro and T.~Vachaspati, {\em Phys. Rept.} {\bf 327}, 347  (2000),
  \href{http://arxiv.org/abs/hep-ph/9904229}{{\sffamily arXiv:hep-ph/9904229
  [hep-ph]}}.

\bibitem{Saurabh:2019rrp}
A.~Saurabh and T.~Vachaspati, {\em Phil. Trans. Roy. Soc. Lond. A} {\bf 377},
  20190143  (2019), \href{http://arxiv.org/abs/1904.02257}{{\sffamily
  arXiv:1904.02257 [hep-th]}}.

\bibitem{Hook:2017vyc}
A.~Hook and J.~Huang, {\em Phys. Rev. D} {\bf 96},   055010  (2017),
  \href{http://arxiv.org/abs/1705.01107}{{\sffamily arXiv:1705.01107
  [hep-ph]}}.

\bibitem{Chandra:2019dnf}
P.~P. Chandra, M.~Korwar and A.~M. Thalapillil, {\em Phys. Rev. D} {\bf 101},
  075028  (2020), \href{http://arxiv.org/abs/1909.12855}{{\sffamily
  arXiv:1909.12855 [hep-ph]}}.

\bibitem{Bellini:2020kub}
M.~Bellini, {\em Phys. Dark Univ.} {\bf 30},   100693  (2020),
  \href{http://arxiv.org/abs/2005.07260}{{\sffamily arXiv:2005.07260 [gr-qc]}}.

\bibitem{Fring:2020xpi}
A.~Fring and T.~Taira, {\em Phys. Lett. B} {\bf 807},   135583  (2020),
  \href{http://arxiv.org/abs/2006.02718}{{\sffamily arXiv:2006.02718
  [hep-th]}}.

\bibitem{Bender:2005tb}
C.~M. Bender, {\em Contemp. Phys.} {\bf 46}, 277  (2005),
  \href{http://arxiv.org/abs/quant-ph/0501052}{{\sffamily
  arXiv:quant-ph/0501052}}.

\bibitem{Bender:2005hf}
C.~M. Bender, H.~Jones and R.~Rivers, {\em Phys. Lett. B} {\bf 625}, 333
  (2005), \href{http://arxiv.org/abs/hep-th/0508105}{{\sffamily
  arXiv:hep-th/0508105}}.

\bibitem{Boulware:1976tv}
D.~G. Boulware, L.~S. Brown, R.~N. Cahn, S.~Ellis and C.-k. Lee, {\em Phys.
  Rev. D} {\bf 14},   2708  (1976).

\bibitem{Gamberg:1999hq}
L.~P. Gamberg and K.~A. Milton, {\em Phys. Rev. D} {\bf 61},   075013  (2000),
  \href{http://arxiv.org/abs/hep-ph/9910526}{{\sffamily arXiv:hep-ph/9910526}}.

\bibitem{Lechner:1999ga}
K.~Lechner and P.~Marchetti, {\em Nucl. Phys. B} {\bf 569}, 529  (2000),
  \href{http://arxiv.org/abs/hep-th/9906079}{{\sffamily arXiv:hep-th/9906079}}.

\bibitem{Schwinger:1976fr}
J.~S. Schwinger, K.~A. Milton, W.-y. Tsai, L.~L. DeRaad, Jr. and D.~C. Clark,
  {\em Annals Phys.} {\bf 101},   451  (1976), [,550(1976)].

\bibitem{Kurochkin:2006jr}
{\relax Yu}.~Kurochkin, I.~Satsunkevich, D.~Shoukavy, N.~Rusakovich and {\relax
  Yu}.~Kulchitsky, {\em Mod. Phys. Lett. A} {\bf 21}, 2873  (2006).

\bibitem{Dougall:2007tt}
T.~Dougall and S.~D. Wick, {\em Eur. Phys. J. A} {\bf 39}, 213  (2009),
  \href{http://arxiv.org/abs/0706.1042}{{\sffamily arXiv:0706.1042 [hep-ph]}}.

\bibitem{Epele:2012jn}
L.~N. Epele, H.~Fanchiotti, C.~A.~G. Canal, V.~A. Mitsou and V.~Vento, {\em
  Eur. Phys. J. Plus} {\bf 127},  ~60  (2012),
  \href{http://arxiv.org/abs/1205.6120}{{\sffamily arXiv:1205.6120 [hep-ph]}}.

\bibitem{Baines:2018ltl}
S.~Baines, N.~E. Mavromatos, V.~A. Mitsou, J.~L. Pinfold and A.~Santra, {\em
  Eur. Phys. J. C} {\bf 78},   966  (2018),
  \href{http://arxiv.org/abs/1808.08942}{{\sffamily arXiv:1808.08942
  [hep-ph]}}, [Erratum: {\it Eur. Phys. J. C} {\bf 79}, 166 (2019)].

\bibitem{Alexandre:2019iub}
J.~Alexandre and N.~E. Mavromatos, {\em Phys. Rev. D} {\bf 100},   096005
  (2019), \href{http://arxiv.org/abs/1906.08738}{{\sffamily arXiv:1906.08738
  [hep-ph]}}.

\bibitem{Terning:2018udc}
J.~Terning and C.~B. Verhaaren, {\em JHEP} {\bf 03},   177  (2019),
  \href{http://arxiv.org/abs/1809.05102}{{\sffamily arXiv:1809.05102
  [hep-th]}}.

\bibitem{Peskin:1995ev}
M.~E. Peskin and D.~V. Schroeder, {\em {An Introduction to quantum field
  theory}} (Addison-Wesley, Reading, USA, 1995).

\bibitem{Vento:2018sog}
V.~Vento, {\em Universe} {\bf 4},   117  (2018).

\bibitem{Vento:2019auh}
V.~Vento and M.~Traini, {\em Eur. Phys. J. C} {\bf 80},  ~62  (2020),
  \href{http://arxiv.org/abs/1909.03952}{{\sffamily arXiv:1909.03952
  [hep-ph]}}.

\bibitem{Schwinger:1966nj}
J.~S. Schwinger, {\em Phys. Rev.} {\bf 144}, 1087  (1966).

\bibitem{Schwinger:1966zza}
J.~Schwinger, {\em Phys. Rev.} {\bf 151}, 1048  (1966).

\bibitem{Schwinger:1966zzb}
J.~Schwinger, {\em Phys. Rev.} {\bf 151}, 1055  (1966).

\bibitem{Deans:1981qs}
W.~Deans, {\em Nucl. Phys. B} {\bf 197}, 307  (1982).

\bibitem{Panagiotakopoulos:1982fp}
C.~Panagiotakopoulos, {\em J. Phys. A} {\bf 16},   133  (1983).

\bibitem{Panagiotakopoulos:1982ne}
C.~Panagiotakopoulos, {\em Nucl. Phys. B} {\bf 212}, 118  (1983).

\bibitem{Coleman:1982cx}
S.~R. Coleman, {\it {The Magnetic Monopole Fifty Years Later}}, in {\em {Les
  Houches Summer School of Theoretical Physics: Laser-Plasma Interactions}\/},
  (1982).
\newblock pp. 461--552.

\bibitem{Calucci:1982wy}
G.~Calucci and R.~Jengo, {\em Nucl. Phys. B} {\bf 223}, 501  (1983).

\bibitem{Calucci:1982fm}
G.~Calucci, R.~Jengo and M.~Vallon, {\em Nucl. Phys. B} {\bf 211}, 77  (1983).

\bibitem{Goebel:1983we}
C.~Goebel and M.~Thomaz, {\em Phys. Rev. D} {\bf 30},   823  (1984).

\bibitem{Tolkachev:1992qx}
E.~Tolkachev and Y.~Shnir, {\em Sov. J. Nucl. Phys.} {\bf 55}, 1596  (1992).

\bibitem{Blagojevic:1985sh}
M.~Blagojevic and P.~Senjanovic, {\em Phys. Rept.} {\bf 157},   233  (1988).

\bibitem{Drukier:1981fq}
A.~K. Drukier and S.~Nussinov, {\em Phys. Rev. Lett.} {\bf 49},   102  (1982).

\bibitem{Schwinger:1951nm}
J.~S. Schwinger, {\em Phys. Rev.} {\bf 82}, 664  (1951).

\bibitem{Affleck:1981ag}
I.~K. Affleck and N.~S. Manton, {\em Nucl. Phys. B} {\bf 194}, 38  (1982).

\bibitem{Gould:2019myj}
O.~Gould, D.~L.~J. Ho and A.~Rajantie, {\em Phys. Rev. D} {\bf 100},   015041
  (2019), \href{http://arxiv.org/abs/1902.04388}{{\sffamily arXiv:1902.04388
  [hep-th]}}.

\bibitem{Ho:2019ads}
D.~L.~J. Ho and A.~Rajantie, {\em Phys. Rev. D} {\bf 101},   055003  (2020),
  \href{http://arxiv.org/abs/1911.06088}{{\sffamily arXiv:1911.06088
  [hep-th]}}.

\bibitem{Zeldovich:1978wj}
{\relax Ya}.~B. Zeldovich and M.~{\relax Yu}. Khlopov, {\em Phys. Lett.} {\bf
  79B}, 239  (1978).

\bibitem{Hill:1982iq}
C.~T. Hill, {\em Nucl. Phys. B} {\bf 224}, 469  (1983).

\bibitem{Dubrovich:2002gp}
V.~K. Dubrovich, {\em Grav. Cosmol. Suppl.} {\bf 8N1}, 122  (2002).

\bibitem{Vento:2007vy}
V.~Vento, {\em Int. J. Mod. Phys. A} {\bf 23}, 4023  (2008),
  \href{http://arxiv.org/abs/0709.0470}{{\sffamily arXiv:0709.0470
  [astro-ph]}}.

\bibitem{Epele:2007ic}
L.~N. Epele, H.~Fanchiotti, C.~A. Garcia~Canal and V.~Vento, {\em Eur. Phys. J.
  C} {\bf 56}, 87  (2008),
  \href{http://arxiv.org/abs/hep-ph/0701133}{{\sffamily arXiv:hep-ph/0701133
  [hep-ph]}}.

\bibitem{Epele:2008un}
L.~N. Epele, H.~Fanchiotti, C.~A.~G. Canal and V.~Vento, {\em Eur. Phys. J. C}
  {\bf 62}, 587  (2009), \href{http://arxiv.org/abs/0809.0272}{{\sffamily
  arXiv:0809.0272 [hep-ph]}}.

\bibitem{Barrie:2016wxf}
N.~D. Barrie, A.~Sugamoto and K.~Yamashita, {\em PTEP} {\bf 2016},   113B02
  (2016), \href{http://arxiv.org/abs/1607.03987}{{\sffamily arXiv:1607.03987
  [hep-ph]}}.

\bibitem{Reis:2017rvb}
J.~T. Reis and W.~K. Sauter, {\em Phys. Rev. D} {\bf 96},   075031  (2017),
  \href{http://arxiv.org/abs/1707.04170}{{\sffamily arXiv:1707.04170
  [hep-ph]}}.

\bibitem{Fanchiotti:2017nkk}
H.~Fanchiotti, C.~A. Garcia~Canal and V.~Vento, {\em Int. J. Mod. Phys. A} {\bf
  32},   1750202  (2017), \href{http://arxiv.org/abs/1703.06649}{{\sffamily
  arXiv:1703.06649 [hep-ph]}}.

\bibitem{Murayama:2009nj}
H.~Murayama and J.~Shu, {\em Phys. Lett. B} {\bf 686}, 162  (2010),
  \href{http://arxiv.org/abs/0905.1720}{{\sffamily arXiv:0905.1720 [hep-ph]}}.

\bibitem{Kibble:1976sj}
T.~Kibble, {\em J. Phys. A} {\bf 9}, 1387  (1976).

\bibitem{Zurek:1985qw}
W.~Zurek, {\em Nature} {\bf 317}, 505  (1985).

\bibitem{Stojkovic:2004hz}
D.~Stojkovic and K.~Freese, {\em Phys. Lett. B} {\bf 606}, 251  (2005),
  \href{http://arxiv.org/abs/hep-ph/0403248}{{\sffamily arXiv:hep-ph/0403248}}.

\bibitem{Dvali:1997sa}
G.~Dvali, H.~Liu and T.~Vachaspati, {\em Phys. Rev. Lett.} {\bf 80}, 2281
  (1998), \href{http://arxiv.org/abs/hep-ph/9710301}{{\sffamily
  arXiv:hep-ph/9710301}}.

\bibitem{Pogosian:1999zi}
L.~Pogosian and T.~Vachaspati, {\em Phys. Rev. D} {\bf 62},   105005  (2000),
  \href{http://arxiv.org/abs/hep-ph/9909543}{{\sffamily arXiv:hep-ph/9909543}}.

\bibitem{Pogosian:2000xv}
L.~Pogosian and T.~Vachaspati, {\em Phys. Rev. D} {\bf 62},   123506  (2000),
  \href{http://arxiv.org/abs/hep-ph/0007045}{{\sffamily arXiv:hep-ph/0007045}}.

\bibitem{Alexander:2000yx}
S.~Alexander, R.~H. Brandenberger and R.~Easther  (2000),
  \href{http://arxiv.org/abs/hep-ph/0008014}{{\sffamily arXiv:hep-ph/0008014}}.

\bibitem{Stojkovic:2005zh}
D.~Stojkovic, K.~Freese and G.~D. Starkman, {\em Phys. Rev. D} {\bf 72},
  045012  (2005), \href{http://arxiv.org/abs/hep-ph/0505026}{{\sffamily
  arXiv:hep-ph/0505026}}.

\bibitem{Dvali:1998pa}
G.~Dvali and S.~Tye, {\em Phys. Lett. B} {\bf 450}, 72  (1999),
  \href{http://arxiv.org/abs/hep-ph/9812483}{{\sffamily arXiv:hep-ph/9812483}}.

\bibitem{Burgess:2001fx}
C.~Burgess, M.~Majumdar, D.~Nolte, F.~Quevedo, G.~Rajesh and R.-J. Zhang, {\em
  JHEP} {\bf 07},   047  (2001),
  \href{http://arxiv.org/abs/hep-th/0105204}{{\sffamily arXiv:hep-th/0105204}}.

\bibitem{Dvali:2003zj}
G.~Dvali and A.~Vilenkin, {\em JCAP} {\bf 03},   010  (2004),
  \href{http://arxiv.org/abs/hep-th/0312007}{{\sffamily arXiv:hep-th/0312007}}.

\bibitem{Avgoustidis:2005vm}
A.~Avgoustidis and E.~Shellard, {\em JHEP} {\bf 08},   092  (2005),
  \href{http://arxiv.org/abs/hep-ph/0504049}{{\sffamily arXiv:hep-ph/0504049}}.

\bibitem{Lakes:2004rc}
R.~Lakes, {\em Phys. Lett. A} {\bf 329}, 298  (2004),
  \href{http://arxiv.org/abs/physics/0405148}{{\sffamily
  arXiv:physics/0405148}}.

\bibitem{Zhang:2019ona}
X.~Zhang and Y.~Yang, {\em Nucl. Phys. B} {\bf 951},   114851  (2020),
  \href{http://arxiv.org/abs/1909.03864}{{\sffamily arXiv:1909.03864
  [math-ph]}}.

\bibitem{Daido:2019tbm}
R.~Daido, S.-Y. Ho and F.~Takahashi, {\em JHEP} {\bf 01},   185  (2020),
  \href{http://arxiv.org/abs/1909.03627}{{\sffamily arXiv:1909.03627
  [hep-ph]}}.

\bibitem{Sato:2018nqy}
R.~Sato, F.~Takahashi and M.~Yamada, {\em Phys. Rev. D} {\bf 98},   043535
  (2018), \href{http://arxiv.org/abs/1805.10533}{{\sffamily arXiv:1805.10533
  [hep-ph]}}.

\bibitem{Baek:2013dwa}
S.~Baek, P.~Ko and W.-I. Park, {\em JCAP} {\bf 10},   067  (2014),
  \href{http://arxiv.org/abs/1311.1035}{{\sffamily arXiv:1311.1035 [hep-ph]}}.

\bibitem{Sousa:2009is}
L.~Sousa and P.~Avelino, {\em Phys. Lett. B} {\bf 689}, 145  (2010),
  \href{http://arxiv.org/abs/0911.3902}{{\sffamily arXiv:0911.3902
  [astro-ph.CO]}}.

\bibitem{Rahaman:2006kw}
F.~Rahaman, M.~Kalam, R.~Mondal and B.~Raychaudhuri, {\em Mod. Phys. Lett. A}
  {\bf 22}, 971  (2007), \href{http://arxiv.org/abs/gr-qc/0607125}{{\sffamily
  arXiv:gr-qc/0607125}}.

\bibitem{Terning:2018lsv}
J.~Terning and C.~B. Verhaaren, {\em JHEP} {\bf 12},   123  (2018),
  \href{http://arxiv.org/abs/1808.09459}{{\sffamily arXiv:1808.09459
  [hep-th]}}.

\bibitem{Tanabashi:2018oca}
Particle Data Group Collaboration, M.~Tanabashi {\em et~al.}, {\em Phys. Rev.
  D} {\bf 98},   030001  (2018).

\bibitem{Cecchini:2016vrw}
S.~Cecchini, L.~Patrizii, Z.~Sahnoun, G.~Sirri and V.~Togo  (2016),
  \href{http://arxiv.org/abs/1606.01220}{{\sffamily arXiv:1606.01220
  [physics.ins-det]}}.

\bibitem{Drell:1982zy}
S.~D. Drell, N.~M. Kroll, M.~T. Mueller, S.~J. Parke and M.~H. Ruderman, {\em
  Phys. Rev. Lett.} {\bf 50}, 644  (1983).

\bibitem{NIKEZIC200451}
D.~Nikezic and K.~Yu, {\em Materials Science and Engineering: R: Reports} {\bf
  46}, 51   (2004).

\bibitem{Cecchini:1995rw}
S.~Cecchini {\em et~al.}, {\em Nuovo Cim. A} {\bf 109}, 1119  (1996).

\bibitem{Patrizii:2010jla}
L.~Patrizii and Z.~Sahnoun, {\em Rev. Mex. Fis. Suppl} {\bf 56},  ~9  (2010).

\bibitem{Ostrovskiy:2014hfa}
I.~Ostrovskiy and J.~Pinfold  (2014),
  \href{http://arxiv.org/abs/1410.5521}{{\sffamily arXiv:1410.5521
  [physics.ins-det]}}.

\bibitem{Houston:2018rvz}
N.~Houston, T.~Li and C.~Sun, {\em JCAP} {\bf 1810},   034  (2018),
  \href{http://arxiv.org/abs/1803.02835}{{\sffamily arXiv:1803.02835
  [hep-ph]}}.

\bibitem{Cabrera:1982gz}
B.~Cabrera, {\em Phys. Rev. Lett.} {\bf 48}, 1378  (1982).

\bibitem{Huber:1990an}
M.~E. Huber, B.~Cabrera, M.~A. Taber and R.~D. Gardner, {\em Phys. Rev. Lett.}
  {\bf 64}, 835  (1990).

\bibitem{Caplin:1986kw}
A.~D. Caplin, M.~Hardiman, M.~Koratzinos and J.~C. Schouten, {\em Nature} {\bf
  321}, 402  (1986).

\bibitem{Price:1975zt}
P.~B. Price, E.~K. Shirk, W.~Z. Osborne and L.~S. Pinsky, {\em Phys. Rev.
  Lett.} {\bf 35}, 487  (1975).

\bibitem{Alvarez:1975gm}
L.~W. Alvarez, {\it {Analysis of a Reported Magnetic Monopole}}, in {\em {In
  *Goldhaber, A.S. (ed.), Trower, W.P. (ed.): Magnetic monopoles* 111-117. (In
  *Stanford 1975, Symposium on lepton and photon interactions at high energies*
  967-979) and Preprint - Alvarez, L.W. (rec.Sep.75) 23 p.B}\/},  (1975).

\bibitem{Fowler:1975db}
P.~H. Fowler, {\it {Have We Seen the Track of a Magnetic Monopole?}}, in {\em
  {14th International Cosmic Ray Conference (ICRC 1975) Munich, Germany, August
  15-29, 1975}\/},  (1975).
\newblock pp. 4049--4064.

\bibitem{Eberhard:1971re}
P.~H. Eberhard, R.~R. Ross, L.~W. Alvarez and R.~D. Watt, {\em Phys. Rev. D}
  {\bf 4},   3260  (1971).

\bibitem{Ross:1973it}
R.~R. Ross, P.~H. Eberhard, L.~W. Alvarez and R.~D. Watt, {\em Phys. Rev. D}
  {\bf 8},   698  (1973).

\bibitem{Kovalik:1986zz}
J.~M. Kovalik and J.~L. Kirschvink, {\em Phys. Rev. A} {\bf 33}, 1183  (1986).

\bibitem{Jeon:1995rf}
H.~Jeon and M.~J. Longo, {\em Phys. Rev. Lett.} {\bf 75}, 1443  (1995),
  \href{http://arxiv.org/abs/hep-ex/9508003}{{\sffamily arXiv:hep-ex/9508003
  [hep-ex]}}, [Erratum: {\it Phys. Rev. Lett.} {\bf 76}, 159 (1996)].

\bibitem{Ebisu:1986dw}
T.~Ebisu and T.~Watanabe, {\em Phys. Rev. D} {\bf 36}, 3359  (1987).

\bibitem{Fleischer:1969mj}
R.~Fleischer, I.~Jacobs, W.~Schwarz, P.~Price and H.~Goodell, {\em Phys. Rev.}
  {\bf 177}, 2029  (1969).

\bibitem{Fleischer:1970zy}
R.~Fleischer, H.~Hart, I.~Jacobs, P.~Price, W.~Schwarz and F.~Aumento, {\em
  Phys. Rev.} {\bf 184}, 1393  (1969).

\bibitem{Kolm:1971xb}
H.~H. Kolm, F.~Villa and A.~Odian, {\em Phys. Rev. D} {\bf 4},   1285  (1971).

\bibitem{Carrigan:1975bk}
J.~Carrigan, Richard~A., F.~Nezrick and B.~Strauss, {\em Phys. Rev. D} {\bf
  13},   1823  (1976).

\bibitem{Price:1983ax}
P.~B. Price, S.-l. Guo, S.~P. Ahlen and R.~L. Fleischer, {\em Phys. Rev. Lett.}
  {\bf 52},   1265  (1984).

\bibitem{Patrizii:2019eud}
L.~Patrizii, Z.~Sahnoun and V.~Togo, {\em Phil.\ Trans.\ Roy.\ Soc.\ Lond.\ A}
  {\bf 377},   20180328  (2019).

\bibitem{Spurio:2019oaq}
M.~Spurio, {\em Searches for Magnetic Monopoles and Other Stable Massive
  Particles}, in {\em Probing Particle Physics with Neutrino Telescopes\/},
  (World Scientific, Singapore), ch.~11, pp. 353--400.

\bibitem{Orito:1990ny}
S.~Orito {\em et~al.}, {\em Phys. Rev. Lett.} {\bf 66}, 1951  (1991).

\bibitem{Novoseltsev:2006mw}
{\relax Yu}.~F. Novoseltsev, M.~M. Boliev, A.~V. Butkevich, S.~P. Mikheev and
  V.~B. Petkov, {\em Nucl. Phys. Proc. Suppl.} {\bf 151}, 337  (2006).

\bibitem{Thron:1992ri}
Soudan-2 Collaboration, J.~L. Thron {\em et~al.}, {\em Phys. Rev. D} {\bf 46},
  4846  (1992).

\bibitem{Krishnaswamy:1984fu}
M.~R. Krishnaswamy, M.~G.~K. Menon, N.~K. Mondal, V.~S. Narasimham, B.~V.
  Sreekantan, Y.~Hayashi, N.~Ito, S.~Kawakami and S.~Miyake, {\em Phys. Lett.}
  {\bf 142B}, 99  (1984).

\bibitem{Balestra:2008ps}
S.~Balestra {\em et~al.}, {\em Eur. Phys. J. C} {\bf 55}, 57  (2008),
  \href{http://arxiv.org/abs/0801.4913}{{\sffamily arXiv:0801.4913 [hep-ex]}}.

\bibitem{Ambrosio:2002qq}
MACRO Collaboration, M.~Ambrosio {\em et~al.}, {\em Eur. Phys. J. C} {\bf 25},
  511  (2002), \href{http://arxiv.org/abs/hep-ex/0207020}{{\sffamily
  arXiv:hep-ex/0207020 [hep-ex]}}.

\bibitem{Ambrosio:2002mb}
MACRO Collaboration, M.~Ambrosio {\em et~al.}, {\em Nucl. Instrum. Meth. A}
  {\bf 486}, 663  (2002).

\bibitem{Ahlen:1993mg}
MACRO Collaboration, S.~Ahlen {\em et~al.}, {\em Phys. Rev. Lett.} {\bf 72},
  608  (1994).

\bibitem{Ambrosio:1997cd}
MACRO Collaboration, M.~Ambrosio {\em et~al.}, {\em Phys. Lett. B} {\bf 406},
  249  (1997).

\bibitem{Ambrosio:2001ai}
MACRO Collaboration, M.~Ambrosio {\em et~al.}, {\em Astropart. Phys.} {\bf 18},
  27  (2002), \href{http://arxiv.org/abs/hep-ex/0110083}{{\sffamily
  arXiv:hep-ex/0110083}}.

\bibitem{Fleischer:1971bd}
R.~Fleischer, H.~Hart, G.~Nichols and P.~Price, {\em Phys. Rev. D} {\bf 4}, 24
  (1971).

\bibitem{Bartlett:1981hy}
D.~F. Bartlett, D.~Soo, R.~L. Fleischer, H.~R. {Hart, Jr.} and
  A.~Mogro-Campero, {\em Phys. Rev. D} {\bf 24},   612  (1981).

\bibitem{Doke:1983if}
T.~Doke {\em et~al.}, {\em Phys. Lett. B} {\bf 129}, 370  (1983).

\bibitem{Battistoni:1983ka}
G.~Battistoni {\em et~al.}, {\em Phys. Lett. B} {\bf 133}, 454  (1983).

\bibitem{Wang:2015ery}
Z.~Wang, {Search for Magnetic Monopoles with the NO$\nu$A Far Detector}, PhD
  thesis, Virginia U., (2015).

\bibitem{Antipin:2007zz}
BAIKAL Collaboration, K.~Antipin {\em et~al.}, {\em Astropart. Phys.} {\bf 29},
  366  (2008).

\bibitem{Abbasi:2010zz}
R.~Abbasi {\em et~al.}, {\em Eur. Phys. J. C} {\bf 69}, 361  (2010).

\bibitem{AdrianMartinez:2011xr}
ANTARES Collaboration, S.~Adrian-Martinez {\em et~al.}, {\em Astropart. Phys.}
  {\bf 35}, 634  (2012), \href{http://arxiv.org/abs/1110.2656}{{\sffamily
  arXiv:1110.2656 [astro-ph.HE]}}.

\bibitem{Albert:2017fud}
ANTARES Collaboration, A.~Albert {\em et~al.}, {\em JHEP} {\bf 07},   054
  (2017), \href{http://arxiv.org/abs/1703.00424}{{\sffamily arXiv:1703.00424
  [astro-ph.HE]}}.

\bibitem{Abbasi:2012eda}
IceCube Collaboration, R.~Abbasi {\em et~al.}, {\em Phys. Rev. D} {\bf 87},
  022001  (2013), \href{http://arxiv.org/abs/1208.4861}{{\sffamily
  arXiv:1208.4861 [astro-ph.HE]}}.

\bibitem{Aartsen:2015exf}
IceCube Collaboration, M.~G. Aartsen {\em et~al.}, {\em Eur. Phys. J. C} {\bf
  76},   133  (2016), \href{http://arxiv.org/abs/1511.01350}{{\sffamily
  arXiv:1511.01350 [astro-ph.HE]}}.

\bibitem{ObertackePollmann:2016uvi}
IceCube Collaboration, A.~Obertacke~Pollmann, {\em EPJ Web Conf.} {\bf 164},
  07019  (2017), \href{http://arxiv.org/abs/1610.06397}{{\sffamily
  arXiv:1610.06397 [astro-ph.IM]}}.

\bibitem{Pollmann:2015nmo}
A.~Pollmann, {Search for mildly relativistic magnetic monopoles with IceCube},
  PhD thesis, Wuppertal U., (2015-11-04).

\bibitem{Niessen:2001jn}
P.~Niessen, {Search for relativistic magnetic monopoles with the AMANDA
  detector}, PhD thesis, Humboldt U., Berlin, (2001).

\bibitem{Aartsen:2014awd}
IceCube Collaboration, M.~G. Aartsen {\em et~al.}, {\em Eur. Phys. J. C} {\bf
  74},   2938  (2014), \href{http://arxiv.org/abs/1402.3460}{{\sffamily
  arXiv:1402.3460 [astro-ph.CO]}}, [Erratum: {\it Eur. Phys. J. C} {\bf 79},
  124 (2019)].

\bibitem{Hogan:2008sx}
D.~P. Hogan, D.~Z. Besson, J.~P. Ralston, I.~Kravchenko and D.~Seckel, {\em
  Phys. Rev. D} {\bf 78},   075031  (2008),
  \href{http://arxiv.org/abs/0806.2129}{{\sffamily arXiv:0806.2129
  [astro-ph]}}.

\bibitem{Detrixhe:2010xi}
ANITA-II Collaboration, M.~Detrixhe {\em et~al.}, {\em Phys. Rev. D} {\bf 83},
   023513  (2011), \href{http://arxiv.org/abs/1008.1282}{{\sffamily
  arXiv:1008.1282 [astro-ph.HE]}}.

\bibitem{Aab:2016poe}
Pierre Auger Collaboration, A.~Aab {\em et~al.}, {\em Phys. Rev. D} {\bf 94},
  082002  (2016), \href{http://arxiv.org/abs/1609.04451}{{\sffamily
  arXiv:1609.04451 [astro-ph.HE]}}.

\bibitem{Pollmann:2018ihz}
IceCube Collaboration, A.~Pollmann, {\em EPJ Web Conf.} {\bf 168},   04010
  (2018).

\bibitem{Bartelt:1986cv}
J.~E. Bartelt {\em et~al.}, {\em Phys. Rev. D} {\bf 36},   1990  (1987),
  [Erratum: {\it Phys. Rev. D} {\bf 40}, 1701 (1989)].

\bibitem{Ambrosio:2002qu}
MACRO Collaboration, M.~Ambrosio {\em et~al.}, {\em Eur. Phys. J. C} {\bf 26},
  163  (2002), \href{http://arxiv.org/abs/hep-ex/0207024}{{\sffamily
  arXiv:hep-ex/0207024 [hep-ex]}}.

\bibitem{BeckerSzendy:1994wb}
R.~Becker-Szendy {\em et~al.}, {\em Phys. Rev. D} {\bf 49}, 2169  (1994).

\bibitem{Balkanov:1997da}
Baikal Collaboration, V.~A. Balkanov {\em et~al.}, {\em Prog. Part. Nucl.
  Phys.} {\bf 40}, 391  (1998),
  \href{http://arxiv.org/abs/astro-ph/9801044}{{\sffamily
  arXiv:astro-ph/9801044 [astro-ph]}}.

\bibitem{Ueno:2012md}
Super-Kamiokande Collaboration, K.~Ueno {\em et~al.}, {\em Astropart. Phys.}
  {\bf 36}, 131  (2012), \href{http://arxiv.org/abs/1203.0940}{{\sffamily
  arXiv:1203.0940 [hep-ex]}}.

\bibitem{Lee:2018pag}
L.~Lee, C.~Ohm, A.~Soffer and T.-T. Yu, {\em Prog. Part. Nucl. Phys.} {\bf
  106}, 210  (2019), \href{http://arxiv.org/abs/1810.12602}{{\sffamily
  arXiv:1810.12602 [hep-ph]}}.

\bibitem{Alimena:2019zri}
J.~Alimena {\em et~al.}  (2019),
  \href{http://arxiv.org/abs/1903.04497}{{\sffamily arXiv:1903.04497
  [hep-ex]}}.

\bibitem{DeRujula:1994nf}
A.~De~Rujula, {\em Nucl. Phys. B} {\bf 435}, 257  (1995),
  \href{http://arxiv.org/abs/hep-th/9405191}{{\sffamily arXiv:hep-th/9405191
  [hep-th]}}.

\bibitem{Ginzburg:1982fk}
I.~Ginzburg and S.~Panfil, {\em Sov.\ J.\ Nucl.\ Phys.} {\bf 36},   850
  (1982).

\bibitem{Abbott:1998mw}
D0 Collaboration, B.~Abbott {\em et~al.}, {\em Phys. Rev. Lett.} {\bf 81}, 524
  (1998), \href{http://arxiv.org/abs/hep-ex/9803023}{{\sffamily
  arXiv:hep-ex/9803023 [hep-ex]}}.

\bibitem{Acciarri:1994gb}
L3 Collaboration, M.~Acciarri {\em et~al.}, {\em Phys. Lett. B} {\bf 345}, 609
  (1995).

\bibitem{Gamberg:1998xf}
L.~P. Gamberg, G.~R. Kalbfleisch and K.~A. Milton  (1998),
  \href{http://arxiv.org/abs/hep-ph/9805365}{{\sffamily arXiv:hep-ph/9805365
  [hep-ph]}}.

\bibitem{Ginzburg:1998vb}
I.~F. Ginzburg and A.~Schiller, {\em Phys. Rev. D} {\bf 57}, 6599  (1998),
  \href{http://arxiv.org/abs/hep-ph/9802310}{{\sffamily arXiv:hep-ph/9802310
  [hep-ph]}}.

\bibitem{Ginzburg:1999ej}
I.~F. Ginzburg and A.~Schiller, {\em Phys. Rev. D} {\bf 60},   075016  (1999),
  \href{http://arxiv.org/abs/hep-ph/9903314}{{\sffamily arXiv:hep-ph/9903314
  [hep-ph]}}.

\bibitem{Epele:2016wps}
L.~N. Epele, H.~Fanchiotti, C.~A.~G. Canal, V.~A. Mitsou and V.~Vento  (2016),
  \href{http://arxiv.org/abs/1607.05592}{{\sffamily arXiv:1607.05592
  [hep-ph]}}.

\bibitem{Harland-Lang:2018iur}
L.~A. Harland-Lang, V.~A. Khoze and M.~G. Ryskin, {\em Eur. Phys. J. C} {\bf
  79},  ~39  (2019), \href{http://arxiv.org/abs/1810.06567}{{\sffamily
  arXiv:1810.06567 [hep-ph]}}.

\bibitem{Bruce:2018yzs}
R.~Bruce {\em et~al.}  (2018),
  \href{http://arxiv.org/abs/1812.07688}{{\sffamily arXiv:1812.07688
  [hep-ph]}}.

\bibitem{Kazama:1976fm}
Y.~Kazama, C.~N. Yang and A.~S. Goldhaber, {\em Phys. Rev. D} {\bf 15}, 2287
  (1977).

\bibitem{Abulencia:2005hb}
CDF Collaboration, A.~Abulencia {\em et~al.}, {\em Phys. Rev. Lett.} {\bf 96},
   201801  (2006), \href{http://arxiv.org/abs/hep-ex/0509015}{{\sffamily
  arXiv:hep-ex/0509015 [hep-ex]}}.

\bibitem{Kalbfleisch:2000iz}
G.~R. Kalbfleisch, K.~A. Milton, M.~G. Strauss, L.~P. Gamberg, E.~H. Smith and
  W.~Luo, {\em Phys. Rev. Lett.} {\bf 85}, 5292  (2000),
  \href{http://arxiv.org/abs/hep-ex/0005005}{{\sffamily arXiv:hep-ex/0005005
  [hep-ex]}}.

\bibitem{Kalbfleisch:2003yt}
G.~R. Kalbfleisch, W.~Luo, K.~A. Milton, E.~H. Smith and M.~G. Strauss, {\em
  Phys. Rev. D} {\bf 69},   052002  (2004),
  \href{http://arxiv.org/abs/hep-ex/0306045}{{\sffamily arXiv:hep-ex/0306045
  [hep-ex]}}.

\bibitem{Mulhearn:2004kw}
M.~J. Mulhearn, {A Direct Search for Dirac Magnetic Monopoles}, PhD thesis,
  MIT, (2004).

\bibitem{Price:1987py}
P.~B. Price, G.-X. Ren and K.~Kinoshita, {\em Phys. Rev. Lett.} {\bf 59}, 2523
  (1987).

\bibitem{Price:1990in}
P.~Price, G.-R. Jing and K.~Kinoshita, {\em Phys. Rev. Lett.} {\bf 65}, 149
  (1990).

\bibitem{Bertani:1990tq}
M.~Bertani {\em et~al.}, {\em Europhys. Lett.} {\bf 12}, 613  (1990).

\bibitem{Hoffmann:1978mp}
H.~Hoffmann, G.~Kantardjian, S.~Di~Liberto, F.~Meddi, G.~Romano and G.~Rosa,
  {\em Lett. Nuovo Cim.} {\bf 23}, 357  (1978).

\bibitem{Aubert:1982zi}
B.~Aubert, P.~Musset, M.~Price and J.~P. Vialle, {\em Phys. Lett.} {\bf 120B},
   465  (1983).

\bibitem{Carrigan:1973mw}
R.~A. Carrigan, F.~A. Nezrick and B.~P. Strauss, {\em Phys. Rev. D} {\bf 8},
  3717  (1973).

\bibitem{Carrigan:1974un}
R.~A. Carrigan, Jr., F.~A. Nezrick and B.~P. Strauss, {\em Phys. Rev. D} {\bf
  10},   3867  (1974).

\bibitem{Carrigan:1977ku}
J.~Carrigan, Richard~A., B.~Strauss and G.~Giacomelli, {\em Phys. Rev. D} {\bf
  17},   1754  (1978).

\bibitem{Abbiendi:2007ab}
OPAL Collaboration, G.~Abbiendi {\em et~al.}, {\em Phys. Lett. B} {\bf 663}, 37
   (2008), \href{http://arxiv.org/abs/0707.0404}{{\sffamily arXiv:0707.0404
  [hep-ex]}}.

\bibitem{Pinfold:1993mq}
J.~L. Pinfold, R.~Du, K.~Kinoshita, B.~Lorazo, M.~Regimbald and B.~Price, {\em
  Phys. Lett. B} {\bf 316}, 407  (1993).

\bibitem{Kinoshita:1992wd}
K.~Kinoshita, R.~Du, G.~Giacomelli, L.~Patrizii, F.~Predieri, P.~Serra,
  M.~Spurio and J.~L. Pinfold, {\em Phys. Rev. D} {\bf 46}, R881  (1992).

\bibitem{Kinoshita:1982mv}
K.~Kinoshita, P.~Price and D.~Fryberger, {\em Phys. Rev. Lett.} {\bf 48},  ~77
  (1982).

\bibitem{Fryberger:1983fa}
D.~Fryberger, T.~Coan, K.~Kinoshita and P.~Price, {\em Phys. Rev. D} {\bf 29},
   1524  (1984).

\bibitem{Musset:1983ii}
P.~Musset, M.~Price and E.~Lohrmann, {\em Phys. Lett.} {\bf 128B}, 333  (1983).

\bibitem{Kinoshita:1988cn}
K.~Kinoshita, M.~Fujii, K.~Nakajima, P.~Price and S.~Tasaka, {\em Phys. Rev.
  Lett.} {\bf 60},   1610  (1988).

\bibitem{Kinoshita:1989cb}
K.~Kinoshita, M.~Fujii, K.~Nakajima, P.~B. Price and S.~Tasaka, {\em Phys.
  Lett. B} {\bf 228}, 543  (1989).

\bibitem{Gentile:1986sf}
CLEO Collaboration, T.~Gentile {\em et~al.}, {\em Phys. Rev. D} {\bf 35},
  1081  (1987).

\bibitem{Braunschweig:1988uc}
TASSO Collaboration, W.~Braunschweig {\em et~al.}, {\em Z. Phys. C} {\bf 38},
  543  (1988).

\bibitem{Aktas:2004qd}
H1 Collaboration, A.~Aktas {\em et~al.}, {\em Eur. Phys. J. C} {\bf 41}, 133
  (2005), \href{http://arxiv.org/abs/hep-ex/0501039}{{\sffamily
  arXiv:hep-ex/0501039 [hep-ex]}}.

\bibitem{Aad:2015zhl}
ATLAS and CMS Collaborations, G.~Aad {\em et~al.}, {\em Phys. Rev. Lett.} {\bf
  114},   191803  (2015), \href{http://arxiv.org/abs/1503.07589}{{\sffamily
  arXiv:1503.07589 [hep-ex]}}.

\bibitem{Akesson:2001su}
T.~Akesson {\em et~al.}, {\em Nucl. Instrum. Meth. A} {\bf 474}, 172  (2001).

\bibitem{Aaboud:2017odu}
ATLAS Collaboration, M.~Aaboud {\em et~al.}, {\em JINST} {\bf 12},   P05002
  (2017), \href{http://arxiv.org/abs/1702.06473}{{\sffamily arXiv:1702.06473
  [hep-ex]}}.

\bibitem{Aad:2012qi}
ATLAS Collaboration, G.~Aad {\em et~al.}, {\em Phys. Rev. Lett.} {\bf 109},
  261803  (2012), \href{http://arxiv.org/abs/1207.6411}{{\sffamily
  arXiv:1207.6411 [hep-ex]}}.

\bibitem{Aad:2015kta}
ATLAS Collaboration, G.~Aad {\em et~al.}, {\em Phys. Rev. D} {\bf 93},   052009
   (2016), \href{http://arxiv.org/abs/1509.08059}{{\sffamily arXiv:1509.08059
  [hep-ex]}}.

\bibitem{Aad:2019pfm}
ATLAS Collaboration, G.~Aad {\em et~al.}, {\em Phys. Rev. Lett.} {\bf 124},
  031802  (2020), \href{http://arxiv.org/abs/1905.10130}{{\sffamily
  arXiv:1905.10130 [hep-ex]}}.

\bibitem{Aad:2010ai}
ATLAS Collaboration, G.~Aad {\em et~al.}, {\em Eur. Phys. J. C} {\bf 70}, 723
  (2010), \href{http://arxiv.org/abs/0912.2642}{{\sffamily arXiv:0912.2642
  [physics.ins-det]}}.

\bibitem{DeRoeck:2011aa}
A.~De~Roeck, A.~Katre, P.~Mermod, D.~Milstead and T.~Sloan, {\em Eur. Phys. J.
  C} {\bf 72},   1985  (2012), \href{http://arxiv.org/abs/1112.2999}{{\sffamily
  arXiv:1112.2999 [hep-ph]}}.

\bibitem{Sakurai:2019bac}
K.~Sakurai, D.~Felea, J.~Mamuzic, N.~E. Mavromatos, V.~A. Mitsou, J.~L.
  Pinfold, R.~Ruiz~de Austri, A.~Santra and O.~Vives, {\it {SUSY discovery
  prospects with MoEDAL}}, in {\em {6th Symposium on Prospects in the Physics
  of Discrete Symmetries (DISCRETE 2018) Vienna, Austria, November 26-30,
  2018}\/},  (2019).
\newblock \href{http://arxiv.org/abs/1903.11022}{{\sffamily arXiv:1903.11022
  [hep-ph]}}.

\bibitem{Felea:2020cvf}
D.~Felea, J.~Mamuzic, R.~{Mase\l ek}, N.~E. Mavromatos, V.~A. Mitsou, J.~L.
  Pinfold, R.~Ruiz~de Austri, K.~Sakurai, A.~Santra and O.~Vives, {\em Eur.
  Phys. J. C} ,   to appear  (2020),
  \href{http://arxiv.org/abs/2001.05980}{{\sffamily arXiv:2001.05980
  [hep-ph]}}.

\bibitem{Acharya:2020uwc}
B.~Acharya, A.~De~Roeck, J.~Ellis, D.~Ghosh, R.~{Mase\l ek}, G.~Panizzo,
  J.~Pinfold, K.~Sakurai, A.~Shaa and A.~Wall  (2020),
  \href{http://arxiv.org/abs/2004.11305}{{\sffamily arXiv:2004.11305
  [hep-ph]}}.

\bibitem{DeRoeck:2012wua}
A.~De~Roeck, H.~P. {H\"{a}chler}, A.~M. Hirt, M.~D. Joergensen, A.~Katre,
  P.~Mermod, D.~Milstead and T.~Sloan, {\em Eur. Phys. J. C} {\bf 72},   2212
  (2012).

\bibitem{MoEDAL:2016jlb}
MoEDAL Collaboration, B.~Acharya {\em et~al.}, {\em JHEP} {\bf 08},   067
  (2016), \href{http://arxiv.org/abs/1604.06645}{{\sffamily arXiv:1604.06645
  [hep-ex]}}.

\bibitem{Acharya:2016ukt}
MoEDAL Collaboration, B.~Acharya {\em et~al.}, {\em Phys. Rev. Lett.} {\bf
  118},   061801  (2017), \href{http://arxiv.org/abs/1611.06817}{{\sffamily
  arXiv:1611.06817 [hep-ex]}}.

\bibitem{Acharya:2017cio}
MoEDAL Collaboration, B.~Acharya {\em et~al.}, {\em Phys. Lett. B} {\bf 782},
  510  (2018), \href{http://arxiv.org/abs/1712.09849}{{\sffamily
  arXiv:1712.09849 [hep-ex]}}.

\bibitem{Acharya:2019vtb}
MoEDAL Collaboration, B.~Acharya {\em et~al.}, {\em Phys. Rev. Lett.} {\bf
  123},   021802  (2019), \href{http://arxiv.org/abs/1903.08491}{{\sffamily
  arXiv:1903.08491 [hep-ex]}}.

\bibitem{Acharya:2020bed}
MoEDAL Collaboration, B.~Acharya {\em et~al.}  (2020),
  \href{http://arxiv.org/abs/2002.00861}{{\sffamily arXiv:2002.00861
  [hep-ex]}}.

\bibitem{Rajantie:2016paj}
A.~Rajantie, {\em Phys.\ Today} {\bf 69}, 40  (2016).

\bibitem{pdg}
{Particle Data Group}, {\it {The Review of Particle Physics}}
  \url{http://pdg.lbl.gov/},  (2019).

\bibitem{beampipe}
{ De~Roeck,~A., Mermod,~P., Pinfold.~J. and others}, {\it {Searching for
  trapped magnetic monopoles in LHC accelerator material}}
  \url{https://indico.cern.ch/event/623746/attachments/1427507/2190853/expression_of_interest_beam_pipe.pdf},
   (2017).

\bibitem{beampipe-cc}
{\em {CMS beam pipe to be mined for monopoles}}, \mbox{{\it CERN Courier }{\bf
  59(2)},
  8}~{\url{https://cerncourier.com/a/cms-beam-pipe-to-be-mined-for-monopoles/}}
   (2019).

\bibitem{LeBreton:2019lpq}
KM3NeT Collaboration, R.~Le~Breton, {\em Nucl. Instrum. Meth. A} {\bf 936}, 204
   (2019).

\bibitem{Blaufuss:2017wzl}
E.~Blaufuss and A.~Karle, {\em {A Next-Generation IceCube: The IceCube-Gen2
  Facility for High-Energy Neutrino Astronomy}} 2017, pp. 183--197.

\bibitem{Dash:2014fba}
N.~Dash, V.~M. Datar and G.~Majumder, {\em Astropart. Phys.} {\bf 70}, 33
  (2015), \href{http://arxiv.org/abs/1406.3938}{{\sffamily arXiv:1406.3938
  [physics.ins-det]}}.

\bibitem{Pinfold:2019zwp}
J.~L. Pinfold, {\em Phil. Trans. Roy. Soc. Lond. A} {\bf 377},   20190382
  (2019).

\end{thebibliography}

\end{document}